\pdfoutput=1
\documentclass[11pt,two-sided]{article}
\usepackage[mathscr]{eucal}
\usepackage{enumerate}
\usepackage{epsfig,amsfonts}
\usepackage{pifont}
\usepackage{amsmath}
\usepackage{amsthm,amssymb}
\usepackage{bm,bbm}
\usepackage{graphicx}
\usepackage{relsize}
\usepackage{hhline}
\usepackage{cite}
\usepackage{psfrag}
\usepackage{mathrsfs} 
\usepackage{bm}
\usepackage{array}
\usepackage{tikz}
\usetikzlibrary{arrows}
\usetikzlibrary{arrows.meta}
\usetikzlibrary{calc}
\usetikzlibrary{fadings}
\usepackage{textcomp}
\usepackage{hyperref}
\usepackage{mathrsfs} 
\usepackage{simplewick}
\usepackage{lettrine}
\usepackage{tcolorbox}
\tcbuselibrary{breakable}
\usepackage[abspath]{currfile}

\usepackage{hyperref}
\hypersetup{breaklinks=true,   
           colorlinks=true,
     linkcolor=blue,
     filecolor=blue,
     citecolor = blue,      
     urlcolor=blue,
} 

\DeclareMathAlphabet{\mathpzc}{OT1}{pzc}{m}{it}
\DeclareMathAlphabet\mathbfcal{OMS}{cmsy}{b}{n} 
\makeatletter
\@addtoreset{equation}{section}
\makeatother
\newcounter{app}
\newcounter{sapp}[app]
\def\theapp{\Alph{app}}
\newcommand{\app}[1]{
\refstepcounter{app}{\vspace{7mm}
\noindent\Large\bf Appendix
\theapp.
 \ #1 \par \vspace{5mm}}
\setcounter{equation}{0}
\def\theequation{\Alph{app}.\arabic{equation}}}
\def\thesapp{\Alph{app}.\arabic{sapp}}
\newcommand{\sapp}[1]{\par \refstepcounter{sapp}{\vspace{3mm}
\noindent\large\bf \thesapp\ 
#1 \par \vspace{3mm}}
\def\theequation{\Alph{app}.\arabic{equation}}}

\urldef{\pfaff}\url{https://www.biodiversitylibrary.org/item/113775#page/53/mode/1up}

\renewcommand{\theequation}{\thesection.\arabic{equation}}

\usepackage{cancel}
\newcommand{\ds}{\displaystyle}

\def\bea{\begin{eqnarray}}
\def\eea{\end{eqnarray}}
\def\be{\begin{equation}}
\def\ee{\end{equation}}
\def\beq{\begin{equation}}
\def\eeq{\end{equation}}
\def\hf{\frac{1}{2}}

\def\be{\begin{equation}}
\def\ee{\end{equation}}
\def\bdm{\begin{displaymath}}
\def\edm{\end{displaymath}}
\def\bea{\begin{eqnarray}}
\def\eea{\end{eqnarray}}

\def\pb{{\bar p}}
\def\zb{{\bar z}}

\def\XXint#1#2#3{{\setbox0=\hbox{$#1{#2#3}{\int}$}
    \vcenter{\hbox{$#2#3$}}\kern-.5\wd0}}

\newcommand{\re}{\mbox{e}}

\newcommand{\Ecal}{{\mathcal E}}

\newcommand{\Hcal}{{\mathcal H}}

\newcommand{\Tib}{{\mathbf T}}
\newcommand{\Tbb}{{\mathbb T}}
\newcommand{\Cbb}{{\mathbb C}}
\newcommand{\Rbb}{{\mathbb R}}
\newcommand{\Zbb}{{\mathbb Z}}

\newcommand{\Qib}{{\bf Q}}

\def\pd{\partial}

\def\Wc{{\mathcal W}}
\def\Sc{{\mathcal S}}
\def\Rc{{\mathcal R}}
\def\xb{{\boldsymbol x}}
\def\yb{{\boldsymbol y}}
\def\zb{{\boldsymbol z}}

\def\as{{\mathsf a}}
\def\bs{{\mathsf b}}
\def\cs{{\mathsf c}}
\def\dds{{\mathsf d}}
\def\es{{\mathsf e}}
\def\fs{{\mathsf f}}

\def\dla{\langle{\kern-.1em} \langle}
\def\dra{\rangle{\kern-.1em} \rangle}

\usepackage[font=small,labelfont=bf,format=hang]{caption}
\def\Fs{{\mathsf F}}
\def\xu{x}
\def\xd{x'}
\def\yr{y}
\def\yl{y'}
\def\GRC{lightgray}
\def\PAC{blue!60!black}
\def\STE{blue!60!black}
\def\GR{red!70!blue}
\begin{document}

\begin{titlepage}
\vglue 2 cm

\begin{center}
\begin{LARGE}

{\bf An Ising-type formulation of 
the six-vertex 
 model}
\end{LARGE}

\vspace{1.3cm}
\begin{large}

{\bf Vladimir V. Bazhanov$^{1}$ and Sergey M. Sergeev$^{1}$}

\end{large}

\vspace{1.cm}
$^1$ Department of Fundamental and Theoretical Physics,
         Research School of Physics,\\
    Australian National University, Canberra, ACT 2601, Australia\\\ \\

\vspace{.2cm}

\vspace{1cm}
\end{center}

\begin{center}

\parbox{14.9cm}{%
{\bf Abstract}. 
We show that the celebrated six-vertex model of statistical mechanics
(along with its multistate generalizations) can be reformulated as 
an Ising-type model with only a two-spin interaction. Such a reformulation
unravels remarkable factorization properties for row to row transfer
matrices, allowing one to uniformly derive all functional relations for their
eigenvalues and present the coordinate Bethe ansatz for the
eigenvectors for all higher spin generalizations of the six-vertex model. 
The possibility of the Ising-type formulation of these models raises
questions about the precedence of the traditional quantum group 
description of the vertex models. Indeed, the role of a
primary integrability condition is now played by the star-triangle
relation, which is not entirely natural in the standard quantum group setting, 
but implies the vertex-type Yang-Baxter equation and commutativity
of transfer matrices as simple corollaries. 
As a mathematical identity the emerging star-triangle
relation is equivalent to the Pfaff-Saalsch\"utz-Jackson summation formula, 
originally discovered by J.~F.~Pfaff in 1797. Plausibly, all
vertex models associated with quantized affine Lie algebras and superalgebras  
can be reformulated as Ising-type models with only two-spin interactions.}

\end{center}

\vskip2cm

\vfill

\end{titlepage}
\setcounter{page}{2}

\setcounter{tocdepth}{2}
\tableofcontents
\newpage
\section{Introduction}
There are various types of integrable 
models of statistical mechanics
\cite{Baxter:1982zz} where interacting ``spins'' are assigned to
different parts of the square lattice. These include the
\emph{Ising-type\/} (edge-interaction), \emph{vertex\/} and
\emph{iteraction-round-a-face\/} (IRF) models, where the Boltzmann
weights of local spin configurations are attributed to the edges,
vertices or faces of the lattice, respectively. The integrability of
these models requires the local Boltzmann weights to satisfy a
relevant type of the Yang-Baxter equation
\cite{McGuire:1964,Yang:1967,Baxter:1972}, which for the 
Ising-type models usually takes the form of Onsager's ``star-triangle
relation'' \cite{Onsager:1944}. 

The vertex type solutions of the Yang-Baxter equation (YBE), commonly called
$R$-matrices, are well understood. 
Their classification is based on the theory of Quantum Groups 
\cite{Drinfeld:1987,Jimbo:1986a,Faddeev:1987ih}. 
In particular, there are
infinite series of the so-called trigonometric solutions 
\cite{Cherednik:1980ey,Perk:1981,Zamolodchikov:1980ku,
IK81,Belavin:1982,Bazhanov:1984gu,Jimbo:1986,
Leites:1984pt,Bazhanov:1986av,
Kirillov:1987,
Khoroshkin:1991cmp,Warnaar:1992,
Delius:1994,Boos:2014,Chicherin:2014fqa},
connected with the evaluation representations of 
quantized affine Lie algebras. 
Similar considerations also apply to the corresponding IRF
models, which are closely related to the vertex models through 
the IRF-vertex correspondence \cite{Bax73a,Bax73b,Jimbo:1988}. 
Here we consider the case of the $U_q(\widehat{sl}(2))$ algebra,  
where $q$ is a deformation (or ``anisotropy'') parameter. 
The standard 
trigonometric $R$-matrix ${\cal R}(\lambda\,|\,s_1,s_2)$ in this case 
depends on a
spectral variable $\lambda$ and parameters $s_1$, $s_2$, which define two
highest weight representations of the algebra $U_q({sl(2)})$. The
associated lattice model is 
usually referred to as ``a higher spin generalization of the 6-vertex
model''. 

Contrary to this holistic ``quantum group'' picture, 
the algebraic nature of the star-triangle
relation and associated Ising-type models 
is much less clear. 
Of course, these models are connected to the vertex and IRF type models. 
For instance, there is a ``star-square'' transformation (see
\eqref{tworel} below) between the Ising-type and IRF models, which has
been applied to some specific models 
\cite{Jungling:1975,Baxter:1986df,Boos:1997}. 
However, these examples in no way suggest 
how to construct an Ising-type formulation for an arbitrary integrable
IRF model,
in particular, for the IRF version of the higher spin generalization of
the 6-vertex model,
mentioned above.

Fortunately, some missing pieces  
in understanding of relevant algebraic structures are coming from
the 3-dimensional interpretation
\cite{Bazhanov:1992jqa,Bazhanov:2005as}  
of the Yang-Baxter equation based on Zamolodchikov's tetrahedron equation 
\cite{Zamolodchikov:1980rus,Zamolodchikov:1981kf,Baxter:1983qc} and
properties of its solutions \cite{Baxter:1986phd,Sergeev:1995rt,
Sergeev:1999,Sergeev:2001,BMS08a,Mangazeev:2013spa}.
In fact, the present paper is
motivated by a remarkable result of Mangazeev
\cite{Mangazeev:2014gwa}, 
who used the above 3D
interpretation to construct a very simple representation for the
$U_q(\widehat{sl}(2))$ trigonometric 
$R$-matrix ${\cal R}(\lambda\,|\,s_1,s_2)$ expressing it via a
terminating basic hypergeometric series ${}_4\varphi_3$. We found that
this result allows one to construct the desired Ising-type formulation of the
higher spin 6-vertex model with arbitrary values of $\lambda$, $s_1$ and $s_2$.
In general, this corresponds to infinite-dimensional representations,
so that the lattice spins take arbitrary non-negative integer
values. However, when the parameters $2s_1$ and $2s_2$ take
non-negative integer values all equations admit
a finite-dimensional reduction, with the spins taking a 
finite number of values. 
In particular, when $s_1=s_2=\frac{1}{2}$ the model reduces
to the usual 6-vertex model with two spin states.

To illustrate the last point 
consider the standard vertex configurations of the usual 6-vertex
model, which are shown below in two ways: (i) by the edge arrangements 
(with bold and thin edges) and (ii) by the corresponding integer ``heights''
arrangements, $a\in{\mathbb Z}$, as in the unrestricted 
solid-on-solid (SOS) model,\footnote{%
The rules are: (a) two faces separated by the thin edge have the same
height, (b) the face height increases by $+1$ when crossing a bold
edge from right to left or from bottom to top. Note, that this trivial
vertex-SOS correspondence is different from that used in \cite{Bax73a,Bax73b}.}
\be\label{figsix}
\def\scl{1.0}
\begin{tikzpicture}[scale=\scl,baseline=(current  bounding  box.center)]
\draw [ultra thick, fill, black] (1.5,1.5) circle [radius=0.03];
\draw [thin] (1.5,0.5) -- (1.5,2.5); 
\draw [thin] (0.5,1.5) -- (2.5,1.5); 
\draw [ultra thick, fill,white] (1,2) circle [radius=0.1]; 
\node at (1.,2) {$a$};
\draw [ultra thick, fill,white] (2,2) circle [radius=0.1]; 
\node at (2,2.0) {$a$};
\draw [ultra thick, fill,white] (1,1) circle [radius=0.1]; 
\node at (1,1) {$a$};
\draw [ultra thick, fill,white] (2,1) circle [radius=0.1]; 
\node at (2.0,1.) {$a$};
\node[below] at (1.5,.4) {$\omega_1$}; 
\node at (2.8,1.5) {$\ \ \ $};
\end{tikzpicture}
\begin{tikzpicture}[scale=\scl,baseline=(current  bounding  box.center)]
\draw [ultra thick, fill, black] (1.5,1.5) circle [radius=0.03];
\draw [ultra thick] (1.5,0.5) -- (1.5,2.5); 
\draw [ultra thick] (0.5,1.5) -- (2.5,1.5); 
\draw [ultra thick, fill,white] (1,2) circle [radius=0.1]; 
\node at (.9,2) {$a+2$};
\draw [ultra thick, fill,white] (2,2) circle [radius=0.1]; 
\node at (2.1,2.0) {$a+1$};
\draw [ultra thick, fill,white] (1,1) circle [radius=0.1]; 
\node at (.9,1) {$a+1$};
\draw [ultra thick, fill,white] (2,1) circle [radius=0.1]; 
\node at (2.1,1.) {$a$};
\node[below] at (1.5,.4) {$\omega_2$}; 
\node at (2.8,1.5) {$\ \ \ $};
\end{tikzpicture}
\begin{tikzpicture}[scale=\scl,baseline=(current  bounding  box.center)]
\draw [ultra thick, fill, black] (1.5,1.5) circle [radius=0.03];
\draw [ultra thick] (1.5,0.5) -- (1.5,2.5); 
\draw [thin] (0.5,1.5) -- (2.5,1.5); 
\draw [ultra thick, fill,white] (1,2) circle [radius=0.1]; 
\node at (.9,2) {$a+1$};
\draw [ultra thick, fill,white] (2,2) circle [radius=0.1]; 
\node at (2,2.0) {$a$};
\draw [ultra thick, fill,white] (1,1) circle [radius=0.1]; 
\node at (.9,1) {$a+1$};
\draw [ultra thick, fill,white] (2,1) circle [radius=0.1]; 
\node at (2.0,1.) {$a$};
\node[below] at (1.5,.4) {$\omega_3$}; 
\node at (2.8,1.5) {$\ \ \ $};
\end{tikzpicture}
\begin{tikzpicture}[scale=\scl,baseline=(current  bounding  box.center)]
\draw [ultra thick, fill, black] (1.5,1.5) circle [radius=0.03];
\draw [thin] (1.5,0.5) -- (1.5,2.5); 
\draw [ultra thick] (0.5,1.5) -- (2.5,1.5); 
\draw [ultra thick, fill,white] (1,2) circle [radius=0.1]; 
\node at (.9,2) {$a+1$};
\draw [ultra thick, fill,white] (2,2) circle [radius=0.1]; 
\node at (2.1,2.0) {$a+1$};
\draw [ultra thick, fill,white] (1,1) circle [radius=0.1]; 
\node at (.9,1) {$a$};
\draw [ultra thick, fill,white] (2,1) circle [radius=0.1]; 
\node at (2.1,1.) {$a$};
\node[below] at (1.5,.4) {$\omega_4$}; 
\node at (2.8,1.5) {$\ \ \ $};
\end{tikzpicture}
\begin{tikzpicture}[scale=\scl,baseline=(current  bounding  box.center)]
\draw [ultra thick, fill, black] (1.5,1.5) circle [radius=0.03];
\draw [ultra thick] (1.5,0.5) -- (1.5,1.5); 
\draw [thin] (0.5,1.5) -- (1.5,1.5); 
\draw [thin] (1.5,1.5) -- (1.5,2.5); 
\draw [ultra thick] (1.5,1.5) -- (2.5,1.5); 
\draw [ultra thick, fill,white] (1,2) circle [radius=0.1]; 
\node at (.9,2) {$a+1$};
\draw [ultra thick, fill,white] (2,2) circle [radius=0.1]; 
\node at (2.1,2.) {$a+1$};
\draw [ultra thick, fill,white] (1,1) circle [radius=0.1]; 
\node at (.9,1) {$a+1$};
\draw [ultra thick, fill,white] (2,1) circle [radius=0.1]; 
\node at (2.0,1.) {$a$};
\node[below] at (1.5,.4) {$\omega_5$}; 
\node at (2.8,1.5) {$\ \ \ $};
\end{tikzpicture}
\begin{tikzpicture}[scale=\scl,baseline=(current  bounding  box.center)]
\draw [ultra thick, fill, black] (1.5,1.5) circle [radius=0.03];
\draw [thin] (1.5,0.5) -- (1.5,1.5); 
\draw [ultra thick] (0.5,1.5) -- (1.5,1.5); 
\draw [ultra thick] (1.5,1.5) -- (1.5,2.5); 
\draw [thin] (1.5,1.5) -- (2.5,1.5); 
 
\draw [ultra thick, fill,white] (1,2) circle [radius=0.1]; 
\node at (.9,2) {$a+1$};
\draw [ultra thick, fill,white] (2,2) circle [radius=0.1]; 
\node at (2.1,2.0) {$a$};
\draw [ultra thick, fill,white] (1,1) circle [radius=0.1]; 
\node at (.9,1) {$a$};
\draw [ultra thick, fill,white] (2,1) circle [radius=0.1]; 
\node at (2.1,1.) {$a$};
\node[below] at (1.5,.4) {$\omega_6$}; 
\end{tikzpicture}
\ee
where $\omega_1,\omega_2,\ldots,\omega_6$ denote the corresponding
Boltzmann weights. For all other configurations the weights are
assumed to vanish identically. Below we assume that 
$\omega_5=\omega_6$, noting that this does not really reduce generality, since
for periodic boundary conditions the weights $\omega_5$, $\omega_6$
always appear in pairs.  Now we replace the
lattice vertices
by the faces of the dual square lattice and go over from the SOS to the IRF
formulation of the model, as shown below,  
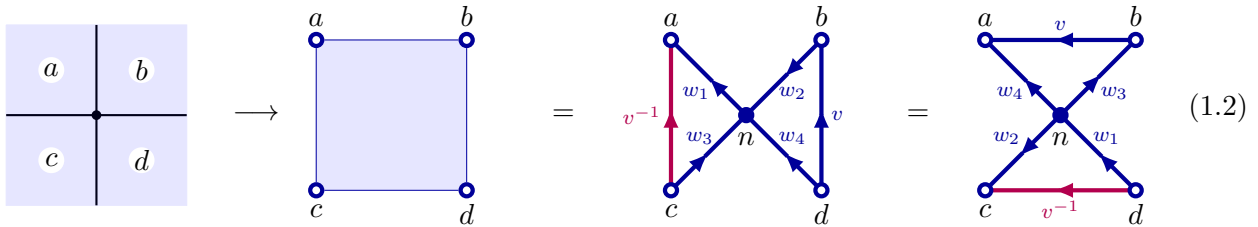
\be\label{star6v}
%
\begin{tikzpicture}[scale=1.2,baseline=(current  bounding  box.center)]
\draw [ultra thick, fill, black] (1.5,1.5) circle [radius=0.03];
\draw [thick] (1.5,0.5) -- (1.5,2.5); 
\draw [thick] (0.5,1.5) -- (2.5,1.5); 
\draw [fill, opacity=.1, blue] (0.5,2.5) -- (0.5,0.5) -- (2.5,0.5) --
(2.5,2.5) -- (0.5,2.5);
\draw [ultra thick, fill,white] (1,2) circle [radius=0.12]; 
\node at (1.,2) {$a$};
\draw [ultra thick, fill,white] (2,2) circle [radius=0.12]; 
\node at (2,2.0) {$b$};
\draw [ultra thick, fill,white] (1,1) circle [radius=0.12]; 
\node at (1,1) {$c$};
\draw [ultra thick, fill,white] (2,1) circle [radius=0.12]; 
\node at (2.0,1.) {$d$};
\node [right] at (2.9,1.5) {$\longrightarrow\ $};
\end{tikzpicture}
\begin{tikzpicture}[scale=1.0,baseline=(current  bounding  box.center)]
\draw [very thin, \STE] (0.5,2.5) -- (0.5,0.5) -- (2.5,0.5) --
(2.5,2.5) -- (0.5,2.5);
\draw [fill, opacity=.1, blue] (0.5,2.5) -- (0.5,0.5) -- (2.5,0.5) --
(2.5,2.5) -- (0.5,2.5);

\draw [ultra thick, fill,white] (0.5,2.5) circle [radius=0.08]; 
\draw [ultra thick, fill,white] (0.5,0.5) circle [radius=0.08];
\draw [ultra thick, fill,white] (2.5,2.5) circle [radius=0.08];
\draw [ultra thick, fill,white] (2.5,0.5) circle [radius=0.08];

\draw [ultra thick, \STE] (0.5,2.5) circle [radius=0.08]; \node[above] at (0.5,2.55) {$a$};
\draw [ultra thick, \STE] (0.5,0.5) circle [radius=0.08];\node[below] at (0.5,0.45) {$c$};
\draw [ultra thick, \STE] (2.5,2.5) circle [radius=0.08];\node[above] at (2.5,2.55) {$b$};
\draw [ultra thick, \STE] (2.5,0.5) circle [radius=0.08];\node[below] at (2.5,0.45) {$d$};
\node [right] at (3.5,1.5) {$=\quad$};
\end{tikzpicture}
%
%
\begin{tikzpicture}[scale=1.,baseline=(current  bounding  box.center)]
\draw [-latex, \PAC, ultra thick] (1.43,1.57) -- (1,2); \draw [\PAC, ultra thick] (1.05,1.95) -- (0.57,2.43);
\draw [-latex, \PAC, ultra thick] (2.43,2.43) -- (2,2); \draw [\PAC, ultra thick] (2.05,2.05) -- (1.57,1.57);
\draw [-latex, \PAC, ultra thick] (0.57,0.57) -- (1,1); \draw [\PAC, ultra thick] (.95,.95) -- (1.43,1.43);
\draw [-latex, \PAC, ultra thick] (2.43,0.57) -- (2.,1.); \draw [\PAC, ultra thick] (2.05,.95) -- (1.57,1.43);
\draw [-latex, \PAC, ultra thick] (2.5,0.58) -- (2.5,1.6); \draw [\PAC, ultra thick] (2.5,1.5) -- (2.5,2.42);
\draw [-latex, \GR, ultra thick] (0.5,0.58) -- (0.5,1.6); \draw [\GR, ultra thick] (0.5,1.5) -- (0.5,2.42);
\node[\PAC,left] at (1.15,1.8) {\scriptsize$w_1$};
\node[\PAC,left] at (1.2,1.2) {\scriptsize$w_3$};
\node[\PAC,right] at (1.8,1.2) {\scriptsize$w_4$};
\node[\PAC,right] at (1.8,1.8) {\scriptsize$w_2$};
\node [\PAC,right] at (2.5,1.5) {\scriptsize $v$};
\node [\GR,left] at (0.5,1.5) {\scriptsize $v^{-1}$};

\draw [ultra thick, \STE] (0.5,2.5) circle [radius=0.08]; \node[above] at (0.5,2.55) {$a$};
\draw [ultra thick, \STE] (0.5,0.5) circle [radius=0.08];\node[below] at (0.5,0.45) {$c$};
\draw [ultra thick, \STE] (2.5,2.5) circle [radius=0.08];\node[above] at (2.5,2.55) {$b$};
\draw [ultra thick, \STE] (2.5,0.5) circle [radius=0.08];\node[below] at (2.5,0.45) {$d$};
\draw [ultra thick, fill, \STE] (1.5,1.5) circle [radius=0.08];\node[below] at (1.5,1.4) {$n$};
\node [right] at (3.5,1.5) {$=\quad$};
\end{tikzpicture}
\begin{tikzpicture}[scale=1.,baseline=(current  bounding  box.center)]
\draw [-latex, \PAC, ultra thick] (1.43,1.57) -- (1,2); \draw [\PAC, ultra thick] (1.05,1.95) -- (0.57,2.43);
\draw [ \PAC, ultra thick] (2.43,2.43) -- (2,2); \draw [-latex,\PAC, ultra thick]  (1.57,1.57) --(2.05,2.05) ;
\draw [ \PAC, ultra thick] (0.57,0.57) -- (1,1); \draw [-latex,\PAC, ultra thick] (1.43,1.43)--(.95,.95);
\draw [-latex, \PAC, ultra thick] (2.43,0.57) -- (2.,1.); \draw [\PAC, ultra thick] (2.05,.95) -- (1.57,1.43);
\draw [-latex, \PAC, ultra thick] (2.42,2.5) -- (1.4,2.5); \draw [\PAC, ultra thick] (1.5,2.5) -- (0.58,2.5);
\draw [-latex, \GR, ultra thick] (2.42,0.5) -- (1.4,0.5); \draw [\GR, ultra thick] (1.5,0.5) -- (0.58,0.5);

\node [\PAC,above] at (1.5,2.5) {\scriptsize $v$};
\node [\GR,below] at (1.5,0.5) {\scriptsize $v^{-1}$};

\node[\PAC,left] at (1.15,1.8) {\scriptsize$w_4$};
\node[\PAC,left=1mm] at (1.2,1.2) {\scriptsize$w_2$};
\node[\PAC,right] at (1.8,1.2) {\scriptsize$w_1$};
\node[\PAC,right=1mm] at (1.8,1.8) {\scriptsize$w_3$};

\draw [ultra thick, \STE] (0.5,2.5) circle [radius=0.08]; \node[above] at (0.5,2.55) {$a$};
\draw [ultra thick, \STE] (0.5,0.5) circle [radius=0.08];\node[below] at (0.5,0.45) {$c$};
\draw [ultra thick, \STE] (2.5,2.5) circle [radius=0.08];\node[above] at (2.5,2.55) {$b$};
\draw [ultra thick, \STE] (2.5,0.5) circle [radius=0.08];\node[below] at (2.5,0.45) {$d$};
\draw [ultra thick, fill, \STE] (1.5,1.5) circle [radius=0.08];\node[below] at (1.5,1.4) {$n$};
\end{tikzpicture}   
\ee
It is not difficult to check that the resulting IRF weights can
be represented 
in two equivalent forms
\be\label{tworel}
\begin{array}{rcl}
\ds {\cal W}(a,b,c,d)&=&\ds
\omega_1\, \frac{v(b-d)}{v(a-c)}\sum_{n=\max(b,c)}^a\,w_1(a-n)\,w_2(n-b)\,
w_3(n-c)\,w_4(n-d)\\[.8cm]
&=&\;\ds
\omega_1\,
\frac{v(a-b)}{v(c-d)}\ \ \sum_{n=d}^{\min(b,c)}\;w_4(a-n)\,w_3(b-n)\, 
w_2(c-n)\,w_1(n-d)\,,
\end{array}
\ee
as Boltzmann weights of four-edge stars of an Ising-type model with four
different types of oriented diagonal edges.  The corresponding edge weights 
$w_i(a-b)$, where $i=1,2,3,4$, depend on the difference of
spins $a$ and $b$ at the ends of the edge
(with the arrow pointing from $b$ to $a$).
These weights vanish for negative arguments 
\be\label{wzero}
w_i(n)=0\,,\qquad n<0\,,\qquad i=1,2,3,4\,,
\ee
therefore both summations over the central spins in \eqref{star6v} 
are restricted to finite intervals. Moreover, there are weights
$v(a-b)$ and $v(a-b)^{-1}$ associated with horizontal and vertical
edges in \eqref{star6v}. These weights only enter into the equivalence
transformation factors in \eqref{tworel}, which  
cancel out in the partition function for periodic boundary conditions.

The only edge weights
that are required to reproduce \eqref{figsix} are given by
\be
\begin{array}{l}
w_1(0)=w_2(0)=w_3(0)=w_4(0)=1\,,\\[.5cm] 
w_1(1)=\epsilon \,
(\omega_5^2-\omega_3^{}\omega_4^{})/(\omega_1\omega_5)\,,\quad
w_2(1)=\epsilon\,\omega_3/\omega_5\,\quad w_3(1)=\epsilon\,\omega_4/\omega_5\,,\quad 
w_4(1)=\omega_5/(\epsilon \,\omega_1)\,,\\[.5cm]
w_4(2)= \omega_5^2\,
(\omega_1^{}\omega_2^{}+\omega_3^{}\omega_4^{}-\omega_5^2)/(\epsilon^{2}\,\omega_1^2\, \omega_3^{}\,\omega_4^{})\,,\qquad \epsilon=v(1)/v(0)\;.
\end{array}
\ee
Evidently, the weights ${\cal W}(a,b,c,d)$ only depend on spin
differences. For instance, if one sets $d=0$ then 
the central spin $n$ 
in \eqref{tworel} could only take three values $0,1,2$. Finally note,
that Eq.\eqref{tworel} relating the Ising-type and IRF models is
usually called the ``star-square transformation'', while the second
line in \eqref{tworel} relating the Boltzmann weights of two different
four-edge stars is called the ``star-star relation''.

Let us now return to the general higher spin 6-vertex
model, discussed above. Its transformation to an 
Ising-type model follows exactly the same steps. First one replaces the
vertex model by a SOS and then by an IRF model, as explained
above. 
Note that the edge indices in the vertex model (as well as the face
spin differences across the edge) can now take arbitrary non-negative
integer values. 
Next, the resulting IRF weights are represented as
Boltzmann weights of the first four-edge star in \eqref{tworel} 
with certain edge weights $w_i(a-b)$ and $v(a-b)$ 
defined for all non-negative 
spin differences $a-b\in {\mathbb Z}_{\ge0}$ 
but having the same vanishing conditions \eqref{wzero}. As a result
the range of summation over the central spin is finite and 
determined by the same
inequalities as in \eqref{tworel}.
Essentially, this representation is equivalent to the Mangazeev's formula
for the $R$-matrix ${\cal R}(\lambda\,|\,s_1,s_2)$ for the higher spin
6-vertex model.

The existence of the Ising-type formulation
for this case
raises questions about the primacy of the quantum group structures in
the description of vertex models. It suggests, that 
the Yang-Baxter equations for 
${\cal R}(\lambda\,|\,s_1,s_2)$ may now be derived from
simpler relations imposed on the edge weights. Indeed, we show that
it is a corollary of the star-star relation, which has the same form
\eqref{tworel} as in the two-state
case. It is presented in Sect.\ref{star-section}, see Fig.~\ref{sspic} therein.
The proof of this relation
is based on the second Sears's transformation formula
\cite{Sears:1951} for the basic
hypergeometric series ${}_4\varphi_3$.

The star-star relation as a primary integrablity condition has
appeared previously \cite{Bazhanov:1992jqa} in the context of 
the $sl(n)$-generalized
chiral Potts model \cite{BKMS,Date:1990bs}. 
For all other known integrable Ising-type models the star-star
relation could also be derived (see e.g., \cite{Baxter:1986df}), 
but it is a corollary of a much simpler star-triangle relation.
Therefore, even though we already have
a proof of the star-star
relation, mentioned above, it is
important to understand whether there are simpler integrability
conditions in the 6-vertex model and its generalizations, 
in particular, is there a star-triangle relation? 

Starting with Mangazeev's
${}_4\varphi_3$ formula for the $R$-matrix it was natural to first conduct the
search in the area of the basic hypergeometric series, especially because
it is extremely well documented, see, e.g., \cite{Andrews:1999,
Gasper:2004,Gasper:1995}. A review of this literature has eventually 
revealed that the required star-triangle relation can be derived from
1910 Jackson's $q$-analog \cite{Jackson:1910} 
of the Pfaff-Saalsch\"utz summation formula, 
originally
discovered by Johann Friedrich
Pfaff in 1797 \cite{Pfaff:1797}\footnote{%
The paper by Pfaff was part of a 1793 collection of the ``Analytical
Observations to L.Euler's Institute of Integral Calculus'' which was
published under the heading ``History 1793'' 
in ``New Transactions of Imperial Academy of Sciences in
St.Petersburg'', volume {\bf 11}. This volume is listed as published in 1793,
but Pfaff's paper is marked as ``Presented to Academy 14 January
1797''. This is probably the reason of a frequent 
volume-year mismatch in referencing
this paper, though after two hundred years the exact publication date
might not be that important.}
 and rediscovered by 
Saalsch\"utz in 1890 \cite{Saalschutz:1890}. An interesting feature of
this star-triangle relation is that it does not have an apparent
``difference property'', i.e., some of its edge weights depend on two
spectral variables rather than their ratio (we use multiplicative 
spectral variables). Nevertheless, we show that this relation implies 
the star-star relation and the Yang-Baxter equation and therefore  
does, indeed, play the role of a primary integrability condition
in the higher spin generalization of the 6-vertex model. 

The organization of the paper is as follows. 
In Sect.~2 we formulate
a new inhomogeneous Ising-type model on the square lattice. 
In Sect.~3,  we present an equivalent IRF model and discuss its
integrability conditions, including the Yang-Baxter equation,
the star-star and the star-triangle relations. Connection to the
6-vertex model and its higher spin generalization is discussed in
Sect.~4, where the most important finite-dimensional
reductions of the model are also studied. 
Factorized $R$-matrices are considered in
Sect.~5. Next we derive all functional relations for transfer 
matrices and Baxter's ${\bf Q}$-operators (Sect.6) and present the
Bethe ansatz solution (Sect 7.) for a general
inhomogeneous model. The column homogeneous case is considered in
Sect.~8. Some generalizations and extensions of the results to other
models are discussed in Sect.~9. In the Conclusion we briefly summarize
a few important aspects of our work and mention some possible applications. 

\section{An Ising-type model\label{isingtype}}
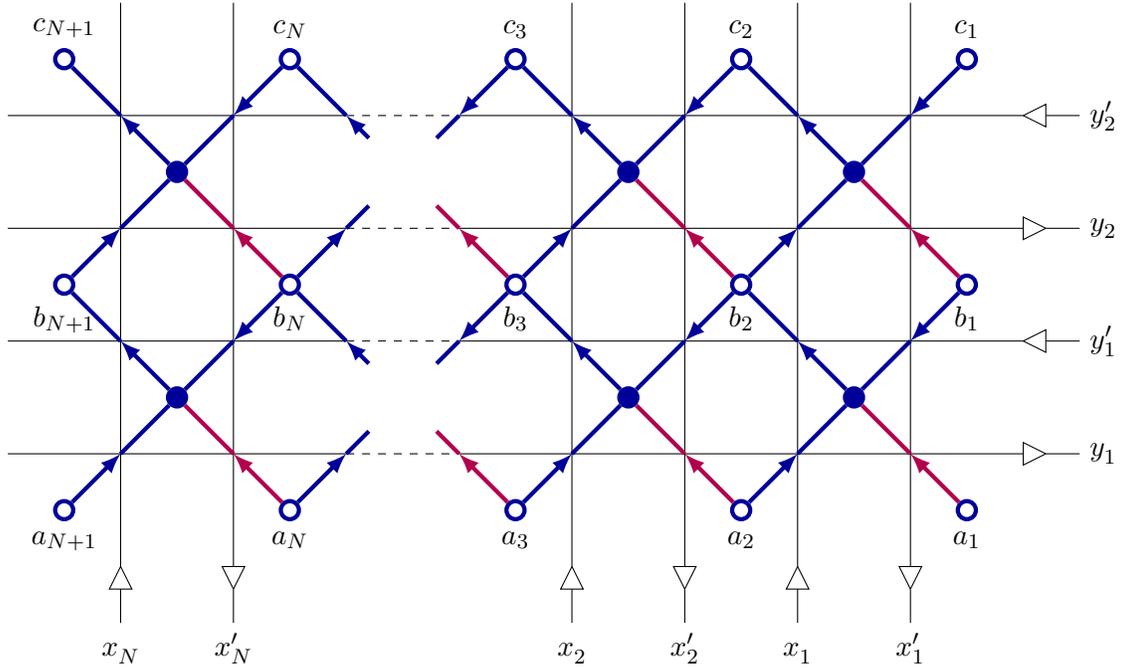
\begin{figure}[ht]
\def\GRC{lightgray}
\def\PAC{blue!60!black}
\def\STE{blue!60!black}
\def\GR{red!70!blue}
\begin{center}
\begin{tikzpicture}[scale=1.5]
\draw [thin] (0,1) -- (3,1); \draw [thin, dashed] (3,1) -- (4,1); \draw [thin] (4,1) -- (9,1); 
\draw [thin] (9.2,1) -- (9,1.1) -- (9.0,0.9) -- (9.2,1); 
\draw [thin] (9.2,1) -- (9.5,1); \node [right] at (9.5,1) {$y_1^{}$};
\draw [thin] (0,2) -- (3,2); \draw [thin, dashed] (3,2) -- (4,2); \draw [thin] (4,2) -- (9,2); 
\draw [thin] (9,2) -- (9.2,2.1) -- (9.2,1.9) -- (9,2); 
\draw [thin] (9.2,2) -- (9.5,2); \node [right] at (9.5,2) {$y_1'$};
\draw [thin] (0,3) -- (3,3); \draw [thin, dashed] (3,3) -- (4,3); \draw [thin] (4,3) -- (9,3); 
\draw [thin] (9.2,3) -- (9,3.1) -- (9.0,2.9) -- (9.2,3); 
\draw [thin] (9.2,3) -- (9.5,3); \node [right] at (9.5,3) {$y_2^{}$};
\draw [thin] (0,4) -- (3,4); \draw [thin, dashed] (3,4) -- (4,4); \draw [thin] (4,4) -- (9,4); 
\draw [thin] (9,4) -- (9.2,4.1) -- (9.2,3.9) -- (9,4); 
\draw [thin] (9.2,4) -- (9.5,4); \node [right] at (9.5,4) {$y_2'$};
\draw [thin] (1,0) -- (1,5); \draw [thin] (1,0) -- (1.1,-0.2) -- (0.9,-0.2) -- (1,0); \draw [thin] (1,-0.2) -- (1,-0.5); 
\node [below] at (1,-0.5) {$x_N^{\phantom{\prime}}$};
\draw [thin] (2,0) -- (2,5); \draw [thin] (2,-0.2) -- (2.1,0) -- (1.9,0) -- (2,-0.2); \draw [thin] (2,-0.2) -- (2,-0.5); 
\node [below] at (2,-0.5) {$x_N'$};
\draw [thin] (5,0) -- (5,5); \draw [thin] (5,0) -- (5.1,-0.2) -- (4.9,-0.2) -- (5,0); \draw [thin] (5,-0.2) -- (5,-0.5); 
\node [below] at (5,-0.5) {$x_2^{\phantom{\prime}}$};
\draw [thin] (6,0) -- (6,5); \draw [thin] (6,-0.2) -- (6.1,0) -- (5.9,0) -- (6,-0.2); \draw [thin] (6,-0.2) -- (6,-0.5); 
\node [below] at (6,-0.5) {$x_2'$};
\draw [thin] (7,0) -- (7,5); \draw [thin] (7,0) -- (7.1,-0.2) -- (6.9,-0.2) -- (7,0); \draw [thin] (7,-0.2) -- (7,-0.5); 
\node [below] at (7,-0.5) {$x_1^{\phantom{\prime}}$};
\draw [thin] (8,0) -- (8,5); \draw [thin] (8,-0.2) -- (8.1,0) -- (7.9,0) -- (8,-0.2); \draw [thin] (8,-0.2) -- (8,-0.5); 
\node [below] at (8,-0.5) {$x_1'$};
\draw [\STE, ultra thick] (0.5,0.5) circle [radius=0.08];
\draw [\STE, ultra thick] (2.5,0.5) circle [radius=0.08];
\draw [\STE, ultra thick] (4.5,0.5) circle [radius=0.08];
\draw [\STE, ultra thick] (6.5,0.5) circle [radius=0.08];
\draw [\STE, ultra thick] (8.5,0.5) circle [radius=0.08];
\draw [\STE, ultra thick] (0.5,2.5) circle [radius=0.08];
\draw [\STE, ultra thick] (2.5,2.5) circle [radius=0.08];
\draw [\STE, ultra thick] (4.5,2.5) circle [radius=0.08];
\draw [\STE, ultra thick] (6.5,2.5) circle [radius=0.08];
\draw [\STE, ultra thick] (8.5,2.5) circle [radius=0.08];
\draw [\STE, ultra thick] (0.5,4.5) circle [radius=0.08];
\draw [\STE, ultra thick] (2.5,4.5) circle [radius=0.08];
\draw [\STE, ultra thick] (4.5,4.5) circle [radius=0.08];
\draw [\STE, ultra thick] (6.5,4.5) circle [radius=0.08];
\draw [\STE, ultra thick] (8.5,4.5) circle [radius=0.08];
\draw [\STE, fill, ultra thick] (1.5,1.5) circle [radius=0.08];
\draw [\STE, fill, ultra thick] (5.5,1.5) circle [radius=0.08];
\draw [\STE, fill, ultra thick] (7.5,1.5) circle [radius=0.08];
\draw [\STE, fill, ultra thick] (1.5,3.5) circle [radius=0.08];
\draw [\STE, fill, ultra thick] (5.5,3.5) circle [radius=0.08];
\draw [\STE, fill, ultra thick] (7.5,3.5) circle [radius=0.08];
\draw [-latex, \PAC, ultra thick] (0.57,0.57) -- (1,1); \draw [\PAC, ultra thick] (1,1) -- (1.43,1.43);
\draw [-latex, \GR, ultra thick] (2.43,0.57) -- (2,1); \draw [\GR, ultra thick] (2,1) -- (1.57,1.43);
\draw [-latex, \PAC, ultra thick] (1.43,1.57) -- (1,2); \draw [\PAC, ultra thick] (1,2) -- (0.57,2.43);
\draw [-latex, \PAC, ultra thick] (2.43,2.43) -- (2,2); \draw [\PAC, ultra thick] (2,2) -- (1.57,1.57);
\draw [-latex, \PAC, ultra thick] (0.57,2.57) -- (1,3); \draw [\PAC, ultra thick] (1,3) -- (1.43,3.43);
\draw [-latex, \GR, ultra thick] (2.43,2.57) -- (2,3); \draw [\GR, ultra thick] (2,3) -- (1.57,3.43);
\draw [-latex, \PAC, ultra thick] (1.43,3.57) -- (1,4); \draw [\PAC, ultra thick] (1,4) -- (0.57,4.43);
\draw [-latex, \PAC, ultra thick] (2.43,4.43) -- (2,4); \draw [\PAC, ultra thick] (2,4) -- (1.57,3.57);
\draw [-latex, \PAC, ultra thick] (4.57,0.57) -- (5,1); \draw [\PAC, ultra thick] (5,1) -- (5.43,1.43);
\draw [-latex, \GR, ultra thick] (6.43,0.57) -- (6,1); \draw [\GR, ultra thick] (6,1) -- (5.57,1.43);
\draw [-latex, \PAC, ultra thick] (5.43,1.57) -- (5,2); \draw [\PAC, ultra thick] (5,2) -- (4.57,2.43);
\draw [-latex, \PAC, ultra thick] (6.43,2.43) -- (6,2); \draw [\PAC, ultra thick] (6,2) -- (5.57,1.57);
\draw [-latex, \PAC, ultra thick] (4.57,2.57) -- (5,3); \draw [\PAC, ultra thick] (5,3) -- (5.43,3.43);
\draw [-latex, \GR, ultra thick] (6.43,2.57) -- (6,3); \draw [\GR, ultra thick] (6,3) -- (5.57,3.43);
\draw [-latex, \PAC, ultra thick] (5.43,3.57) -- (5,4); \draw [\PAC, ultra thick] (5,4) -- (4.57,4.43);
\draw [-latex, \PAC, ultra thick] (6.43,4.43) -- (6,4); \draw [\PAC, ultra thick] (6,4) -- (5.57,3.57);
\draw [-latex, \PAC, ultra thick] (6.57,0.57) -- (7,1); \draw [\PAC, ultra thick] (7,1) -- (7.43,1.43);
\draw [-latex, \GR, ultra thick] (8.43,0.57) -- (8,1); \draw [\GR, ultra thick] (8,1) -- (7.57,1.43);
\draw [-latex, \PAC, ultra thick] (7.43,1.57) -- (7,2); \draw [\PAC, ultra thick] (7,2) -- (6.57,2.43);
\draw [-latex, \PAC, ultra thick] (8.43,2.43) -- (8,2); \draw [\PAC, ultra thick] (8,2) -- (7.57,1.57);
\draw [-latex, \PAC, ultra thick] (6.57,2.57) -- (7,3); \draw [\PAC, ultra thick] (7,3) -- (7.43,3.43);
\draw [-latex, \GR, ultra thick] (8.43,2.57) -- (8,3); \draw [\GR, ultra thick] (8,3) -- (7.57,3.43);
\draw [-latex, \PAC, ultra thick] (7.43,3.57) -- (7,4); \draw [\PAC, ultra thick] (7,4) -- (6.57,4.43);
\draw [-latex, \PAC, ultra thick] (8.43,4.43) -- (8,4); \draw [\PAC, ultra thick] (8,4) -- (7.57,3.57);
\draw [-latex, \PAC, ultra thick] (2.57,0.57) -- (3,1); \draw [\PAC, ultra thick] (3,1) -- (3.2,1.2); 
\draw [-latex, \GR, ultra thick] (4.43,0.57) -- (4,1); \draw [\GR, ultra thick] (4,1) -- (3.8,1.2);
\draw [-latex, \PAC, ultra thick] (3.2,1.8) -- (3,2); \draw [\PAC, ultra thick] (3,2) -- (2.57,2.43); 
\draw [-latex, \PAC, ultra thick] (4.43,2.43) -- (4,2); \draw [\PAC, ultra thick] (4,2) -- (3.8,1.8);
\draw [-latex, \PAC, ultra thick] (2.57,2.57) -- (3,3); \draw [\PAC, ultra thick] (3,3) -- (3.2,3.2); 
\draw [-latex, \GR, ultra thick] (4.43,2.57) -- (4,3); \draw [\GR, ultra thick] (4,3) -- (3.8,3.2);
\draw [-latex, \PAC, ultra thick] (3.2,3.8) -- (3,4); \draw [\PAC, ultra thick] (3,4) -- (2.57,4.43); 
\draw [-latex, \PAC, ultra thick] (4.43,4.43) -- (4,4); \draw [\PAC, ultra thick] (4,4) -- (3.8,3.8);
%
%
\node [below] at (8.5,0.4) {$a_1$};
\node [below] at (6.5,0.4) {$a_2$};
\node [below] at (4.5,0.4) {$a_3$};
\node [below] at (2.5,0.4) {$a_N$};
\node [below] at (0.5,0.4) {$a_{N+1}$};
\node [below] at (8.5,2.4) {$b_1$};
\node [below] at (6.5,2.4) {$b_2$};
\node [below] at (4.5,2.4) {$b_3$};
\node [below] at (2.5,2.4) {$b_N$};
\node [below] at (0.5,2.4) {$b_{N+1}$};
\node [above] at (8.5,4.6) {$c_1$};
\node [above] at (6.5,4.6) {$c_2$};
\node [above] at (4.5,4.6) {$c_3$};
\node [above] at (2.5,4.6) {$c_N$};
\node [above] at (0.5,4.6) {$c_{N+1}$};
\end{tikzpicture}
\end{center}
%
\caption{\small 
An oriented square lattice and its medial lattice shown by thick and
thin lines, respectively. All edges and lines are oriented as
indicated by arrows. There are two types of edges distinguished by relative 
orientation of the edge and the associated thin lines, passing through
the edge, shown in \eqref{twotypes}.
\label{fig-lattice}}
\end{figure}

Here we introduce a new two-dimensional solvable edge-interaction  
model. The model can be formulated on  rather general planar
graphs, however, for the purposes of this presentation it is convenient to
take a regular square lattice. 
Consider an oriented square lattice 
drawn diagonally as in
Fig.~\ref{fig-lattice}. The edges of the lattice 
are shown by thick lines and the sites 
are shown with either open or filled circles in a
checkerboard order. We will refer to the latter as to the 
``white'' or ``black'' 
sites, respectively.
The edges of the lattice are oriented as indicated by arrows, 
namely, all the SE-NW edges
have the same (SE-NW) direction, while the SW-NE edges are oriented 
in a checkerboard order, always pointing towards the black sites.   

\bigskip

The thin vertical and horizontal
lines in Fig.~\ref{fig-lattice} represent the {\em medial}\/ graph 
whose vertices lie on the edges of the original square lattice. 
The lines are directed as shown by arrows. Each of these lines
carries its own parameter (in general complex) which we call 
the ``rapidity'' or  
``spectral variable''. In general, these variables may all be
different for different lines as illustrated in Fig.~\ref{fig-lattice}.
The minimal level of
generality corresponds to the homogeneous case 
when one assigns the same variable $\xu$ to
all lines directed upwards (i.e., when 
$\xu_1=\xu_2=\ldots=\xu_N\equiv\xu$) 
and the same variable $\xd$ to all lines directed
downwards (i.e., $\xd_1=\xd_2=\ldots=\xd_N\equiv\xd$). 
Similarly, one assigns the same variables $\yr$ and $\yl$ to all
thin horizontal lines directed right and left, respectively. 

At each lattice site place an integer spin variable $a\in {\mathbb Z}$. 
Two spins interact only if they are connected with an edge. 
There are two types
of edges distinguished by relative orientation of the edge and the
associated thin lines, passing through the edge,
\begin{equation}
\begin{tikzpicture}
\node [left] at (-0.5,1) {$({i}):$};
\draw [-open triangle 45, black, thin] (0,0) -- (2,2);
\draw [-open triangle 45, black, thin] (2,0) -- (0,2);
\draw [\STE, ultra thick] (1,0) circle [radius=0.09];
\draw [\STE, ultra thick] (1,2) circle [radius=0.09];
\draw [-latex, \PAC, ultra thick] (1,0.12) -- (1,1);
\draw [\PAC, ultra thick] (1,1) -- (1,1.88);
\node [above] at (2,2) {$x$};
\node [above] at (0,2) {$y$};
\node [above] at (1,2.1) {$a$};
\node [below] at (1,-0.1) {$b$};
\node [right] at (2.1,1) {$
\ds =\;V_{x/y}(a-b)\;,
$};
\end{tikzpicture}
\qquad
\begin{tikzpicture}
\node [left] at (-0.5,1) {$({ii}):$};
\draw [-open triangle 45, black, thin] (2,2) -- (0,0);
\draw [-open triangle 45, black, thin] (0,2) -- (2,0);
\draw [\STE, ultra thick] (1,0) circle [radius=0.09];
\draw [\STE, ultra thick] (1,2) circle [radius=0.09];
\draw [-latex, \GR, ultra thick] (1,0.12) -- (1,1); 
\draw [\GR, ultra thick] (1,1) -- (1,1.88);
\node [below] at (2,0) {$x$};
\node [below] at (0,0) {$y$};
\node [above] at (1,2.1) {$a$};
\node [below] at (1,-0.1) {$b$};
\node [right] at (2.1,1) {$
\ds =\;\frac{1}{V_{x/y}(a-b)}\;.
$};
\end{tikzpicture}
\label{twotypes}
\end{equation}
The corresponding Boltzmann weights $V_{x/y}(a-b)$ and
$\big(V_{x/y}(a-b)\big)^{-1}$ depend on the difference of
spins $a$ and $b$ at the ends of the edge 
and on the ratio of two thin line variables $x$ and $y$. 
The edge arrow pointing from $b$ to $a$ 
shows that the argument of the weight function is
$(a-b)$ rather than $(b-a)$. 
For an easier perception the edges are also 
distinguished by their color: the blue
color 
for the edges of type $(i)$ and the red color for those of the type $(ii)$.

The function $V_x(n)$ is defined as 
\be
V_x(n)=\left(\frac{q}{x}\right)^n\,\frac{(x^2;q^2)_n}{(q^2;q^2)_n}\,,\label{Vdef}
\qquad n\in {\mathbb Z}\,,
\ee
where $n$ is an integer, $n\in {\mathbb Z}$,
\be\label{qprod0}
(x;q^2)_n=\prod_{j=0}^{n-1}(1-x\, q^{2j})\,,\qquad
(x;q^2)_{-n}=\frac{1}{(x q^{-2n};q^2)_n}\,,
\ee
and $q$ is an arbitrary parameter of the model. Note that 
\be
V_x(n)\equiv0\,,\qquad \mbox{for} \ n<0\,,\label{vanish}
\ee
and 
\begin{equation}\label{inv}
\sum_{n=-\infty}^\infty V_{a-n}(x^{-1}) V_{n-b}(x) \;=\;\delta_{a,b}\,,
\end{equation}
where, due to 
\eqref{vanish}, the summation is
effectively restricted to a finite interval $a\ge n\ge b$. Obviously, the edge 
weights are ``chiral'', since 
\be
V_x(n)\not=V_x(-n)\,,\qquad n\not=0\,.
\ee
Moreover, the vanishing property \eqref{vanish} is potentially 
dangerous for the type $(ii)$ edges with negative spin difference arguments
in \eqref{twotypes}, since their Boltzmann weights then diverge. 
However, this does not create any real problems, since 
such spin configurations will be excluded, see below.   

Now we need to describe the boundary conditions, which play rather
important role in the definition of the model. 
First, we assume that the values of white spins at the bottom row
of the lattice (see Fig.~\ref{fig-lattice}) are ordered as
\begin{subequations}
\be\label{bc1}
a_1 \le a_2 \le a_3 \le \ldots  
\ee
and so on, starting from right to left
and, similarly, the values of white spins 
at the right vertical boundary of the
lattice are ordered as 
\label{bc}
\be\label{bc2}
a_1 \le b_1 \le c_1 \le \ldots
\ee
\end{subequations}
starting from bottom to top. 

The Boltzmann weight of a particular spin configuration of the 
whole lattice is equal to the product of all the edge weights. 
The lattice spin configuration will be called {\it admissible\/}  if all the 
spin differences appearing in the edge weight functions \eqref{twotypes} 
are non-negative (i.e., for admissible 
configurations the values of spins cannot decrease  
along the edge directions). Remarkably, for the boundary conditions
\eqref{bc} the admissible configurations exhaust all
lattice spin configurations with non-zero Boltzmann weight.
Of course, the exclusion of non-admissible configurations
containing only 
type $(i)$ edges with negative spin arguments is obvious, since their
weights vanish identically due to \eqref{vanish}. However, this is not
obvious for configurations containing type $(ii)$ edges with $a<b$ in
\eqref{twotypes}, since their weights then become infinite. Our statement
follows from the limiting procedure based on the replacement of
\eqref{vanish} by 
\be
V_x(n)=O(\varepsilon)\,,\qquad \varepsilon\to0\,,\qquad n<0\,.\label{vanish2} 
\ee
Then a careful analysis shows, that as a consequence of \eqref{bc}, 
all divergent $1/\varepsilon$ pole
factors
coming from the type $(ii)$ edges are always compensated by zeros in
$\varepsilon$ coming the type $(i)$ edges. Actually, in all cases the
degree of zero is always higher than the degree of pole, so that the Boltzmann
weights of all non-admissible spin configurations with the 
boundary conditions \eqref{bc} do indeed vanish. Some additional 
explanations of this point are given after \eqref{starbound} 
in the next section.

Next, we impose 
(quasi) periodic boundary conditions in the horizontal direction.
The lattice in Fig.~\ref{fig-lattice} has $N$ sites per row. Clearly, 
for consistency with spin difference dependence of the Boltzmann weights, 
one needs to formulate the periodic boundary condition in terms of 
differences of spins on the left and right vertical boundaries. Here we 
choose the simplest option
\be\label{quasi}
b_{N+1}-a_{N+1}=b_1-a_1\,,\qquad 
c_{N+1}-b_{N+1}=c_1-b_1\,,\qquad \mbox{etc.}
\ee
by equating the jumps of boundary spins between successive rows on
both sides of the lattice. If required, the periodic boundary
conditions in the vertical direction should also be defined in terms
differences of boundary spins between successive columns of the lattice.
Finally, introduce the ``boundary fields'' by assigning additional
weights to the spin jumps on the horizontal and vertical boundaries 
of the lattice
\be\label{ffactor}
w_{\rm fields}=\Big( \omega_v^{a_2-a_1} \, \omega_v^{a_3-a_2} \,
\cdots\, \omega_v^{a_{N+1}-a_N} \Big)\times
\Big(\omega_h^{b_1-a_1} \, \omega_h^{c_1-b_1} \, \cdots \,\Big)
\,, 
\ee
where $\omega_h$ and $\omega_v$ are arbitrary parameters.

The partition function is defined as a sum of the 
products of edge weights \eqref{twotypes} and the field factors 
\eqref{ffactor} over all admissible spin configurations,
\be 
Z=\sum_{\mbox{\scriptsize (admissible spins)}}
w_{\rm fields}\ \prod_{\mbox{\scriptsize (edges)}}\ 
 \mbox{ (edge weights \eqref{twotypes})}\,. \label{part1}
\ee
Obviously, the weights 
remain unchanged if all
lattice spins are simultaneously incremented by the same 
integer. Therefore, one 
particular spin, say the spin $a_1$ at the bottom right corner of the
lattice in Fig.~\ref{fig-lattice}, should be set to some fixed value (we
chose $a_1\equiv0$) to 
avoid a repeated counting of equivalent spin configurations.

To summarize, we have defined an Ising-type model on the square
lattice, see Fig.~\ref{fig-lattice}, 
with integer-valued spins at the lattice sites. The spins interact
with their nearest neighbours (two spins interact only if they are
connected by an edge).  
The edge Boltzmann weights 
depend on the difference of spins at the ends of the edge and the
ratio of two spectral variables assigned to the medial graph lines
passing through the edge. There are two types of edges, which are
assigned with different Boltzmann weights defined in \eqref{twotypes},
\eqref{Vdef}. The admissible spin configurations are controlled by the
boundary conditions \eqref{bc}, which play an essential role in the
definition of the model.
In the homogeneous lattice case the model has six
arbitrary, in general complex, parameters: the ``anisotropy'' parameter
$q$, the boundary field parameters $\omega_h, \omega_v$ and three independent
ratios of the four spectral variables $x,x',y,y'$. Note, that this 
counting does not include the trivial overall normalization of the
Boltzmann weights.    
For a fully inhomogeneous lattice there are additional spectral variables: 
there are 
two such variables for each column and two variables for each row (these are
the variables $x_1,x_1',x_2,x_2',\ldots$ and
$y_1,y_1',y_2,y_2',\ldots$, respectively, shown in Fig.~\ref{fig-lattice}).  
Recall, the weights only depend on the ratios of these variables.

\section{Equivalent models and integrability conditions\label{integrability}}

\subsection{Interaction-round-a-face formulation\label{irf-form}}
\begin{figure}[ht]
\begin{center}
\begin{tikzpicture}[scale=1.3]
\draw [black] (1,0.3) -- (1,3); \draw [-open triangle 45, black, thin] (1,0.1) -- (1,0.3); \draw (1,0) -- (1,0.1); 
\node[below] at (1,0) {$x^{\phantom\prime}$};
\draw [black] (2,3) -- (2,0.3); \draw [-open triangle 45, black, thin] (2,0.3) -- (2,0.1); \draw (2,0) -- (2,0.1); 
\node[below] at (2,0) {$x'$};
\draw [black] (2.7,2) -- (0,2); \draw [-open triangle 45, black, thin] (2.9,2) -- (2.7,2); \draw (3,2) -- (2.9,2);
\node[right] at (3,2) {$y'$};
\draw [black] (0,1) -- (2.7,1); \draw [-open triangle 45, black, thin] (2.7,1) -- (2.9,1); \draw (3,1) -- (2.9,1);
\node[right] at (3,1) {$y$};
\draw [-latex, \PAC, ultra thick] (1.43,1.57) -- (1,2); \draw [\PAC, ultra thick] (1,2) -- (0.57,2.43);
\draw [-latex, \PAC, ultra thick] (2.43,2.43) -- (2,2); \draw [\PAC, ultra thick] (2,2) -- (1.57,1.57);
\draw [-latex, \PAC, ultra thick] (0.57,0.57) -- (1,1); \draw [\PAC, ultra thick] (1,1) -- (1.43,1.43);
\draw [-latex, \GR, ultra thick] (2.43,0.57) -- (2,1); \draw [\GR, ultra thick] (2,1) -- (1.57,1.43);
\draw [ultra thick, \STE] (0.5,2.5) circle [radius=0.08]; \node[above] at (0.5,2.55) {$a$};
\draw [ultra thick, \STE] (0.5,0.5) circle [radius=0.08];\node[below] at (0.5,0.45) {$c$};
\draw [ultra thick, \STE] (2.5,2.5) circle [radius=0.08];\node[above] at (2.5,2.55) {$b$};
\draw [ultra thick, \STE] (2.5,0.5) circle [radius=0.08];\node[below] at (2.5,0.45) {$d$};
\draw [ultra thick, fill, \STE] (1.5,1.5) circle [radius=0.08];\node[below] at (1.5,1.4) {$n$};
\node [left] at (0,1.5) {$\Sc(a,b,c,d\,|\,x,x',y,y')=\ \ \ \ $};
\end{tikzpicture}
\caption{An elementary four-edge star centred around a black site.
\label{fig-star}}
\end{center}
\end{figure}
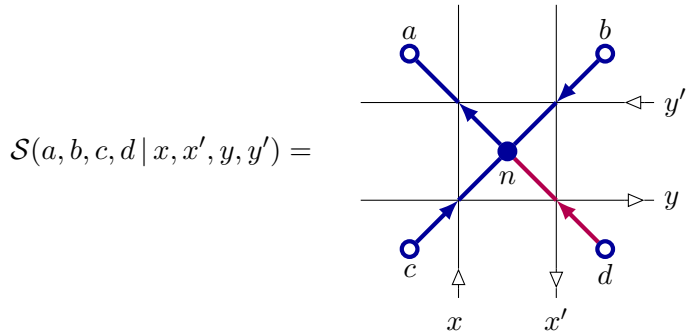

The edge interaction model introduced above could be
reformulated either as a {\em vertex\/} model
or as an {\em interaction-round-a-face\/} (IRF) model\/. In fact, this can be
done in a 
variety of ways. Let us first describe an equivalent IRF model on the square
lattice formed by the white sites in Fig.~\ref{fig-lattice}. 
The lattice in Fig.~\ref{fig-lattice} 
can be divided into 
four-edge ``stars'' with white corner sites, 
such as the star 
shown Fig.~\ref{fig-star}.
Applying the rules \eqref{twotypes} one can write the IRF-type 
Boltzmann weights for this star as
\be\label{irf1}
\Sc(a,b,c,d\,|\,x,x',y,y')=
\sum_{n=\max(b,c)}^{a}\frac{V_{x/y'}(a-n)V_{y'/x'}(n-b)
V_{y/x}(n-c)}{V_{y/x'}(n-d)}\,.
\ee
The weights depend on the four corner spins $a,b,c,d\in {\mathbb Z}$, 
which are supposed to obey the relations
\be\label{starbound}
a\ge b\ge d\,,\qquad a\ge c\ge d\,.
\ee
Outside this domain the weights \eqref{irf1} are assumed to vanish
identically. 
It is important to understand the origin of these
restrictions as well as the range of summation over the central spin
in \eqref{irf1}. Consider the four-edge star at the bottom right
corner of Fig.~\ref{fig-lattice}. The boundary conditions \eqref{bc}
require the ordering of the values of spins $a_2\ge a_1$ 
at the bottom and $b_1\ge a_1$ at the right side of that star. In the
notation of Fig.~\ref{fig-star} this corresponds to 
\be\label{starbound1}
c\ge d\,,\qquad b\ge d \,.
\ee
Let us now examine the sum in \eqref{irf1}. To regularize the summand 
for the case when its denominator vanishes we replace the strict
vanishing condition \eqref{vanish} by the regularized version
\eqref{vanish2}. The function ${V_{y/x'}(n-d)}$ 
in the denominator of \eqref{irf1} vanishes if its argument is
negative, i.e., when $n<d$. However, due to \eqref{starbound1} this
means that in this case $n<b$ as well as $n<c$, therefore, at least 
two factors $V_{y'/x'}(n-b)$ and $V_{y/x}(n-c)$
in the numerator of \eqref{irf1} must vanish, i.e., the degree of zero
in $\varepsilon$ is higher than the degree of the pole. Therefore the
configurations with a vanishing denominator from the type $(ii)$ edges 
do not contribute. Thus, the non-vanishing
contributions to the sum \eqref{irf1} could only arise when
$n\ge\max(b,c)$. Next, note that 
the first factor $V_{x/y'}(a-n)$ in the numerator 
is non-zero only when $a\ge n$. This explains the range of summation
in  \eqref{irf1} as well as the domain 
\eqref{starbound} where the star weights \eqref{irf1} do not
vanish. 

Let us now return to the bottom right star in
Fig.~\ref{fig-lattice}. The conditions \eqref{starbound} imply an
ordering of spins at the top and left sides of that star,
\be
b_2\ge b_1\,,\qquad b_2\ge a_2\,.
\ee     
Combining these with \eqref{bc} we can now apply the same reasonings
to the stars which are either directly above or to
the left of the bottom right corner star, that we started with. 
In this way the ordering of white spins in the horizontal and vertical
directions, which is initially only implied by \eqref{bc} to the
boundary, will propagate to the whole lattice.\footnote{%
It is not difficult to
see that this result is essentially equivalent to the statement 
about the admissible spin configurations with non-zero Boltzmann weights
for the edge-interaction formulation presented in the previous
section.
}   
Thus, we have reformulated our model as an IRF model on the square
lattice formed by the white sites, with Boltzmann weights given by
\eqref{irf1}, \eqref{starbound}.
The partition function \eqref{part1} of the edge-interaction model from the
previous Section can now be equivalently rewritten 
\be\label{part2}
Z=\sum_{\mbox{\scriptsize(white\ spins)}}
w_{\rm fields}
\prod_{\mbox{\scriptsize(black-center\ stars)}}
\Sc(a,b,c,d\,|\,x,x',y,y')\,. 
\ee

\subsection{Commuting transfer matrices}

Below it will be convenient to use the IRF weights 
\be\label{W-def}
\Wc(a,b,c,d\,|\,\,x,x',y,y')=\frac{V_{y/y'}(b-d)}{V_{y/y'}(a-c)}\,
\ \Sc(a,b,c,d\,|\,x,x',y,y')\,,
\ee
which only differ from \eqref{irf1} by simple equivalence transformation
factors. 
Consider the transfer matrix of the edge-interaction
model between the two bottom rows of white spins in
Fig.~\ref{fig-lattice}, with the quasi-periodic
boundary conditions \eqref{quasi},
\be\label{quasi1}
b_{N+1}-a_{N+1}=b_1-a_1\,.
\ee
Its matrix elements can be written using
the IRF weights \eqref{W-def}, 
\be\label{tmat}
\big({\Tbb}(y,y'\,|\,x_N,x_N',\ldots,x_1,x_1')\big)_{a_{N+1},a_{N},\ldots,a_{1}}^{b_{N+1},\,b_{N},\ldots,b_{1}}=\omega_h^{b_1-a_1}\,\omega_v^{a_{N+1}-a_1}\,\prod_{i=1}^N
\Wc(b_{i+1},b_i,a_{i+1},a_i\,|\,x_i,x'_i,y,y')
\ee
with the $a$'s and $b$'s labeling matrix rows and columns,
respectively. 
The parameters $y_1,y_1'$ from 
Fig.~\ref{fig-lattice} have been replaced by $y,y'$. 
Note, that the equivalence transformation factors,
introduced in \eqref{W-def},
cancel out in the definition of the
transfer matrix (i.e., the expression \eqref{tmat} remains
unchanged, if the weights $\Wc$ therein are replaced by the weights $\Sc$).
Recall, that the weights $\Wc(a,b,c,d)$ take non-zero values only if
the corner spins $a,b,c,d$ lie in the domain \eqref{starbound}. 
As a result the matrix elements \eqref{tmat} are non-zero 
\be\label{nonzero}
\big({\Tbb}(y,y'\,|\,x_N,x_N',\ldots,x_1,x_1')\big)_{a_{N+1},a_{N},\ldots,a_{1}}^{b_{N+1},\,b_{N},\ldots,b_{1}}\not=0\,, 
\qquad \mbox{iff}\qquad a_i\le b_i\,,\qquad 
i=1,2,\ldots,N+1\,,
\ee
only if $a_i\le b_i$ for all $i=1,2,\ldots,N+1$. 

The transfer matrix acts in the infinite-dimensional 
vector space ${\cal H}^{{\rm(IRF)}}_N$ spanned by the vectors $\psi_{\bf a}$ 
labeled by ordered spin sequences 
\be\label{secdef}
\psi_{\bf a}\in {\cal H}^{\rm(IRF)}_N\,,\qquad 
{\bf a}=(a_{N+1},a_N,\ldots,a_{1})\,\qquad \,a_i\in {\mathbb
  Z},\,\qquad 
a_{N+1}\ge a_N\ge \cdots\ge a_{1}\,.
\ee 
As a consequence of \eqref{quasi1}, 
the transfer matrix preserves an integer quantity 
\be\label{down}
M=\sum_{i=1}^N (a_{i+1}-a_{i})
=a_{N+1}-a_1=b_{N+1}-b_1
=\sum_{i=1}^N (b_{i+1}-b_i)\,,
\ee
which is similar to the conserved 
number of ``down arrows'' in the six-vertex model. For this reason the
dependence on the vertical field parameter $\omega_v$ in \eqref{tmat} 
becomes trivial. Therefore, in what follows we set $\omega_v=1$.

The lattice model defined above is integrable. Namely, the transfer matrices
\eqref{tmat} with different values of $y$ and $y'$ (but the same
sets of $\{x\}$  and $\{x'\}$) form a two-parameter commuting family
\be\label{tcomm}
[{\Tbb}(y,y'\,|\,x_N,x_N',\ldots,x_1,x_1'),
{\Tbb}(z,z'\,|\,x_N,x_N',\ldots,x_1,x_1')]=0\,,\qquad \forall\, y,y',z,z'\,.
\ee
This follows in a standard way from the IRF-type Yang-Baxter
equation for the weights \eqref{W-def},
\begin{equation}\label{ybe}
\begin{array}{l}
\ds \sum_{a\in{\mathbb Z}} 
\Wc(g,a,b,f\,|\,{\boldsymbol x},{\boldsymbol y})\ 
\Wc(c,e,g,a\,|\,\xb,\zb)\ 
\Wc(e,d,a,f\,|\,\yb,\zb)\;=\;\\
[3mm]
\ds \qquad\qquad\qquad\qquad\qquad=\;
\sum_{h\in{\mathbb Z}} 
\Wc(c,h,g,b\,|\,\yb,\zb)\ 
\Wc(h , d ,b , f\,|\,\xb,\zb)\ 
\Wc(c , e ,h , d\,|\,\xb,\yb)
\end{array}
\end{equation}
where we have introduced compact notations for pairs of spectral
variables 
\be
\qquad \xb=(x,x')\,,\qquad \yb=(y,y')\,,\qquad \zb=(z,z')\,,\qquad \mbox{etc.}\,,\label{xpair}
\ee
since they 
always appear in pairs.  Note that due to the corner spin constrains 
\eqref{starbound} for the weights \eqref{W-def} 
both summations in \eqref{ybe} are
restricted to finite intervals.

For future references we mention here some elementary symmetry
transformations of the YBE \eqref{ybe}. In particular, it is not 
affected by the re-scaling of the weights 
\be\label{W-tilde}
\Wc(a,b,c,d\,|\,\xb,\yb)\to
\frac{g_{\xb}(c,d)\,
  g_{\yb}(a,c)}{g_\xb(a,b)\,g_{\yb}(b,d)}\,
\Wc(a,b,c,d\,|\,\xb,\yb)\,,
\ee
with arbitrary non-vanishing factors $g_\xb(a,b)$.

\subsection{The star-star relation\label{star-section}}
\begin{figure}[ht]
\begin{center}
\begin{tikzpicture}[scale=1.5,baseline={(0,-.05)}]
\draw [black] (1,0.15) -- (1,3); \draw [-open triangle 45, black, thin] (1,0.0) -- (1,0.15); \draw [black, thin] (1,0.0) -- (1,-0.2); 
\node[below] at (1,-.2) {$x^{\phantom\prime}$};
\draw [black] (2,3) -- (2,0.2); \draw [-open triangle 45, black, thin] (2,0.2) -- (2,0.0); \draw [black, thin] (2,0.0) -- (2,-0.2); 
\node[below] at (2,-.2) {$x'$};
\draw [black] (0,2) to [out=-45,in=180] (1.5,1) to [out=0,in=-135] (3,2); 
\draw [-open triangle 45, black, thin] (3,2) -- (3.1,2.1); \draw [black, thin] (3.1,2.1) -- (3.2,2.2); \node[right] at (3.2,2.2) {$y$};
\draw [black] (3,1) to [out=135,in=0] (1.5,2) to [out=180,in=45] (0,1); 
\draw [-open triangle 45, black, thin] (3.1,0.9) -- (3,1); \draw [black, thin] (3.1,0.9) -- (3.2,0.8);
\node[right] at (3.2,0.8) {$y'$};
\draw [-latex, \PAC, ultra thick] (1.43,1.57) -- (1,2); \draw [\PAC, ultra thick] (1,2) -- (0.57,2.43);
\draw [-latex, \PAC, ultra thick] (2.43,2.43) -- (2,2); \draw [\PAC, ultra thick] (2,2) -- (1.57,1.57);
\draw [-latex, \PAC, ultra thick] (0.57,0.57) -- (1,1); \draw [\PAC, ultra thick] (1,1) -- (1.43,1.43);
\draw [-latex, \GR, ultra thick] (2.43,0.57) -- (2,1); \draw [\GR, ultra thick] (2,1) -- (1.57,1.43);
\draw [-latex, \GR, ultra thick] (0.5,0.58) -- (0.5,1.5); \draw [\GR, ultra thick] (0.5,1.5) -- (0.5,2.42);
\draw [-latex, \PAC, ultra thick] (2.5,0.58) -- (2.5,1.5); \draw [\PAC, ultra thick] (2.5,1.5) -- (2.5,2.43);
\draw [ultra thick, \STE] (0.5,2.5) circle [radius=0.08]; \node[above] at (0.5,2.55) {$a$};
\draw [ultra thick, \STE] (0.5,0.5) circle [radius=0.08];\node[below] at (0.5,0.45) {$c$};
\draw [ultra thick, \STE] (2.5,2.5) circle [radius=0.08];\node[above] at (2.5,2.55) {$b$};
\draw [ultra thick, \STE] (2.5,0.5) circle [radius=0.08];\node[below] at (2.5,0.45) {$d$};
\draw [ultra thick, fill, \STE] (1.5,1.5) circle [radius=0.08];\node[below] at (1.5,1.4) {$n$};
\node [right] at (3.5,1.5) {$=$};
\end{tikzpicture}
\begin{tikzpicture}[scale=1.5,baseline={(0,0)}]
\draw [black] (0,2) -- (2.7,2); \draw [-open triangle 45, black, thin] (2.7,2) -- (2.9,2); \draw (2.9,2) -- (3,2); 
\node[right] at (3,2) {$y$};
\draw [black] (2.7,1) -- (0,1); \draw [-open triangle 45, black, thin] (2.9,1) -- (2.7,1); \draw (2.9,1) -- (3,1);
\node[right] at (3,1) {$y'$};
\draw [black] (1,0) to [out=45,in=-90] (2,1.5) to [out=90,in=-45] (1,3); 
\draw [-open triangle 45, black, thin] (0.8,-0.2) -- (1,0); 
\node[below] at (0.8,-0.2) {$x^{\phantom\prime}$};
\draw [black] (2,3) to [out=-135,in=90] (1,1.5) to [out=-90,in=135] (2,0); 
\draw [-open triangle 45, black, thin] (2,0) -- (2.13,-0.13); \draw (2.13,-0.13) -- (2.2,-0.2);
\node[below] at (2.2,-0.2) {$x'$};
\draw [-latex, \GR, ultra thick] (1.43,1.57) -- (1,2); \draw [\GR, ultra thick] (1,2) -- (0.57,2.43);
\draw [-latex, \PAC, ultra thick] (1.57,1.57) -- (2,2); \draw [\PAC, ultra thick] (2,2) -- (2.43,2.43);
\draw [-latex, \PAC, ultra thick] (1.43,1.43) -- (1,1); \draw [\PAC, ultra thick] (1,1) -- (0.57,0.57);
\draw [-latex, \PAC, ultra thick] (2.43,0.57) -- (2,1); \draw [\PAC, ultra thick] (2,1) -- (1.57,1.43);
\draw [-latex, \PAC, ultra thick] (2.42,2.5) -- (1.5,2.5); \draw [\PAC, ultra thick] (1.5,2.5) -- (0.58,2.5);
\draw [-latex, \GR, ultra thick] (2.42,0.5) -- (1.5,0.5); \draw [\GR, ultra thick] (1.5,0.5) -- (0.58,0.5);
\draw [ultra thick, fill,\STE] (0.5,2.5) circle [radius=0.08]; \node[above] at (0.5,2.55) {$a$};
\draw [ultra thick,fill, \STE] (0.5,0.5) circle [radius=0.08];\node[below] at (0.5,0.45) {$c$};
\draw [ultra thick, fill, \STE] (2.5,2.5) circle [radius=0.08];\node[above] at (2.5,2.55) {$b$};
\draw [ultra thick, fill,\STE] (2.5,0.5) circle [radius=0.08];\node[below] at (2.5,0.45) {$d$};
\draw [ultra thick,  \STE] (1.5,1.5) circle [radius=0.08];\node[below] at (1.5,1.4) {$n$};
\end{tikzpicture}
\end{center}
\caption{A graphical representation of the star-star relation
  \eqref{ssrel}.\label{sspic}} 
\end{figure}
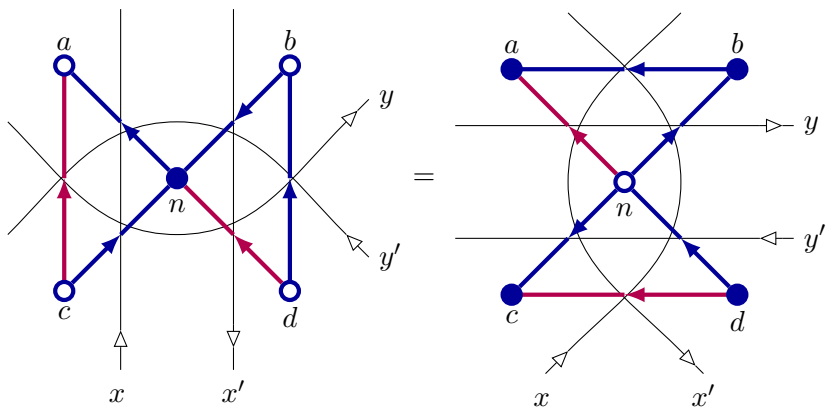

The proof of Yang-Baxter equation \eqref{ybe} 
is based on the so-called ``star-star'' relation 
for the edge weights \eqref{twotypes}, which we present below.
The Boltzmann weights \eqref{W-def}, \eqref{irf1} are associated with
the black-centered four-edge stars of the type shown in Fig.~\ref{fig-star}.
Writing them in full, one gets
\be\label{W-def1}
\Wc(a,b,c,d\,|\,\,x,x',y,y')=\frac{V_{y/y'}(b-d)}{V_{y/y'}(a-c)}\,
\sum_{n=\max(b,c)}^{a}\frac{V_{x/y'}(a-n)V_{y'/x'}(n-b)
V_{y/x}(n-c)}{V_{y/x'}(n-d)}\,.
\ee
Similarly, 
one 
can define IRF-type weights associated with the white-centered stars
in Fig.~\ref{fig-lattice} having black corner spins,
\be\label{W-def2}
{\overline \Wc}(a,b,c,d\,|\,x,x',y,y')
=\frac{V_{x/x'}(a-b)}{V_{x/x'}(c-d)}
\,\sum_{n=d}^{\min(b,c)} \ \frac{V_{x/y'}(n-d)
V_{y'/x'}(c-n) 
V_{y/x}(b-n)}
{V_{y/x'}(a-n)}\,.
\ee
It turns out that the two definitions (with suitably chosen pre-factors 
in \eqref{W-def1} and \eqref{W-def2}) lead to the same result,
\be\label{ssrel}
\Wc(a,b,c,d\,|\,x,x',y,y')=\overline{\Wc}(a,b,c,d\,|\,x,x',y,y')\,.
\ee
This relation is a
statement of equality of the Boltzmann weights of the two ``stars'' shown
graphically in Fig.~\ref{sspic}, 
so it can naturally be called the star-star relation.
A verification of this identity 
is based on a repeated application of the second Sears's 
transformation formula for basic hypergeometric series (see \cite{Sears:1951}).
Details of these calculations together with a subsequent proof of
the Yang-Baxter equation \eqref{ybe} are presented in the Appendix~\ref{appA}.

\subsection{The star-triangle relation}
The ``star-star'' relation as a primary 
integrability condition has previously appeared in
\cite{Baxter:1986phd,Bazhanov:1992jqa} 
for 3-dimensional solvable models
\cite{Zamolodchikov:1981kf,Bazhanov:1992jqa}, which also cover  
the $sl(n)$-generalized
chiral Potts model in 2D \cite{BKMS,Date:1990bs}. 
For all other known integrable 2D edge interaction models the star-star
relation could also be derived (see e.g., \cite{Baxter:1986df}), 
but it is a corollary of a much simpler star-triangle relation.
Therefore, even though we already have
a proof of the star-star
relation \eqref{ssrel}, mentioned above (see Appendix~\ref{appA}), it is
important to understand whether there are simpler integrability
conditions in our model, in particular, the star-triangle relation. 
Taking into account that the topic of the hypergeometric series has
been intensively studied by many mathematician for more than two
centures, it appears that 
the relevant star-triangle relation --- if it exists at all
--- should have already been discovered. Indeed, a purposeful 
study of the literature shows that the desired relation is the 1910
Jackson's $q$-analog \cite{Jackson:1910} 
of the Pfaff-Saalsch\"utz formula, originally
discovered by Pfaff in 1797 \cite{Pfaff:1797} and rediscovered by 
Saalsch\"utz in 1890 \cite{Saalschutz:1890}.  For a brief introduction
see the review article \cite{Gasper:1995}.  

To present the star-triangle relation we need to introduce yet another
edge weight function $W_{x,y}(a-b)$, which depends on the difference 
of edge spins $a$ and $b$ 
and on two spectral variables $x$ and $y$ (but not just on their ratio),
\begin{equation}\label{Wxy}
W_{x,y}(n) 
\;=\; \left(\frac{y}{x}\right)^n \;
\frac{(x^2;q^2)_n}{(y^2;q^2)_n}\;,\qquad n\in{\mathbb Z}\,.
\end{equation}
It possesses the following symmetries
\begin{equation}
W_{x,y}(n) \;=\; W_{q/y,q/x}(-n)\;=\;\frac{1}{W_{y,x}(n)}\;=\;\frac{1}{W_{q/x,q/y}(-n)}\;,\qquad n\in{\mathbb Z}\,.
\end{equation}
Note, in particular, that
\be
W_{x,q}(n)=V_x(n)\,,\qquad W_{q,x}(n)=\frac{1}{V_x(n)}\,.
\ee
The weights $V$ and $W$ satisfy the following two star-triangle
relations, 
\begin{subequations}
\label{str}
\begin{equation}
\sum_d \,V_{y/z}(d-a)\, W_{z,x}(b-d)\, V_{x/y}(c-d)  \;=\;
W_{z,y}(b-c) \,V_{x/z}(c-a)  \,W_{y,x}(b-a)\,,\label{str1}
\end{equation}
and
\begin{equation}
\sum_d V_{y/z}(a-d)\,W_{z,x}(d-b)\,V_{x/y}(d-c)   \;=\;
W_{z,y}(c-b)\, V_{x/z}(a-c)\, W_{y,x}(a-b)\,,\label{str2}
\end{equation}
\end{subequations}
where due to \eqref{vanish} the summations are restricted to finite
intervals.  
The two relations are
corollaries of each other under the substitution
\be\label{str-tr}
b\leftrightarrow c\,,\qquad x\to q/z\,,\qquad y\to q/y\,,\qquad z\to
q/x\,.
\ee
As mathematical identities, 
they are equivalent to the most general form of the 
Pfaff-Saalshch\"utz-Jackson summation
formula for the terminating balanced series
$\phantom{|}_3\varphi_2$, see eq.(3.5.1) in \cite{Gasper:1995}. 
The details of the correspondence are given in Appendix~\ref{appB}, 
where we also
present a graphical representation of the star-triangle relation
\eqref{str} and define yet another integrable edge interaction model 
involving the edge weights \eqref{Wxy}. 
Moreover, in the same Appendix we show that 
this relation 
implies the star-star relation \eqref{ssrel} and, consequently, the
Yang-Baxter equation \eqref{ybe} and commutativity of the transfer
matrices \eqref{tcomm}. Therefore  the star-triangle relation
\eqref{str} does, indeed, play the role of a primary integrability condition
in the model.

\section{Connection to the six-vertex model\label{connection}}
The reader might have already become perplexed 
on how exactly the above constructions are related to the six-vertex model
mentioned in the title of the paper. These connections are discussed below. 
\subsection{Higher spin generalizations of the six-vertex
  model\label{irf-vertex}}
In the previous section we have reformulated our edge interaction
(Ising-type) model as an IRF model on the square lattice. Its face
weights  
depend on differences of corner spins, 
therefore one can easily convert this IRF model 
into a vertex model by replacing the face weights by an equivalent
vertex $R$-matrix, 
\be
{\mathcal R}(x,x',y,y')^{j_1,j_2}_{i_1,i_2}=\delta_{i_1+i_2,j_1+j_2} \,
\Wc(i_1+i_2,j_2,i_1,0\,|x,x',y,y')\,.\label{R-def}
\ee
where the weights $\Wc$ are defined by \eqref{W-def1} (or, equivalently, by 
\eqref{W-def} and \eqref{starbound}).
\begin{figure}[ht!]
\begin{center}
\begin{tikzpicture}[scale=1.,baseline={(0,-.05)}]
\draw [thick, \STE] (0.5,2.5) -- (0.5,0.5) -- (2.5,0.5) --
(2.5,2.5) -- (0.5,2.5);

\draw [ultra thick, fill,white] (0.5,2.5) circle [radius=0.08]; 
\draw [ultra thick, fill,white] (0.5,0.5) circle [radius=0.08];
\draw [ultra thick, fill,white] (2.5,2.5) circle [radius=0.08];
\draw [ultra thick, fill,white] (2.5,0.5) circle [radius=0.08];

\draw [ultra thick, \STE] (0.5,2.5) circle [radius=0.08]; \node[above] at (0.5,2.55) {$a$};
\draw [ultra thick, \STE] (0.5,0.5) circle [radius=0.08];\node[below] at (0.5,0.45) {$c$};
\draw [ultra thick, \STE] (2.5,2.5) circle [radius=0.08];\node[above] at (2.5,2.55) {$b$};
\draw [ultra thick, \STE] (2.5,0.5) circle [radius=0.08];\node[below] at (2.5,0.45) {$d$};
\draw [ultra thick, fill, black] (1.5,1.5) circle [radius=0.08];

\draw [-latex,thick] (1.5,-.2) -- (1.5,.2); 
\draw [thick] (1.5,.0) -- (1.5,3.2); 
\draw [-latex,thick] (-.2,1.5) -- (.2,1.5); 
\draw [thick] (.0,1.5) -- (3.2,1.5); 
\node[right=.5mm] at (1.5,-.0) {$i_1$};
\node[below] at (1.5,-.3) {$(\lambda_1,s_1)$};
\node[left] at (-.4,1.5) {$(\lambda_2,s_2)$};
\node[right=.5mm] at (1.5,3.0) {$j_1$};
\node[below] at (0,1.5) {$i_2$};
\node[below] at (3.0,1.5) {$j_2$};

\end{tikzpicture}
\caption{A typical face of the square lattice
  with corner spins $a$, $b$, $c$, $d$ and the associated vertex of
  the dual lattice with the edge spin variables $i_1$, $j_1$, $i_2$, $j_2$. 
The spectral parameters $(\lambda_1,s_1)$ and $(\lambda_2,s_2)$ are related 
to $(x,x')$ and $(y,y')$ as in \eqref{pardef}.
\label{fig5c}}
\end{center}
\end{figure}
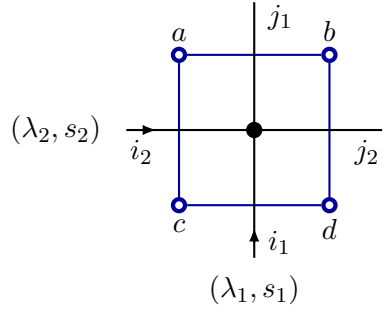
The matrix indices 
\be\label{ij-def}
i_1=c-d\,,\qquad i_2=a-c\,,\qquad j_1=a-b\,,\qquad j_2=b-d\,,
\ee
are defined as differences of corner spins, arranged as in Fig.~\ref{fig5c}. 
Due to \eqref{starbound} these indices take non-negative integer values
\be\label{j-pos}
i_1,i_2,j_1,j_2\ge0\,,\qquad  i_1,i_2,j_1,j_2\in {\mathbb Z}\,.
\ee
The emergence of the conservation law 
\be
i_1+i_2=j_1+j_2\,,\label{ar-cons}
\ee
appearing in the delta-function in \eqref{R-def} is a trivial
consequence of the definitions 
\eqref{ij-def}. Using the two representations for the IRF weights
\eqref{W-def1} and \eqref{W-def2} one can write \eqref{R-def} in the form
\bea\label{R-def1}
{\cal R}(x,x',y,y')^{j_1,j_2}_{i_1,i_2}&=&\delta_{i_1+i_2,j_1+j_2}\ 
\frac{V_{y/y'}(j_2)}{V_{y/y'}(i_2)}\,
\sum_{n=0}^{\min(i_2,j_1)}\frac{V_{x/y'}(n)V_{y'/x'}(j_1-n)
V_{y/x}(i_2-n)}{V_{y/x'}(i_1+i_2-n)}\,\nonumber\\
&&\label{R-def2}\\
&=&\delta_{i_1+i_2,j_1+j_2}\ \frac{V_{x/x'}(j_1)}{V_{x/x'}(i_1)}
\,\sum_{n=0}^{\min(i_1,j_2)} \ \frac{V_{x/y'}(n)
V_{y'/x'}(i_1-n) 
V_{y/x}(j_2-n)}
{V_{y/x'}(i_1+i_2-n)}\,.\nonumber
\eea
With these
notations the Yang-Baxter equation \eqref{ybe} takes the form
\be
\begin{array}{r}
\ds\sum_{j_1,j_2,j_3}
{\cal R}(x,x',y,y')_{i_1,i_2}^{j_1,j_2}\ \ 
{\cal R}(x,x',z,z')_{j_1,i_3}^{k_1,j_3}\ \
{\cal R}(y,y',z,z')_{j_2,j_3}^{k_2,k_3}{\qquad \qquad}
\\[.6cm]
\ds =\sum_{j_1,j_2,j_3}
{\cal R}(y,y',z,z')_{i_2,i_3}^{j_2,j_3}\ \
{\cal R}(x,x',z,z')_{i_1,j_3}^{j_1,k_3}\ \
{\cal R}(x,x',y,y')_{j_1,j_2}^{k_1,k_2}\,.
\end{array}\label{Rybe}
\ee
Note that due to \eqref{j-pos} and \eqref{ar-cons} all summations
here are restricted to finite intervals. It is worth
mentioning some simple symmetries of this equation. 
Obviously, it is invariant under the
simultaneous transposition in all three vector spaces. 
Below we will use this symmetry 
in combination with the rescaling of the weights as in  
\eqref{W-tilde}. It is easy to
check that \eqref{Rybe} is invariant under the transformation
\be\label{R-tilde}
\begin{array}{rcl}
\ds{{\cal R}}(x,x',y,y')_{i_1,i_2}^{j_1,j_2}\to
\ds\widetilde{{\cal R}}(x,x',y,y')_{i_1,i_2}^{j_1,j_2}&=&\ds
\Big(\frac{\ds y^{\prime j_1}x^{j_2}}{\ds y^{i_1}x^{\prime
    i_2}}\Big)^{-1}
 \;\mathcal{R}^T(x,x',y,y')_{i_1,i_2}^{j_1,j_2}\,,
\end{array}
\ee
for all three $R$-matrices (with corresponding substitutions of their
arguments).  Here the superscript $T$
denotes the matrix transposition of $\cal R$ in both spaces
\be \label{T-trans}
{\mathcal R}^{T}(x,x',y,y')_{i_1,i_2}^{j_1,j_2}=
{\mathcal R}(x,x',y,y')^{i_1,i_2}_{j_1,j_2}\,.
\ee
The scaling factor in \eqref{R-tilde} exactly coincides to that in
\eqref{W-tilde} with $g_{\xb}(a, b) = x^b/x'^a$ (taking into account
\eqref{R-def}, \eqref{ij-def}). 

Note, that the $R$-matrix \eqref{R-def2} 
only depends on three independent  
ratios of the four spectral parameters $x,x',y,y'$. Therefore, 
it is useful to
introduce a new set of parameters  
\be\label{pardef}
x=\lambda_1\,q^{-s_1}\,,\quad x'=\lambda_1\,q^{s_1}\,,
\quad y=\lambda_2\,q^{-s_2}\,,\quad y'=\lambda_2\,q^{s_2}\,,
\quad z=\lambda_3\,q^{-s_3}\,,\quad z'=\lambda_3\,q^{s_3}\,.
\ee
Then it is easy to see that \eqref{R-def2} only depends on the
variables $\lambda_1/\lambda_2$, $s_1$ and $s_2$. Therefore we will
write it as 
\be \label{RR-rel}
{\cal R}(\lambda_1/\lambda_2\,|\,s_1,s_2)={\cal R}(x,x',y,y')\,,
\ee
provided the parameters are related as in \eqref{pardef}. 
The Yang-Baxter equation \eqref{Rybe} then takes the form
\be
\begin{array}{r}
\ds\sum_{j_1,j_2,j_3}
{\cal R}(\lambda_1/\lambda_2\,|\,s_1,s_2)_{i_1,i_2}^{j_1,j_2}\ \ 
{\cal R}(\lambda_1/\lambda_3\,|\,s_1,s_3)_{j_1,i_3}^{k_1,j_3}\ \
{\cal R}(\lambda_2/\lambda_3\,|\,s_2,s_3)_{j_2,j_3}^{k_2,k_3}\phantom{\qquad
\qquad }
\\[.6cm]
\ds =\sum_{j_1,j_2,j_3}
{\cal R}(\lambda_2/\lambda_3\,|\,s_2,s_3)_{i_2,i_3}^{j_2,j_3}\ \
{\cal R}(\lambda_1/\lambda_3\,|\,s_1,s_3)_{i_1,j_3}^{j_1,k_3}\ \
{\cal R}(\lambda_1/\lambda_2\,|\,s_1,s_2)_{j_1,j_2}^{k_1,k_2}\,.
\end{array}\label{Rybe2}
\ee
Let us mention also two symmetry relations, 
\bea
{\cal R}(\lambda\,|\,s_1,s_2)_{j_1,j_2}^{i_1,i_2}&=&
\frac{V_{q^{-2s_1}}(i_1) V_{q^{-2s_2}}(i_2)}
{V_{q^{-2s_1}}(j_1) V_{q^{-2s_2}}(j_2)}\ 
{\cal R}(\lambda\,|\,s_1,s_2)_{i_1,i_2}^{j_1,j_2}\,,
\label{sym1}\\[.4cm]
{\cal R}(\lambda\,|\,s_2,s_1)_{i_2,i_1}^{j_2,j_1}&=&
{\cal R}(\lambda\,|\,s_1,s_2)_{i_1,i_2}^{j_1,j_2}\,,\label{sym2}
\eea
and the normalization conditions 
\begin{equation}
\begin{array}{rcl}\ds
\lim_{\lambda\to0}
\,\mathcal{R}(\lambda\,|\,s_1,s_2)_{i_1,i_2}^{j_1,j_2}&=&\delta_{i_1,j_1}
\delta_{i_2,j_2} q^{-2(s_1-j_1)(s_2-j_2)+2s_1s_2}\;,
\\[.4cm]\ds
\lim_{\lambda\to\infty}
\mathcal{R}(\lambda\,|\,s_1,s_2)_{i_1,i_2}^{j_1,j_2}&=&\delta_{i_1,j_1}
\delta_{i_2,j_2} q^{+2(s_1-j_1)(s_2-j_2)-2s_1s_2}\;, 
\end{array}\label{R-norm}
\end{equation}
which simply follow from \eqref{R-def2} for fixed values of $s_1$ and $s_2$.

We state that the $R$-matrix \eqref{R-def2} coincides with Mangazeev's
$R$-matrix \cite{Mangazeev:2014gwa} for the higher spin generalizations of the 
6-vertex model. It is the trigonometric
$R$-matrix associated with two infinite-dimensional evaluation
representations of $U_q(\widehat{sl}(2))$ with the highest weights
$2s_1$ and $2s_2$. The details of the correspondence are presented in
Appendix~\ref{appC}. 

\subsection{Finite-dimensional reductions\label{red}}
As usual we consider the $R$-matrix \eqref{R-def2} as a matrix acting in the
tensor product of two vector spaces, assigning the matrix indices
$i_1,j_1$ and $i_2,j_2$ to the first and second spaces, respectively.
Following the terminology of the QISM we will call them as the ``quantum''
and ``auxiliary'' spaces. Having in mind further applications 
we will discuss various reductions associated with the second
(auxiliary) space, however, in view of the symmetry \eqref{sym2},
the same considerations also apply to the first
space.  
When $2s_2$ is a non-negative integer the corresponding evaluation 
representation of $U_q(\widehat{sl}(2))$ becomes reducible and splits
into a semi-direct 
sum of the $(2s_2+1)$-dimensional and an infinite-dimensional
representations. The $R$-matrix then takes the block-triangular form in the
second space\footnote{%
Note that in  
the considered case the formula \eqref{R-def2} requires a limiting
procedure where a non-integer 
$2s_2$ is continuously
approaching the required integer value. In this way all the matrix
elements \eqref{R-def2} remain finite and well defined.}, 
\be\label{twos2}
2s_2\in{\mathbb
    Z}_{\ge0}\,:\qquad \qquad
\Rc(\lambda\,|\,s_1,s_2)_{i_1,i_2}^{j_1,j_2}=0\,,\qquad 0\le i_2\le 2s_2,
\qquad j_2> 2s_2\,,\qquad \forall\, i_1,j_1\,.
\ee 
To select its finite-dimensional block one just needs
to restrict the indices to a finite set
\be\label{fin}
i_2,j_2\in {\cal I}_{s_2}=\{0,1,2,\ldots,2s_2\}\,.
\ee
Similarly, for the infinite-dimensional block one needs
to select the complimentary set of non-negative integers, 
\be\label{inf}
i_2,j_2\in \overline{\cal I}_{s_2}=\{2s_2+1,2s_2+2,\ldots,\infty\}\,.
\ee
The Yang-Baxter equation \eqref{Rybe2} turns into two equations of the
same form, where 
the indices $i_2,j_2,k_2$ related to the second space are restricted to
either the finite ${\cal I}_{s_2}$ or the infinite set 
$\overline{\cal I}_{s_2}$. Below we consider a few examples of this reduction. 
\subsubsection{Inversion relation, $s_2=0\,.$}
Setting $s_2=0$ in \eqref{pardef}, using the definition
\eqref{R-def2} and the inversion relation 
\eqref{inv} one obtains for the finite-dimensional part \eqref{fin},
\be\label{spin0}
s_2=0:\qquad\qquad
\Rc(\lambda\,|\,s_1,0\,)_{i_1,0}^{j_1,0}=\delta_{i_1,j_1}\,,
\ee
where $\lambda=\lambda_1/\lambda_2$.
For the infinite-dimensional part \eqref{inf} one get an important identity 
\be\label{rel0}
s_2=0:\,\qquad\qquad
\Rc(\lambda\,|\,s_1,0\,)_{i_1,i_2}^{j_1,j_2}=\,\frac{[\lambda q^{-s_1}]\,
[q^{i_2}]}{[\lambda q^{+s_1}]\,[q^{j_2}]}\ 
\Rc(\lambda\,|\,s_1,-1\,)_{i_1,i_2-1}^{j_1,j_2-1}\,,\qquad i_2,j_2\ge1\,,
\ee
relating the $R$-matrices with $s_2=0$ and $s_2=-1$. Here and below we
use the notation
\be
[x]=x-x^{-1}\,.
\ee

\subsubsection{The $L$-operator, $s_2=\frac{1}{2}\,.$\label{Lop-red}}
The next important case to
consider is when $s_2=\frac{1}{2}$ and $s_1$ is arbitrary. 
With the relations \eqref{pardef} this corresponds to $y'=q\,y$.
Introducing the notation
\be\label{Lop-first}
\Rc(x,x',y,q\,y)_{i_1,i_2}^{j_1,j_2}
=\frac{1}{[x'/y]}{\cal L}(x,x',y)_{i_1,i_2}^{j_1,j_2}\,,
\ee
and selecting the
2-dimensional subspace 
${\cal I}_{\frac{1}{2}}$ of the second space  
one obtains from \eqref{R-def2},
\begin{equation}
\bm{\mathcal{L}}(x,x',y)_{i_1}^{j_1}=
\ds\begin{pmatrix}
{\mathcal{L}}_{i_1,0}^{j_1,0}&
\mathcal{L}_{i_1,0}^{j_1,1}\\[.4cm]
\mathcal{L}_{i_1,1}^{j_1,0}&
\mathcal{L}_{i_1,1}^{j_1,1}
\end{pmatrix}=
\ds\begin{pmatrix}
\ \delta_{i_1,j_1} [q^{-i_1}x'/y]&\delta_{i_1,j_1+1} {[q^{i_1}]}\\[.4cm]
\delta_{i_1+1,j_1}[q^{-i_1}x'/x]&
\ \ \delta_{i_1,j_1}[q^{i_1}x/y]\end{pmatrix}\,,
\label{Lmat}
\end{equation}
where $[x]=x-1/x$.  Notice, that here for the arguments of ${\cal
  L}(x,x',y)$ we use the same (original) variables as in ${\cal
  R}(x,x',y,qy)$.   
As usual we consider the above
expressions as matrix elements of the $L$-operator, which is regarded
as a two by two
matrix in the second space, whose elements are operators acting in the 
first space,
\be
\boldsymbol{{\cal L}}(x,x',y)
=\begin{pmatrix}
\mu\, q^{\frac{H}{2}}-\mu^{-1} \,q^{-\frac{H}{2}}&
F\\[3mm]
E& \mu \,q^{-\frac{H}{2}}-\mu^{-1} \,q^{\frac{H}{2}}
\end{pmatrix}\,,\quad q^{2s_1}=x'/x\,,\quad \mu=(x x')^{\hf}/y\,,
\label{Lop}
\ee
where the introduction of the dependent variable $\mu$ was used to
bring the above formula to the standard form.
Here $E,F,H$ stand for the generators of 
the quantum universal enveloping algebra $U_q(sl(2))$ 
\be
[H,E]=2 E\,,\qquad [H,F]=-2F\,,\qquad
[E,F]={(q-q^{-1})}\,(q^H-q^{-H})\,.\label{uqsl2}
\ee
The matrix elements \eqref{Lmat} correspond to the
infinite-dimensional highest weight 
representation $\pi^+_{s_1}$ of this algebra 
(the parameter $s_1$  is defined by $x'/x=q^{2 s_1}$),
\be\label{piplus_s1}
\pi^+_{s_1}[H]\,|j\rangle=(2s_1-2j)\,|j\rangle\,,
\qquad \pi^+_{s_1}[E]\,|j\rangle=[q^{2s_1+1-j}]\,|j-1\rangle\,,\qquad 
\pi^+_{s_1}[F]\,|j\rangle=[q^{j+1}]\,|j+1\rangle\,,
\ee
spanned on the basis vectors $|j\rangle\in {\mathbb C}^{(\infty)}$ 
with $j=0,1,2,\ldots,\infty$.

\bigskip
For the infinite-dimensional part of the $R$-matrix 
with the indices $i_2,j_2\ge2$ one gets the relation,
\be\label{rel1}
s_2={\textstyle\frac{1}{2}}:\,\qquad
\Rc\big(\lambda\,|\,s_1,{\textstyle\frac{1}{2}}\,\big)_{i_1,i_2}^{j_1,j_2}
=\,\frac{[\lambda q^{-s_1-\frac{1}{2}}]\,
[\lambda q^{-s_1+\frac{1}{2}}]\,
[q^{i_2}]\,[q^{i_2-1}]}{[\lambda q^{+s_1-\frac{1}{2}}]\,[\lambda
    q^{+s_1+\frac{1}{2}}]\,
[q^{j_2}]\,[q^{j_2-1}]}\ 
\Rc\big(\lambda\,|\,s_1,-{\textstyle\frac{3}{2}}\,\big)_{i_1,i_2-2}^{j_1,j_2-2}\,,
\qquad i_2,j_2\ge2\,,
\ee
which connect the $R$-matrices with $s_2=\frac{1}{2}$ and $s_2=-\frac{3}{2}$. 
Note, that here we use the notations for ${\cal
  R}(\lambda\,|\,s_1,s_2)=\Rc(x,x',y,y')$,
as it is defined in \eqref{RR-rel} and \eqref{pardef}.

\subsubsection{Six-vertex $R$-matrix, $s_1=s_2=\frac{1}{2}\,.$ }
Consider the case when
$s_1=s_2=\frac{1}{2}$. Setting 
\be \label{Rhh}
{\cal
  R}(\lambda\,|\,{\textstyle\frac{1}{2}},{\textstyle\frac{1}{2}})
=\frac{1}{[\lambda q]}\,
{\cal R}^{\rm (6v)}(\lambda)\,,
\ee
and 
using \eqref{R-def2}, \eqref{pardef}, 
one gets the standard $R$-matrix of the 6-vertex model,
\be
\begin{array}{ll}
\Rc^{\rm (6v)}(\lambda)_{00}^{00}=\Rc^{\rm (6v)}(\lambda)_{11}^{11}=
\rho(\lambda)\,(\lambda q-\lambda^{-1} q^{-1})\,,\qquad\qquad 
&\Rc^{\rm (6v)}(\lambda)_{01}^{10}=\Rc^{\rm (6v)}(\lambda)_{10}^{01}=
\rho(\lambda)\,(q-q^{-1})\,,\\[.6cm]
\Rc^{\rm (6v)}(\lambda)_{01}^{01}=\Rc^{\rm (6v)}(\lambda)_{10}^{10}=
\rho(\lambda)\,(\lambda -\lambda^{-1})\,,
&
\end{array}\label{R6v}
\ee
where $\lambda=\lambda_1/\lambda_2$ and $\rho(\lambda)\equiv1$.
Alternatively, one could get the same expression from \eqref{Lop}
by choosing the 2-dimensional representation of \eqref{uqsl2} with
$s_1=\frac{1}{2}$.  

\subsubsection{Infinite-dimensional $R$-matrices, $2s_2\in{\mathbb
    Z}_{\ge0}\,. $}
Here we present the generalization of the relations \eqref{rel0},
\eqref{rel1},
\be
\Rc\big(\lambda\,|\,s_1,s_2\,\big)_{i_1,i_2}^{j_1,j_2}
=\,\frac{V_{y/x}(2s_2+1)\,V_{q^{-2i_2}}(2s_2+1)}
{V_{y/x'}(2s_2+1)\,V_{q^{-2j_2}}(2s_2+1)}\ 
\Rc\big(\lambda\,|\,s_1,-s_2-1\,\big)_{i_1,i_2-2s_2-1}^{j_1,j_2-2s_2-1}\,,
\qquad i_2,j_2\ge2s_2+1\,,\label{twos2inf}
\ee
valid for $2s_2\in{\mathbb Z}_{\ge0}$ and arbitrary $s_1$. The
factor in the RHS is expressed via the weight function \eqref{Vdef}, 
the parameter relations \eqref{pardef} are assumed and
$\lambda=\lambda_1/\lambda_2$. 

\subsection{Symmetry transformation}
Below we will also use an alternative variant of the $L$-operator
originating from the $R$-matrix \eqref{R-tilde},
\be
\widetilde{\Rc}(x,x',y,q\,y)_{i_1,i_2}^{j_1,j_2}
=\frac{1}{[x'/y]}\widetilde{{\cal
    L}}(x,x',y)_{i_1,i_2}^{j_1,j_2}\,,\qquad i_2,j_2=0,1\,.
\ee
It is simply related to \eqref{Lop-first} (note the matrix transposition in both spaces)
\be\label{Lop-second}
\ds\widetilde{{\cal L}}(x,x',y)_{i_1,i_2}^{j_1,j_2}=\ds
\Big(\frac{q^{j_1}y^{j_1} x^{j_2} }{y^{i_1}{x'}^{i_2} }\Big)^{-1}\,
 {\cal
  L}(x,x',y)_{j_1,j_2}^{i_1,i_2}\,.
\ee
Writing $\widetilde{\bm{\mathcal{L}}}$ as a two by two matrix 
in the second space, but keeping the matrix indices of the first space, 
one gets
\begin{equation}
\widetilde{\bm{\mathcal{L}}}(x,x',y)_{i_1}^{j_1}=\begin{pmatrix}
\ds
q^{-i_1}\,[\mu\,q^{s_1-j_1}]\,\delta_{i_1,j_1}
& \ds  \mu^{-1}\,q^{s_1-j_1}\,[q^{2s_1-j_1}]
\,\delta_{i_1,j_1+1}\\[7mm]
\ds\mu\,q^{s_1-j_1}\,[q^{j_1}]\,\delta_{i_1,i_1-1}
&
\ds  q^{2s_1-i_1}\,[\mu\,q^{-s_1+i_1}]\,\delta_{i_1,j_1}
\end{pmatrix}\label{Lmat4}
\end{equation}
Here we use the same notations as in \eqref{pardef} and $\mu=\lambda_1
q^{\frac{1}{2}}/\lambda_2$. Similar to \eqref{Lop} we can rewrite 
\eqref{Lmat4} in the operator form
\be
\widetilde{\bm{\mathcal{L}}}\;=\;
\left(\begin{array}{cc}
\ds q^{-s_1}\,\big(\mu \,q^H -\mu^{-1}\big) & \ds \mu^{-1}\,F\,q^{\frac{H}{2}} \\
[5mm]
\ds \mu\,E\,q^{\frac{H}{2}} & \ds q^{+s_1}\,\big(\mu - \mu^{-1}\, q^H\big)
\end{array}\right)
\end{equation}
where $E,F,H$ denote the generators
of the quantum universal enveloping algebra $U_q(sl(2))$, defined by
\eqref{uqsl2}. The matrix elements \eqref{Lmat4} correspond to the
infinite-dimensional representation 
\be\label{rep-first}
\pi^+_{s_1}[H]\,|j\rangle=(2s_1-2j)\,|j\rangle\,,
\qquad \pi^+_{s_1}[E]\,|j\rangle=[q^{j}]\,|j-1\rangle\,,\qquad 
\pi^+_{s_1}[F]\,|j\rangle=[q^{2s_1-j}]\,|j+1\rangle\,,
\ee
spanned on the basis vectors $|j\rangle$ with $j=0,1,2,\ldots,\infty$.
This representation is equivalent to \eqref{piplus_s1}.

It is instructive to rewrite \eqref{Lmat4} in terms of generators of the 
Weyl algebra
\be\label{weyl}
\boldsymbol{u}\,\boldsymbol{v}\;=\;q^2\,\boldsymbol{v}\,\boldsymbol{u}\;.
\end{equation}
Choosing the representation
\begin{equation}\label{uvj}
\boldsymbol{u}\,|\, j\, \rangle \;=\; |\, j\, \rangle\, q^{2s_1-2j}\;,\quad
\boldsymbol{v}\,|\, j\, \rangle \;=\; |\, j-1\, \rangle\;,
\end{equation}
one gets
\begin{equation}\label{L-form}
\widetilde{\bm{\mathcal{L}}}\;=\;\left(\begin{array}{cc}
\ds \mu \, q^{-s_1} \, \boldsymbol{u} \,-\, \mu^{-1} \, q^{-s_1} & \ds 
\mu^{-1} \, q^{s_1} \, \boldsymbol{v}^{-1} \, ( \boldsymbol{u}\,-\, q^{-2s_1} ) \\
\\
\ds \mu \, q^{s_1} \, \boldsymbol{v} \, ( 1\,-\,q^{-2s_1} \, \boldsymbol{u} ) & 
\ds \mu \, q^{s_1} \,-\, \mu^{-1} \, q^{s_1} \, \boldsymbol{u}
\end{array}\right)\;.
\end{equation}

\subsection{New solutions of the Yang-Baxter equation}
The existence of the star-triangle relation \eqref{str} 
allows one in the standard
way to construct new solutions of the Yang-Baxter equation.
For completeness we present them in the Appendix~\ref{sappb}, 
though they are not relevant to the main topic of this paper.
From the quantum group point of view the $R$-matrix \eqref{R-def1} 
intertwines two infinite-dimensional 
{\em evaluation highest weight\/} representations of
the $U_q(\widehat{sl}(2))$ algebra. The new $R$-matrices 
presented in Appendix~\ref{sappb} are related to infinite-dimensional
representations without highest and lowest weights.

\section{Six-vertex model as a descendant of the six-vertex 
model\label{descen}}

The Ising-type model presented above is inspired by a  
remarkable structure of Mangazeev’s $R$-matrix 
\cite{Mangazeev:2014gwa} for the higher spin
generalizations of the 6-vertex model, defined in \eqref{R-def2} (see 
Appendix~C for connections with the notations of \cite{Mangazeev:2014gwa}). 
In this section we demonstrate how the observed Ising-type structure can
be discovered by elementary calculations, based on the idea of
factorized $R$-matrices and Baxter's ``propagation
through the vertex'' techniques \cite{Baxter:1982zz}. Previously these
methods were used in \cite{Bazhanov:1989nc}
for the chiral Potts model \cite{Au-Yang:1987syu,McCoy:1987pt,Baxter:1987eq}
which is related to cyclic representations of $U_q(\widehat{sl}(2))$ with $q$
being a root of unity. In this way the chiral Potts model (which is an
Ising-type model) has turned out 
to be  a ``descendant'' of the six-vertex model. 
Here we show that the same
methods are equally powerful for arbitrary values of $q$ in the context
of the six-vertex model itself.

\subsection{Factorized $R$-matrices}

\begin{figure}[ht]
\begin{center}
\begin{tikzpicture}[scale=.95]
\draw [black, thin] (1,4.3) -- (1,-0.0); \draw [-open triangle 45, black, thin] (1,0) -- (1,-0.3); \draw [thin] (1,-0.3) -- (1,-0.5);
\draw [black, thin] (3,-0.0) -- (3,4.3); \draw [-open triangle 45, black, thin] (3,-0.3) -- (3,0); \draw [thin] (3,-0.3) -- (3,-0.5);
\draw [black, thin] (-0.3,1) -- (4.0,1); \draw [-open triangle 45, black, thin] (4,1) -- (4.3,1); \draw [thin] (4.3,1) -- (4.5,1);
\draw [black, thin] (4.0,3) -- (-0.3,3); \draw [-open triangle 45, black, thin] (4.3,3) -- (4,3); \draw [thin] (4.3,3) -- (4.5,3);
\draw [\STE, ultra thick, fill] (0,2) circle [radius=0.09];
\draw [\STE, ultra thick, fill] (4,2) circle [radius=0.09];
\draw [\STE, ultra thick] (2,0) circle [radius=0.09];
\draw [\STE, ultra thick] (2,4) circle [radius=0.09];
\draw [-latex, \GR, ultra thick] (1.92,0.08) -- (1,1);  \draw[\GR, ultra thick] (1,1) -- (0.08,1.92);
\draw [-latex, \PAC, ultra thick] (2.08,0.08) -- (3,1); \draw[\PAC, ultra thick] (3,1) -- (3.95,1.92);
\draw [-latex, \PAC, ultra thick] (3.92,2.08) -- (3,3); \draw[\PAC, ultra thick] (3,3) -- (2.08,3.92);
\draw [-latex, \PAC, ultra thick] (1.92,3.92) -- (1,3); \draw[\PAC, ultra thick] (1,3) -- (0.08,2.08);
\node [right] at (4.6,3) {$y'$};
\node [right] at (4.6,1) {$y$};
\node [below] at (3,-0.5) {$x^{\phantom\prime}$};
\node [below] at (1,-0.5) {$x'$};
\node [above] at (2,4.1) {$a'$};
\node [below] at (2,-0.1) {$a$};
\node [right] at (4.1,2) {$b'$};
\node [left] at (-0.1,2) {$b$};
\end{tikzpicture}
\hspace{25mm}
\begin{tikzpicture}[scale=.95]
\draw [black, thin] (3,4.3) -- (3,-0.0); \draw [-open triangle 45, black, thin] (3,0) -- (3,-0.3); \draw [thin] (3,-0.3) -- (3,-0.5);
\draw [black, thin] (1,-0.0) -- (1,4.3); \draw [-open triangle 45, black, thin] (1,-0.3) -- (1,0); \draw [thin] (1,-0.3) -- (1,-0.5);
\draw [black, thin] (-0.3,3) -- (4.0,3); \draw [-open triangle 45, black, thin] (4,3) -- (4.3,3); \draw [thin] (4.3,3) -- (4.5,3);
\draw [black, thin] (4.0,1) -- (-0.3,1); \draw [-open triangle 45, black, thin] (4.3,1) -- (4,1); \draw [thin] (4.3,1) -- (4.5,1);
\draw [\STE, ultra thick, fill] (0,2) circle [radius=0.09];
\draw [\STE, ultra thick, fill] (4,2) circle [radius=0.09];
\draw [\STE, ultra thick] (2,0) circle [radius=0.09];
\draw [\STE, ultra thick] (2,4) circle [radius=0.09];
\draw [-latex, \PAC, ultra thick] (1.92,0.08) -- (1,1);  \draw[\PAC, ultra thick] (1,1) -- (0.08,1.92);
\draw [\PAC, ultra thick] (2.08,0.08) -- (3,1); \draw[-latex, \PAC, ultra thick] (3.95,1.92) -- (3,1);
\draw [-latex, \GR, ultra thick] (3.92,2.08) -- (3,3); \draw[\GR, ultra thick] (3,3) -- (2.08,3.92);
\draw [\PAC, ultra thick] (1.92,3.92) -- (1,3); \draw[-latex, \PAC, ultra thick] (0.08,2.08) -- (1,3);
\node [right] at (4.6,3) {$y$};
\node [right] at (4.6,1) {$y'$};
\node [below] at (3,-0.5) {$x'$};
\node [below] at (1,-0.5) {$x^{\phantom\prime}$};
\node [above] at (2,4.1) {$a'$};
\node [below] at (2,-0.1) {$a$};
\node [right] at (4.1,2) {$b'$};
\node [left] at (-0.1,2) {$b$};
\end{tikzpicture}
\end{center}
\caption{Graphical representation of the factorized $R$-matrces. 
Eq.\eqref{def3} corresponds to the left diagram.}
\label{fig-box}
\end{figure}
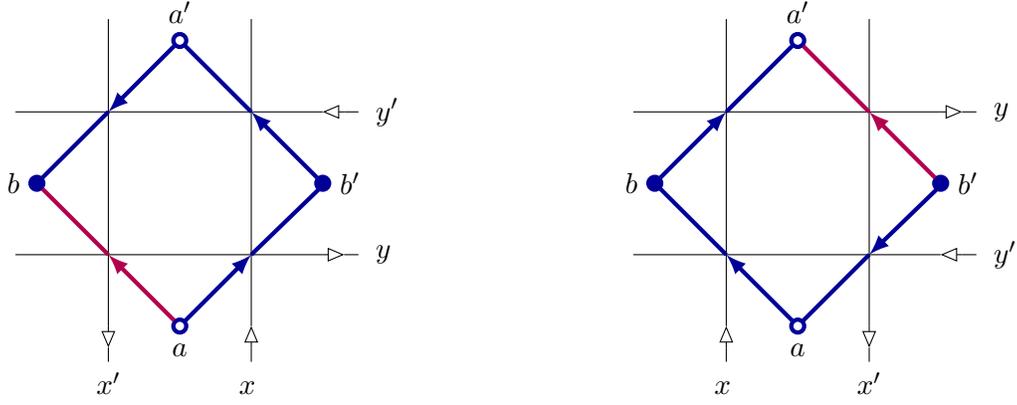

In Sect.~\ref{irf-form} and \ref{irf-vertex},
we have reformulated our Ising-type model first as 
an IRF model and then as 
a vertex model. It is easy to see that to within boundary
effects the same model can also be reformulated as yet another
vertex model. Indeed, 
the square lattice in Fig.~\ref{fig-lattice} 
can be formed by periodic translations of
the ``box diagrams'' shown in Fig.~\ref{fig-box}. 
One can choose any
of these two diagrams, for definiteness we take the left diagram in
Fig.~\ref{fig-box}.    
The Boltzmann weight of this box can be conveniently associated with
an $R$-matrix
\begin{equation}\label{def3}
{\mathbb R}(\xb,\yb)_{a\,,\,b}^{a',\,b'}\;=\;
\frac{V_{y'/x'}(b-a') V_{x/y'}(a'-b') V_{y/x}(b'-a)}{V_{y/x'}(b-a)}\;,
\end{equation}
where $\xb=(x,x')$ and $\yb=(y,y')$ stand for the pairs of spectral
variables in the vertical and 
horizontal directions. The spins $a,b,a',b'\in{\mathbb Z}$ are
supposed to satisfy the relations 
\be\label{dom2}
b\ge a'\ge b'\ge a\,.
\ee
Outside this domain the RHS of \eqref{def3} is 
assumed to vanish identically. 
The above inequalities ensure that all the spin differences appearing
in \eqref{def3} 
are non-negative, as it is required for admissible lattice spin 
configurations defined in Sect.~\ref{isingtype}.

Naturally, as the reader might have expected, the $R$-matrix
\eqref{def3} satisfies the vertex form Yang-Baxter equation 
\begin{equation}\label{ybe4}
\sum_{a',\,b',\,c'}\Rbb(\xb,\yb)_{a\,,\,b}^{a',\,b'} \,
\Rbb(\xb,\zb)_{a'\;,\,c}^{a'',\,c'}\,
\Rbb(\yb,\zb)_{b'\;,\,c'}^{b'',\,c''}
\;=\;
\sum_{a',\,b',\,c'}
\Rbb(\yb,\zb)_{b\,,\,c}^{b',\,c'}\,
\Rbb(\xb,\zb)_{a\,,\,c'}^{a',\,c''}\,
\Rbb(\xb,\yb)_{a'\;,\,b'}^{a'',\,b''}\;.
\end{equation}
A sketch of the proof, based on the star-star relation \eqref{ssrel}, 
is presented
in the Appendix~\ref{box-YBE}. Note also, that 
apparently related factorized $R$-matrices
in an operator form were previously 
constructed in \cite{Faddeev:1994,Derkachev:2006}.

At first sight the above equation \eqref{ybe4} looks identical to
\eqref{Rybe} (apart from the relabelling of spin indices). The
difference is in the range of allowed values of these indices.
For \eqref{Rybe} the indices
can take arbitrary non-negative integer values and the restrictions on
summations are determined by the conservation laws of the type 
\eqref{ar-cons} for
every vertex. Contrary to this the external indices in \eqref{ybe4} 
take arbitrary integer values, satisfying the inequalities 
\be
a\le c''\le b'' \le a'' \le c\,,\qquad a\le b \le c\,,
\ee
resulting from the the conditions
\eqref{dom2} for each $R$-matrix.
By the same reason the summation indices in  \eqref{ybe4}
are restricted by  
\begin{equation}
a\leq b'\leq c''\;,\quad b'' \leq c' \leq a''\;,\quad b'\leq a' 
\leq \min (c',b) 
\end{equation}
for the LHS and
\begin{equation}
c''\leq a'\leq b''\;,\quad a'' \leq b' \leq c\;,\quad \max (a',b) \leq
c' \leq b' 
\end{equation}
for the RHS. Note, that all the summations in \eqref{ybe4} go over
finite intervals.   

A remarkable feature of the $R$-matrix \eqref{def3} is that its 
matrix elements factorize into a product of factors depending 
only on two spins connected by an edge.
On the other hand the $R$-matrices \eqref{def3} and 
\eqref{R-def2} describe the same model and, it is natural to expect
that they should be related 
to each other by a linear transformation.\footnote{%
In the case when $q$ is a root
of unity relations of this type were obtained in
\cite{Baxter:1986df,Boos:1997}.} The corresponding relation is given
by a discrete Fourier transform 
\begin{equation}\label{fourier}
\sum_{a',b'\in{\mathbb Z}}\mathbb{R}(\xb,\yb)_{a\,,\,b}^{a',\,b'}\ 
\Big(\frac{x}{x'}q^{2j_1}\Big)^{\varepsilon a'} 
\Big(\frac{y}{y'}q^{2j_2}\Big)^{\varepsilon b'}\;=\;
\sum_{i_1,i_2\ge0}\Big(\frac{x}{x'}q^{2i_1}\Big)^{\varepsilon a}
 \Big(\frac{y}{y'}q^{2i_2}\Big)^{\varepsilon b}\, 
\Big(\frac{\ds y^{\prime j_1}x^{j_2}}{\ds y^{i_1}x^{\prime
    i_2}}\Big)^{\varepsilon}
 \;\mathcal{R}^T(\xb,\yb)_{i_1,i_2}^{j_1,j_2}
\end{equation}
where $\xb=(x,x')$, $\yb=(y,y')$, $\varepsilon=\pm1$,  and the superscript $T$
denotes 
the matrix transposition in both spaces
\be\label{Rtr}
{\cal R}^T(\xb,\yb)_{i_1,i_2}^{j_1,j_2}=
{\cal R}(\xb,\yb)_{j_1,j_2}^{i_1,i_2}\,.
\ee
Note, that Eq.\eqref{fourier} contains 
two different (but equivalent) relations with $\varepsilon=+1$ and 
$\varepsilon=-1$. A proof of \eqref{fourier} is presented below. 
Separate considerations of the second diagram (on the right side of
Fig.~\ref{fig-box}) are really not required, since its Boltzmann weight is
expressed through \eqref{def3} as ${\mathbb R}(\xb,\yb)_{-a',-b'}^{-a,\,\,-b}$ and all relevant formulae can be obtained by simple symmetry transformations.

\subsection{The $RLL$-relations.}

As a preparation to the proof of \eqref{fourier} consider a few important  
cases of the Yang-Baxter equation \eqref{Rybe}, 
which are sufficient for an {\em ab initio} calculation of 
the $R$-matrix \eqref{R-def2}. 
First, quote the defining relation for the $L$-operator,  
\be
\begin{array}{r}
\ds\sum_{j_1,j_2,j_3}
{\cal L}(x,x',y)_{i_1,i_2}^{j_1,j_2}\ \ 
{\cal L}(x,x',z)_{j_1,i_3}^{k_1,j_3}\ \
{\cal R}^{\rm (6v)}(y/z)_{j_2,j_3}^{k_2,k_3}\phantom{\qquad
\qquad }
\\[.6cm]
\ds =\sum_{j_1,j_2,j_3}
{\cal R}^{\rm (6v)}(y/z)_{i_2,i_3}^{j_2,j_3}\ \
{\cal L}(x,x',z)_{i_1,j_3}^{j_1,k_3}\ \
{\cal L}(x,x',y)_{j_1,j_2}^{k_1,k_2}
\end{array}\label{LLR1}
\ee
where ${\cal R}^{\rm (6v)}$ is given by \eqref{R6v}. 
This equation is obtained from \eqref{Rybe} via the
$2$-dimensional reduction (see \eqref{Lop-first} in Sect.~\ref{Lop-red}) 
in the second and third spaces when $y'=q y$
and $z'=q z$. The indices $i_2,j_2,k_2,i_3,j_3,k_3=0,1$ take two values, while 
the indices $i_1,j_1,k_1\ge0$ take arbitrary non-negative integer values.
It is well known
that with the substitution \eqref{Lop} the above equation just reduces to 
the commutation relations \eqref{uqsl2} of the algebra $U_q(sl(2))$. 
Thus, any representation of this algebra gives a solution of
\eqref{LLR1}. 
Here we choose the representation \eqref{piplus_s1},
which leads to the expression \eqref{Lmat} for the matrix elements of 
${{\cal L}}(x,x',y)_{i_1,i_2}^{j_1,j_2}$. 

Next, consider the $2$-dimensional reduction of \eqref{Rybe} in the
third space, when $z'=q z$,
\be
\begin{array}{r}
\ds\sum_{j_1,j_2\in{\mathbb Z}_{\ge0}}\ 
\bm{{\cal L}}(y,y',z)_{i_2}^{j_2}\ 
\bm{{\cal L}}(x,x',z)_{i_1}^{j_1}\ 
{\cal R}(x,x',y,y')_{j_1,j_2}^{k_1,k_2}\phantom{\qquad
\qquad }
\\[.6cm]
\ds =\sum_{j_1,j_2}
\ {\cal R}(x,x',y,y')_{i_1,i_2}^{j_1,j_2}\ 
\bm{{\cal L}}(x,x',z)_{j_1}^{k_1}\ 
\bm{{\cal L}}(y,y',z)_{j_2}^{k_2}\,.
\end{array}\label{LLR2}
\ee
Note, that the boldface
$
\bm{{\cal L}}(x,x',z)_{i}^{j}$ and  
$\bm{{\cal L}}(y,y',z)_{i}^{j}$ 
are defined by \eqref{Lmat}.
They are considered as
two-by-two matrices and the above equation
involves their matrix product.  
As usual we call Eq.\eqref{LLR2} the ``$RLL$-relation''. 
Once the $L$-operators $
\bm{{\cal L}}(x,x',z)$ and  
$\bm{{\cal L}}(y,y',z)$ are fixed by \eqref{Lmat} 
this relation determines ${\cal R}(x,x',y,y')$ to within an overall
normalization (see, e.g., Appendix~C of ref.\cite{Mangazeev:2014gwa}).
Furthermore,  
by using \eqref{fourier} this relation can
be transformed to another $RLL$-relation, which uniquely determines the
factorized $R$-matrix \eqref{def3}. 

To proceed, first note the Yang-Baxter equations \eqref{LLR1} and \eqref{LLR2} are invariant under the substitution 
\be\label{t-trans}
\bm{{\cal R}}\to 
\widetilde{\bm{{\cal R}}}\,,\qquad\qquad
\bm{{\cal L}}\to 
\widetilde{\bm{{\cal L}}}\,,\qquad\qquad
\bm{{\cal R}}^{(6v)}\to 
\widetilde{\bm{{\cal R}}}^{(6v)}\,,\qquad\qquad
\ee
where the modified $R$-matrices are defined by \eqref{R-tilde},
\eqref{Lop-second} and 
\be\label{R6v-tilde}
\widetilde{\mathcal R}^{(6v)}(\lambda)_{i_1,i_2}^{j_1,j_2}=
q^{i_2-j_1}\,\lambda^{i_2-j_2}\,
{\mathcal R}^{(6v)}(\lambda)_{i_1,i_2}^{j_1,j_2}\,.
\ee
As before, we will consider $\widetilde{\bm{{\cal
      L}}}_i^j$, defined by \eqref{Lmat4},
as a two-by-two matrix in the second space,
with operator entries acting in the first space 
(the indices $i,j$ label their elements). 
Changing basis in this space via the formula
\be\label{L-trans}
\sum_{a'}\,\bm{{\mathbbm L}}(x,x',y)_{a}^{a'}\ \Big(\frac{x}{x'} \,q^{2
  j}\Big)^{-a'} =\sum_{i} \Big(\frac{x}{x'}\,q^{2i}\Big)^{-a}\ 
\widetilde{\bm{{\mathcal L}}}(x,x',y)_i^j
\ee
and using \eqref{Lmat4}, 
one obtains the matrix elements of the transformed $L$-operator
\be\label{Lop-new}
\bm{{\mathbbm L}}(x,x',y)_{a}^{a'}=
\begin{pmatrix}\ds
\frac{x}{y}\, \delta_{a,a'-1}-\frac{y}{x'}\,\delta_{a,a'}&
\ds q^{-2a}\Big(\,\frac{y}{x}\,\delta_{a,a'-1}-\frac{y}{x'}\,
\delta_{a,a'}\,\Big)\\[.5cm]
\ds q^{2a}\Big(\,\frac{x'}{y}\,\delta_{a,a'}
- \frac{x}{y}\,\delta_{a,a'-1}\,\Big)& 
\ds \frac{x'}{y}\,\delta_{a,a'}-\frac{y}{x}\,\delta_{a,a'-1} 
\end{pmatrix}\,.
\ee
Evidently, it still has the form \eqref{L-form}, but the Weyl algebra \eqref{weyl} is now realized as
\begin{equation}\label{uva}
\boldsymbol{u}\,|\, a\, \rangle \;=\; |\, a-1\, \rangle \;,\quad
\boldsymbol{v}\,|\, a\, \rangle \;=\; |\, a\, \rangle\, q^{2a}\;.
\end{equation}

Next, taking the ``tilded'' version of \eqref{LLR2}, obtained by the symmetry transformation \eqref{t-trans}, and using \eqref{L-trans} and the variant of \eqref{fourier} with
$\varepsilon=-1$ one obtains
\begin{equation}\label{RLL-irf}
\begin{array}{r}\ds
\sum_{a',b'} \mathbb{R}(x,x',y,y')_{a\,,\,b}^{a',\,b'} \ 
\bm{{\mathbbm L}}(x,x',z)_{a'}^{a''}\  \bm{{\mathbbm
    L}}(y,y',z)_{b'}^{b''}{z\qquad \qquad \qquad}\\[.5cm]
=\ds\sum_{a',b'}
\bm{{\mathbbm{L}}}(y,y',z)_{b}^{b'}\  \bm{{\mathbbm{L}}}(x,x',z)_{a}^{a'}
\ 
\mathbb{R}(x,x',y,y')_{a'\;,\,b'}^{a'',\,b''}\,.
\end{array}
\end{equation}
It should be stressed that this relation is equivalent to
\eqref{LLR2}. It is obtained from the latter by the symmetry transformation 
\eqref{t-trans}, followed by a linear
transformation, changing basis in the first and second vector spaces. 
At this stage we regard \eqref{fourier} 
as the definition of   
$\mathbb{R}(x,x',y,y')$. 
However, remembering that the new $RLL$-relation \eqref{RLL-irf} (with
the $L$-operators specified by \eqref{Lop-new}) determines
$\mathbb{R}(x,x',y,y')$ up to an overall normalization, one concludes that 
in order to prove \eqref{fourier} it suffices to show
that the factorized expression \eqref{def3} satisfies \eqref{RLL-irf}.

\subsection{Propagation through the vertex techniques}
The matrix elements of the $L$-operator \eqref{Lop-new} possess
important factorization properties, which we describe below. Let 
\be
\bm{v}_0=\begin{pmatrix}1\\0\end{pmatrix}\,,
\qquad \bm{v}_1=\begin{pmatrix}0\\1\end{pmatrix}\,,\qquad
\overline{\bm{v}}_i=\bm{v}^t_i\,,\qquad i=0,1\,,\label{v-basis}
\ee
be the two-dimensional basis vectors.
Introduce the following set of vectors labelled by $a,a'\in{\mathbb Z}$ 
\be\label{sos-vertex}
\bm{\Phi}(x,y)_{a}^{a'}=\sum_{i=0,1}\Phi(x,y)_{a,i}^{a'}\, 
\bm{v}_i\,,\qquad
\overline{\bm{\Phi}}(x,y)_{a}^{a'}=
\sum_{i=0,1}\overline{\Phi}(x,y)_{a,i}^{a'}\, 
\overline{\bm{v}}_i\,,
\ee
with the components 
\be\label{Phi-def}
\begin{array}{rcl}
\ds\Phi(x,y)_{a,i}^{a'}&=&\ds\sigma_i\,q^{-\sigma_i a}\,
\big((x/y)^{-\sigma_i}\,\delta_{a,a'}+\delta_{a,a'-1}\big)\,,\\[.3cm]
\ds\overline{\Phi}(x,y)_{a,i}^{a'}&=&\ds\phantom{\sigma_i\,}q^{+\sigma_i a}\,
\big(\delta_{a,a'}-(x/y)^{\sigma_i}\,\delta_{a,a'-1}\big)\,,
\end{array}
\ee
where $i=0,1$ and  $\sigma_0=+1$, $\sigma_1=-1$.
The above vectors satisfy important orthogonality relations
\begin{equation}\label{Phi-ort}
\sum_{a'} \Phi(x,y)_{a,i}^{a'}\,
\overline{\Phi}(x,y)_{a,j}^{a'}\;=\;[y/x]\,
\delta_{i,j}\;,\qquad
\sum_{i} \overline{\Phi}(x,y)_{a,i}^{a'}\,
\Phi(x,y)_{a,i}^{a''}\;=\;[y/x]\,\delta_{a',a''}\;.
\end{equation}
Using the explicit form of the two-by-two $L$-operator \eqref{Lop-new} 
it is easy to check that 
\begin{equation}\label{LL2}
\begin{tikzpicture}[baseline=(current  bounding  box.center)]
\draw [fill, opacity=0.1, blue] (1.5,-1.5) -- (1.5,1.5) -- (2,1.7) -- (2.5,1.5) -- (2.5,-1.5) -- (2,-1.7) -- (1.5,-1.5);
\draw [-open triangle 45, thin] (1.5,1.5) -- (1.5,-1.5); \node [below] at (1.5,-1.5) {$x'$};
\draw [-open triangle 45, thin] (2.5,-1.5) -- (2.5,1.5); \node [above] at (2.5,1.5) {$x$};
\draw [thick] (1.5,0) -- (0.8,0); 
\draw [-latex, thick] (0,0) -- (0.8,0);
\node [left] at (0,0) {$i$}; \node [right] at (4,0) {$i'$};
\node [below] at (0.45,0) {$z$};
\node [above] at (3.3,0) {$z$};
\draw [thick, dashed] (2.5,0) -- (1.5,0); 
\draw [thick] (4,0) -- (3.5,0); 
\draw [-latex,thick] (2.5,0) -- (3.5,0);

\node [anchor=south east] at (1.5,0) {$\Phi$};
\node [anchor=north west] at (2.5,0) {$\overline\Phi$};
\node [below] at (2,-1) {$a$};
\node [above] at (2,1) {$a'$};
\node [left] at (-0.6,0)
      {$\big(\mathbb{L}(x,x',z)_{a}^{a'}\big)_{i,i'}
\;=\;\Phi(x',z)_{a,i}^{a'}\,
\overline{\Phi}(x,z)_{a,i'}^{a'}\;=\;$};
\end{tikzpicture}
\end{equation}
Recall, that the indices $(i,i')=(0,0)$ refer to the top left element 
in \eqref{Lop-new}.  
Here we have used the following graphical notations 
\begin{equation}\label{Psi2}
\begin{tikzpicture}[baseline=(current  bounding  box.center)]
\draw [fill, opacity=0.2, blue,path fading=east] (1.5,0) -- (1.5,3) -- (3.0,3) -- (3.0,0) -- (1.5,0);
\draw [-open triangle 45, black, thin] (1.5,3) -- (1.5,0); \node [above] at (1.5,3) {$x$};
\draw [thick] (1.5,1.5) -- (0.80,1.5); \draw [-latex,thick] (0,1.5) --  
(0.8,1.5);
\node [left] at (0,1.5) {$i$};
\node [below] at (0.45,1.5) {$z$};
\node [anchor = south east] at (1.5,1.5) {$\Phi$};
\draw[dashed, thick] (1.5,1.5) -- (3.0,1.5); \node[right] at (2.0,2.5)
     {$a'$}; \node[right] at (2.0,0.5) {$a$};  
\node[left] at (-0.5,1.5) {$\ds \Phi(x,z)^{a'}_{a,i}\;=$};
\end{tikzpicture}
\qquad\qquad \qquad
\begin{tikzpicture}[baseline=(current  bounding  box.center)]
\draw [fill, opacity=0.2, blue, path fading=west] (1.5,0) -- (1.5,3) -- (0,3) -- (0,0) -- (1.5,0);
\draw [-open triangle 45, black, thin] (1.5,0) -- (1.5,3); \node [above] at (1.5,3) {$x$};
\draw [dashed, thick] (1.5,1.5) -- (0,1.5); \node[right] at (3,1.5) {$i$};
\node [above] at (2.3,1.5) {$z$};
\node [anchor = north west] at (1.5,1.5) {$\overline{\Phi}$};
\draw[thick] (3,1.5) -- (2.5,1.5); \draw [-latex, thick] 
(1.5,1.5) -- (2.5,1.5); 
\node[left] at (1,2.5) {$a'$}; \node[left] at (1,0.5) {$a$};
\node[left] at (0,1.5) {$\ds
  \overline{\Phi}(x,z)_{a,i}^{a'}\;=\ $};
\end{tikzpicture}
\end{equation}
The are two types of spin variables in these pictures. 
The variables $a,a'\in{\mathbb Z}$ are the ``face spins'' 
assigned to the shaded faces, while $i,i'=\pm1$ are the edge
spins (like in vertex models) assigned to the directed solid lines in
the unshaded areas.  Moreover these lines (as well as their dashed
continuations into shaded areas) carry a spectral variable $z$. Another
spectral parameter $x$ is assigned to the directed vertical lines,
which separate the shaded and unshaded areas.

\begin{figure}[ht]
\begin{center}
\begin{tikzpicture}[scale=.9,baseline=(current  bounding  box.center)]
\draw [-open triangle 45, thin] (2,3) -- (2,-1); \node [below] at
(2,-1) {$x'$}; 
\draw [thick] (2,0) -- (-2.,2); 
\draw [-latex, thick] (-2,2) -- (-1.5,1.75); 

\node[left] at (-2,2) {$i_3$};
\draw [dashed, thick] (2,0) -- (3,-0.5);
\node [above] at (-0,1.2) {$\widetilde{\cal R}^{(6v)}$};
\draw [black, ultra thick, fill] (0,1) circle [radius=0.09];
\node [anchor = south east] at (2.05,.2) {$\Phi$};
\draw [thick] (2,2) -- (-2,.0); 
\draw [-latex, thick] (-2,0) -- (-1.5,.25); 
\node[left] at (-2,0) {$i_2$};
\draw [dashed, thick] (2,2) -- (3,2.5);
\node [anchor = south east] at (2.05,1.95) {$\Phi$};
\node [above] at (1,1.5) {$j_2$};
\node [below] at (1,0.5) {$j_3$};
\node [above] at (-1.8,1.3) {$z$};
\node [below] at (-1.8,0.7) {$y$};
\node [right] at (2.25,-0.9) {$d$};
\node [right] at (2.25,1) {$b$};
\node [right] at (2.25,2.9) {$a$};
\draw [fill, opacity=0.25, blue,path fading=east] (2,-1) -- (2,3) -- (3,3.5) -- (3,-1.5) -- (2,-1);
\node [right] at (4,1) {$=$};
\end{tikzpicture}
\hskip 2mm
\begin{tikzpicture}[scale=.9,baseline=(current  bounding  box.center)]
\draw [-open triangle 45, thin] (0,3) -- (0,-1); \node [below] at (0,-1) {$x'$};
\draw [thick] (0,0) -- (-1,-0.5); 
\draw [-latex, thick] (-1,-0.5)--(-0.5,-0.25); 
\node[left] at (-1,2.5) {$i_3$};
\draw [dashed, thick] (0,0) -- (4,2);
\draw [thick] (0,2) -- (-1,2.5); 
\draw [-latex, thick]  (-1,2.5) -- (-0.5,2.25); 
\node[left] at (-1,-0.5) {$i_2$};
\draw [dashed, thick] (0,2) -- (4,0.);
\node [below] at (-0.7,2.3) {$z$};
\node [below] at (-0.7,-0.4) {$y$};
\node [anchor = south east] at (0,0) {$\Phi$};
\node [anchor = south east] at (0,2.2) {$\Phi$};
\node [right] at (0.4,1) {$c$};
\node [above] at (2,2) {$a$};
\node [below] at (2,0.) {$d$};
\node [right] at (3,1) {$b$};
\draw [fill, opacity=0.25, blue,path fading=east] (0,-1) -- (0,3) -- (4,3.5) -- (4,-1.5) -- (0,-1);
\node [above] at (2,1.2) {${\cal R}^{(6v)}$};
\draw [black, ultra thick, fill] (2,1) circle [radius=0.09];
\end{tikzpicture}
\end{center}
\caption{Graphical representation of the 
SOS-vertex duality \eqref{sos-duality} for the 6-vertex model.}
\label{fig-Rpsi-1}
\end{figure}
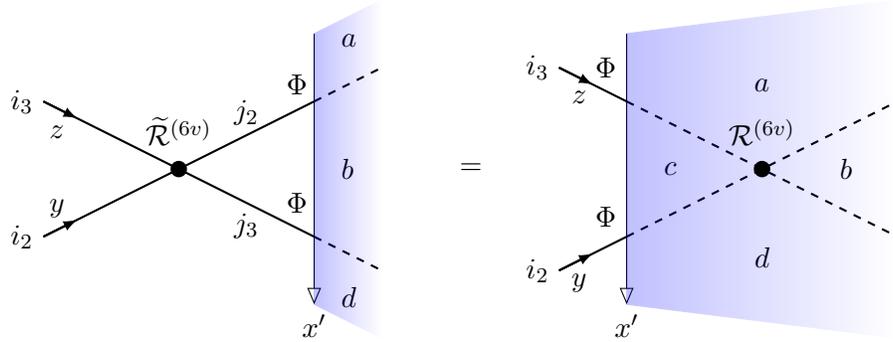

The vectors \eqref{sos-vertex} enter the so-called ``SOS-vertex''
correspondence and ``propagation through the vertex techniques''
invented by Baxter \cite{Baxter:1972,Bax73b} and further developed by
many authors including \cite{Andrews:1984,Date:1986,Krichever:1981}. 
The approach we use here is most similar to that of \cite{BKMS}, 
based on the factorized form of the $L$-operator \eqref{LL2}.   
The key relation of the SOS-vertex correspondence in our case reads
\begin{equation}\label{sos-duality}
\sum_{j_2,j_3=0,1} 
\; \widetilde{\mathcal{R}}^{(6v)}(y/z)_{i_2,i_3}^{j_2,j_3} \,
\Phi(x',z)_{d,j_3}^{b} \,
\Phi(x',y)_{b,j_2}^{a}\;=\;
\sum_{c=d}^a \; \Phi(x',y)_{d,i_2}^{c} \,
\Phi(x',z)_{c,i_3}^{a}\,
\mathcal{R}^{(6v)}(y/z)_{c-d,\;a-c}^{a-b,\;b-d}
\end{equation}
where $\mathcal{R}^{(6v)}$ and   $\widetilde{\mathcal{R}}^{(6v)}$ 
are given by \eqref{R6v} and \eqref{R6v-tilde}, respectively. 
This relation is presented graphically in
Fig.~\ref{fig-Rpsi-1}. Using the orthogonality \eqref{Phi-ort} one can
convert \eqref{sos-duality} into a similar relation containing the
$\overline{\Phi}_a^{a'}$ vectors instead of ${\Phi}_a^{a'}$,
\begin{equation}
\sum_{i_2,i_3=0,1} \; 
\overline{\Phi}(x,y)_{d,i_2}^{c}\,
\overline{\Phi}(x,z)_{c,i_3}^{a} \,
\widetilde{\mathcal{R}}^{(6v)}(y/z)_{i_2,i_3}^{j_2,j_3} 
\;=\;
\sum_{b=d}^a \; 
\mathcal{R}^{(6v)}(y/z)_{c-d,\;a-c}^{a-b,\;b-d} \,
\overline{\Phi}(x,z)_{d,j_3}^{b}\,
\overline{\Phi}(x,y)_{b,j_2}^{a}\,.
\end{equation}
%

Similarly to \eqref{sos-vertex} we need to define additional vectors
\be\label{omega-def}
\begin{array}{rcl}
\ds\Omega(x,y)_{a,i}^{a'}&=&\ds\sigma_i\,q^{-\sigma_i a}\,
\big(\delta_{a,a'}+(y q/x)^{-\sigma_i}\,\delta_{a,a'-1}\big)\,,\\[.3cm]
\ds\overline{\Omega}(x,y)_{a,i}^{a'}&=&\ds\phantom{\sigma_i\,}q^{+\sigma_i a}\,
\big((y/x)^{\sigma_i}\delta_{a,a'}-q^{\sigma_i}\,\delta_{a,a'-1}\big)\,,
\end{array}
\ee
satisfying the following orthogonality
relations
\begin{equation}\label{omega-ort}
\sum_{a} \Omega(x,y)_{a,i}^{a'}\,
\overline{\Omega}(x,y)_{a,i'}^{a'}\;=\;[y/x]\,
\delta_{i,i'}\;,\quad
\sum_{i} \overline{\Omega}(x,y)_{a',i}^{a}\,
\Omega(x,y)_{a'',i}^{a}\;=\;[y/x]\,\delta_{a',a''}\;.
\end{equation}
These vectors are represented graphically as 
\begin{equation}\label{Psi1}
\begin{tikzpicture}[baseline=(current  bounding  box.center)]
\draw [fill, opacity=0.2, blue,path fading=east] (1.5,0) -- (1.5,3) -- (3,3) -- (3,0) -- (1.5,0);
\draw [-open triangle 45, black, thin] (1.5,0) -- (1.5,3); \node [above] at (1.5,3) {$x$};
\draw [thick] (1.5,1.5) -- (0.80,1.5); \draw [-latex,thick] (0,1.5) --  
(0.8,1.5);
\node [left] at (0,1.5) {$i$};
\node [below] at (0.45,1.5) {$z$};
\node [anchor = south east] at (1.5,1.5) {$\Omega$};
\draw[dotted, thick] (1.5,1.5) -- (3,1.5); \node[right] at (2,2.5) {$a'$}; \node[right] at (2,0.5) {$a$};
\node[left] at (-0.5,1.5) {$\ds \Omega(x,z)_{a,i}^{a'}\;=$};
\end{tikzpicture}
\qquad\qquad \qquad
\begin{tikzpicture}[baseline=(current  bounding  box.center)]
\draw [fill, opacity=0.2, blue,path fading=west] (1.5,0) -- (1.5,3) -- (0,3) -- (0,0) -- (1.5,0);
\draw [-open triangle 45, black, thin] (1.5,3) -- (1.5,0); \node
      [above] at (1.5,3) {$x$};
\draw [dotted, thick] (1.5,1.5) -- (0,1.5); \node[right] at (3,1.5) {$i$};
\node [above] at (2.3,1.5) {$z$};
\node [anchor = north west] at (1.5,1.5) {$\overline{\Omega}$};
\draw[thick] (3,1.5) -- (2.5,1.5); \draw [-latex, thick] 
(1.5,1.5) -- (2.5,1.5);

\node[left] at (1,2.5) {$a'$}; \node[left] at (1,0.5) {$a$};
\node[left] at (0,1.5) {$\ds
  \overline{\Omega}(x,z)^{a'}_{a,i}\;=\ $};
\end{tikzpicture}
\end{equation}
Note that here the directions of the
vertical lines, carrying the spectral variable $x$, 
are reversed in comparison with \eqref{Psi2}.

Now, return to Fig.~\ref{fig-lattice} on page \pageref{fig-lattice}
and shade alternatively all faces of the 
medial lattice to cover those containing the white and black sites (the medial
lattice is a square lattice formed by the thin rapidity lines). Then we
can regard the spins in our Ising-type model as face spins and update
the graphical notations \eqref{twotypes} for the edge Boltzmann weights
\begin{equation}
\begin{tikzpicture}
\draw [fill, opacity=.2,blue,path fading=north] 
(0,2) -- (1,1) -- (2,2) --(1,3);
\node [left] at (-0.5,1) {$({i}):$};
\draw [fill, opacity=.2,blue,path fading=south] 
(0,0) -- (1,-1) -- (2,0) --(1,1);
\draw [-open triangle 45, black, thin] (0,0) -- (2,2);
\draw [-open triangle 45, black, thin] (2,0) -- (0,2);
\draw [\STE, ultra thick] (1,0) circle [radius=0.09];
\draw [\STE, ultra thick] (1,2) circle [radius=0.09];
\draw [-latex, \PAC, ultra thick] (1,0.12) -- (1,1);
\draw [\PAC, ultra thick] (1,1) -- (1,1.88);
\node [above] at (2,2) {$x$};
\node [above] at (0,2) {$y$};
\node [above] at (1,2.1) {$a$};
\node [below] at (1,-0.1) {$b$};
\node [right] at (2.1,1) {$
\ds =\;V_{x/y}(a-b)\;,
$};
\end{tikzpicture}
\qquad
\begin{tikzpicture}
\node [left] at (-0.5,1) {$({ii}):$};
\draw [fill, opacity=.2,blue,path fading=north] 
(0,2) -- (1,1) -- (2,2) --(1,3);
\draw [fill, opacity=.2,blue,path fading=south] 
(0,0) -- (1,-1) -- (2,0) --(1,1);
\draw [-open triangle 45, black, thin] (2,2) -- (0,0);
\draw [-open triangle 45, black, thin] (0,2) -- (2,0);
\draw [\STE, ultra thick] (1,0) circle [radius=0.09];
\draw [\STE, ultra thick] (1,2) circle [radius=0.09];
\draw [-latex, \GR, ultra thick] (1,0.12) -- (1,1); 
\draw [\GR, ultra thick] (1,1) -- (1,1.88);
\node [below] at (2,0) {$x$};
\node [below] at (0,0) {$y$};
\node [above] at (1,2.1) {$a$};
\node [below] at (1,-0.1) {$b$};
\node [right] at (2.1,1) {$
\ds =\;\frac{1}{V_{x/y}(a-b)}\;.
$};
\end{tikzpicture}
\label{V-shade}
\end{equation}
where $V_x(n)$ is defined in \eqref{Vdef}.

Now we can present three easily verifiable Yang-Baxter relations
involving the SOS-vertex vectors \eqref{Psi2}, \eqref{Psi1} and the
edge weights \eqref{V-shade}. The first equation is 
\begin{equation}\label{psieq1}
V_{x/y}(a-b) \; \sum_{i} \overline{\Phi}(x,z)_{a,i}^{a'}\, \Phi(y,z)_{b,i}^{b'} \;=\;
\sum_{i} \overline{\Omega}(y,z)_{a,i}^{a'} \,\Omega(x,z)_{b,i}^{b'} \; V_{x/y}(a'-b')\;.
\end{equation}
It has the following graphical representation
%
\begin{equation}\label{peq1}
\begin{tikzpicture}[scale=0.65]
\draw [fill, opacity=0.2, blue, path fading=west] (-2,-1) -- (-0.5,-1) -- (0,0) -- (-2,4) -- (-4,4) -- (-2,-1);
\draw [fill, opacity=0.2, blue, path fading=east] (2,-1) -- (0.5,-1) -- (0,0) -- (2,4) -- (4,4) -- (2,-1);
\draw [-open triangle 45, black, thin] (2,4) -- (-0.5,-1); \node [above] at (2,4) {$y$};
\draw [-open triangle 45, black, thin] (0.5,-1) -- (-2,4); \node [above] at (-2,4) {$x$};
\draw [-latex, thick] (-1,2) -- (0,2); \draw [thick] (1,2) -- (-1,2);
\draw[dashed, thick] (-1,2) -- (-3.2,2);  \draw[dashed, thick] (1,2) -- (3.2,2); \node [below] at (0,2) {$z$};
\draw [\STE, ultra thick, fill] (1.3,0) circle [radius=0.09]; \draw [\STE, ultra thick, fill] (-1.3,0) circle [radius=0.09];
\draw [-latex, \PAC, ultra thick] (1.3,0) -- (0,0); \draw[\PAC, ultra thick] (0,0) -- (-1.3,0);
\node [left] at (-1.35,0) {$a$}; \node[right] at (1.35,0) {$b$};
\node [left] at (-2.1,3) {$a'$}; \node [right] at (2.1,3) {$b'$};
\node [anchor=south west] at (-1,2) {$\overline{\Phi}$};
\node [anchor=south east] at (1,2) {$\Phi$};
\node [right] at (4.5,1.5) {$=$};
\end{tikzpicture}
\hskip 10mm
\begin{tikzpicture}[scale=0.65,baseline=-40]
\draw [fill, opacity=0.2, blue,path fading=west] (-4,-2) -- (-2,-2) -- (0,2) -- (-0.5,3) -- (-2,3) -- (-4,-2);
\draw [fill, opacity=0.2, blue,path fading=east] (4,-2) -- (2,-2) -- (0,2) -- (0.5,3) -- (2,3) -- (4,-2);
\draw [-open triangle 45, black, thin] (0.5,3) -- (-2,-2) ; \node [below] at (-2,-2) {$y$};
\draw [-open triangle 45, black, thin] (2,-2) -- (-0.5,3); \node [below] at (2,-2) {$x$};
\draw [-latex, thick] (-1,0) -- (0,0); \draw [thick] (1,0) -- (-1,0);
\draw[dashed, thick] (-1,0) -- (-3.2,0);  \draw[dashed, thick] (1,0) -- (3.2,0); \node [above] at (0,0) {$z$};
\draw [\STE, ultra thick, fill] (1.3,2) circle [radius=0.08]; \draw [\STE, ultra thick, fill] (-1.3,2) circle [radius=0.09];
\draw [-latex, \PAC, ultra thick] (1.3,2) -- (0,2); \draw[\PAC, ultra thick] (0,2) -- (-1.3,2);
\node [left] at (-1.35,2) {$a'$}; \node[right] at (1.35,2) {$b'$};
\node [left] at (-2.1,-1) {$a$}; \node [right] at (2.1,-1) {$b$};
\node [anchor=north west] at (-1,0) {$\overline{\Omega}$};
\node [anchor=north east] at (1,0) {$\Omega$};
\end{tikzpicture}
\end{equation}
Note, that after a substitution of the definitions \eqref{Psi2}, \eqref{Psi1}
this relation uniquely determines the function $V_x(n)$, defined by
\eqref{Vdef}-\eqref{vanish}, up to an overall normalization.  The next
relation is 
\be
\label{psieq2}
\sum_{b} V_{y/x}(a-b)\  \Phi(x,z)_{b,i}^{c} \  
\overline{\Omega}(y,z)_{b,i'}^{c}\;=\;
\sum_{b} \Phi(y,z)_{a,i}^{b} \ 
\overline{\Omega}(x,z)_{a,i'}^{b}\  
V_{y/x}(b-c)\;,
\ee
It is represented graphically as
%
\begin{equation}\label{peq2}
\begin{tikzpicture}[scale=0.7,baseline=(current  bounding  box.center)]
\draw [fill, opacity=0.25, blue, path fading=north] (0,1) -- (-1.5,4) -- (0,4.5) -- (1.5,4) -- (0,1);
\draw [fill, opacity=0.2, blue,path fading=south] (0,1) -- (1,-1) -- (0,-1.5) -- (-1,-1) -- (0,1);
\draw [-open triangle 45, black, thin] (1.5,4) -- (-1,-1); \node [above] at (1.5,4) {$y$};
\draw [-open triangle 45, black, thin] (-1.5,4) -- (1,-1); \node [above] at (-1.5,4) {$x$};
\draw [dashed, thick] (1,3) -- (0,3); \draw [dashed, thick] (0,3) -- (-1,3);
\draw[-latex, thick] (-2,3) -- (-1.3,3); \draw [thick] (-1.,3) -- (-2,3);  
\draw[-latex, thick] (1,3) -- (1.7,3); \draw [thick] (2.,3) -- (1,3); 
\node [left] at (-2,3) {$i$}; \node [right] at (2,3) {$i'$};
\node [above] at (-1.75,3) {$z$};
\node [above] at (1.6,3) {$z$};
\draw [\STE, ultra thick, fill] (0,-0.3) circle [radius=0.09]; \draw [\STE, ultra thick, fill] (0,2.3) circle [radius=0.09];
\draw [-latex, \PAC, ultra thick] (0,2.3) -- (0,1); \draw [\STE, ultra thick] (0,1) -- (0,-0.3);
\node [below] at (0,-0.35) {$a$}; \node[above] at (0,2.34) {$b$};
\node [above] at (0,3.35) {$c$};
\node [anchor=north east] at (-1,3) {$\Phi$};
\node [anchor=north west] at (1,3) {$\overline{\Omega}$};
\node [right] at (3.5,1.5) {$=$};
\end{tikzpicture}
\hskip 10mm
\begin{tikzpicture}[scale=0.7,baseline=(current  bounding  box.center)]
\draw [fill, opacity=0.2, blue,path fading=north] (0,2) -- (-1,4) -- (0,4.5) -- (1,4) -- (0,2);
\draw [fill, opacity=0.25, blue,path fading=south] (0,2) -- (1.5,-1) -- (0,-1.5) -- (-1.5,-1) -- (0,2);
\draw [-open triangle 45, black, thin] (1,4) -- (-1.5,-1); \node [above] at (1,4) {$y$};
\draw [-open triangle 45, black, thin] (-1,4) -- (1.5,-1); \node [above] at (-1,4) {$x$};
\draw [dashed, thick] (1,0) -- (0,0); \draw [dashed, thick] (0,0) -- (-1,0);
\draw[-latex, thick] (-2,0) -- (-1.3,0); \draw [thick] (-1.,0) -- (-2,0);  
\draw[-latex, thick] (1,0) -- (1.7,0); \draw [thick] (2.,0) -- (1,0); 
\node [left] at (-2,0) {$i$}; \node [right] at (2,0) {$i'$};
\node [below] at (-1.75,0) {$z$};
\node [below] at (1.6,0) {$z$};
\draw [\STE, ultra thick, fill] (0,0.7) circle [radius=0.09]; \draw [\STE, ultra thick, fill] (0,3.3) circle [radius=0.09];
\draw [-latex, \PAC, ultra thick] (0,3.3) -- (0,2); \draw [\STE, ultra thick] (0,2) -- (0,0.7);
\node [below] at (0,0.66) {$b$}; \node[above] at (0,3.35) {$c$};
\node [below] at (0,-0.3) {$a$};
\node [anchor=south east] at (-1,0) {$\Phi$};
\node [anchor=south west] at (1,0) {$\overline{\Omega}$};
\end{tikzpicture}
\end{equation}
%
%
Note the summation over the spin $b$ corresponding to the internal
face. Finally, the third relation reads
\be
\label{psieq3}
\sum_{b} V_{y/x}(b-a)\  \Omega(x,z)_{b,i}^{c} \  
\overline{\Phi}(y,z)_{b,i'}^{c}\;=\;
\sum_{b} \Omega(y,z)_{a,i}^{b} \ 
\overline{\Phi}(x,z)_{a,i'}^{b}\  
V_{y/x}(c-b)\;.
\ee
It is depicted below
%
\begin{equation}\label{peq3}
\begin{tikzpicture}[scale=0.7,baseline=(current  bounding  box.center)]
\draw [fill, opacity=0.25, blue,path fading=north] (0,1) -- (-1.5,4) -- (0,4.5) -- (1.5,4) -- (0,1);
\draw [fill, opacity=0.2, blue,path fading=south] (0,1) -- (1,-1) -- (0,-1.5) -- (-1,-1) -- (0,1);
\draw [-open triangle 45, black, thin] (-1,-1) -- (1.5,4); \node [above] at (1.5,4) {$y$};
\draw [-open triangle 45, black, thin] (1,-1) -- (-1.5,4); \node [above] at (-1.5,4) {$x$};
\draw [dashed, thick] (1,3) -- (-1,3); 
\draw[-latex, thick] (-2,3) -- (-1.3,3); \draw [thick] (-1.,3) -- (-2,3);  
\draw[-latex, thick] (1,3) -- (1.7,3); \draw [thick] (2.,3) -- (1,3); 
\node [left] at (-2,3) {$i$}; \node [right] at (2,3) {$i'$};
\node [above] at (1.6,3) {$z$};
\node [above] at (-1.75,3) {$z$};
\draw [\STE, ultra thick, fill] (0,-0.3) circle [radius=0.09]; \draw [\STE, ultra thick, fill] (0,2.3) circle [radius=0.09];
\draw [-latex, \PAC, ultra thick] (0,-0.3) -- (0,1); \draw [\STE, ultra thick] (0,1) -- (0,2.3);
\node [below] at (0,-0.35) {$a$}; \node[above] at (0,2.34) {$b$};
\node [above] at (0,3.55) {$c$};
\node [anchor=north east] at (-1,3) {$\Omega$};
\node [anchor=north west] at (1,3) {$\overline{\Phi}$};
\node [right] at (3.5,1.5) {$=$};
\end{tikzpicture}
\hskip 10mm
\begin{tikzpicture}[scale=0.7,baseline=(current  bounding  box.center)]
\draw [fill, opacity=0.2, blue,path fading=north] (0,2) -- (-1,4) -- (0,4.5) -- (1,4) -- (0,2);
\draw [fill, opacity=0.25, blue,path fading=south] (0,2) -- (1.5,-1) -- (0,-1.5) -- (-1.5,-1) -- (0,2);
\draw [-open triangle 45, black, thin] (-1.5,-1) -- (1,4); \node [above] at (1,4) {$y$};
\draw [-open triangle 45, black, thin] (1.5,-1) -- (-1,4); \node [above] at (-1,4) {$x$};
\draw [dashed, thick] (1,0) -- (0,0); \draw [dashed, thick] (0,0) -- (-1,0);
\draw[-latex, thick] (-2,0) -- (-1.3,0); \draw [thick] (-1.,0) -- (-2,0); 
\draw[-latex, thick] (1,0) -- (1.7,0); \draw [thick] (2.,0) -- (1,0); 
\node [left] at (-2,0) {$i$}; \node [right] at (2,0) {$i'$};
\node [below] at (-1.75,0) {$z$};
\node [below] at (1.6,0) {$z$};
\draw [\STE, ultra thick, fill] (0,0.7) circle [radius=0.09]; \draw [\STE, ultra thick, fill] (0,3.3) circle [radius=0.09];
\draw [-latex, \PAC, ultra thick] (0,0.7) -- (0,2); \draw [\STE, ultra thick] (0,2) -- (0,3.3);
\node [below] at (0,0.66) {$b$}; \node[above] at (0,3.35) {$c$};
\node [below] at (0,-0.5) {$a$};
\node [anchor=south east] at (-1,0) {$\Omega$};
\node [anchor=south west] at (1,0) {$\overline{\Phi}$};
\end{tikzpicture}
\end{equation}
%
%
%
With the vectors \eqref{Phi-def} and \eqref{omega-def} regarded as an
input, each of the equations \eqref{psieq2} and \eqref{psieq3}
can be considered as an equation for the weight function $V_x(n)$.
Similarly to \eqref{psieq1}, these equations also
lead to the function $V_x(n)$, defined by
\eqref{Vdef}-\eqref{vanish} (up to an overall normalization). All
these equations can be viewed as (Yang-Baxter type) exchange
relations, suitable for defining $Z$-invariant lattices models
\cite{Baxter:1978xr}. The models of this type are formulated 
on irregular lattices
formed by an intersection of straight lines (with at most two lines
intersecting at one point). Their partition functions remain unchanged
upon parallel transport of individual lattice lines. In our
case the lattices shown in \eqref{peq1}, \eqref{peq2} and \eqref{peq3}
are the simplest examples of such lattices. For instance, the lattice
on the left side of \eqref{peq1} is formed by three straight lines
associated with the spectral variables $x$, $y$ and $z$. There are four spins
$a$, $a'$, $b$, $b'$, assigned to external faces and one spin $i$
assigned to the internal edge on the $z$-line. The model contains 
the three-spin interactions, with the Boltzmann 
weights given by \eqref{Phi-def},
\eqref{omega-def}, as well as the two-spin interaction with the
weights \eqref{V-shade}.
The LHS of
\eqref{psieq1} is the partition function of this model for the left lattice
of \eqref{peq1} with fixed boundary spins on the external faces. 
Eq. \eqref{psieq1} states that 
this partition function remains unchanged by
moving the $z$-line down through the intersection point of the other
two lines and transforming the lattice to that shown in the right side of
\eqref{peq1}.  A similar interpretation exists for the other two
relations \eqref{psieq2} and \eqref{psieq3}.

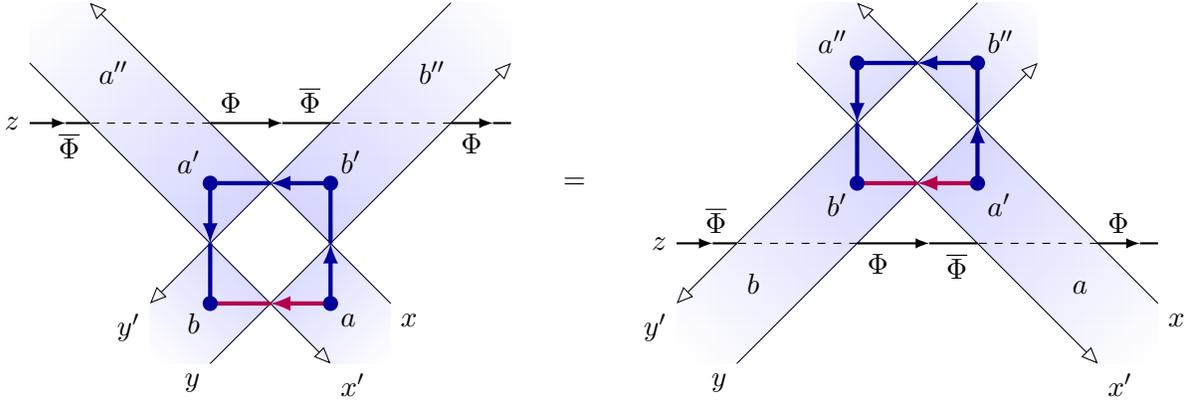
\begin{figure}[ht]
\begin{center}
\begin{tikzpicture}[scale=0.8]
\draw [-open triangle 45, thin, black] (-4,3) -- (1,-2);
\draw [-open triangle 45, thin, black] (2,-1) -- (-3,4);
\draw [-open triangle 45, thin, black] (3,4) -- (-2,-1);
\draw [-open triangle 45, thin, black] (-1,-2) -- (4,3);
\draw [fill, opacity=0.1, blue, path fading = east] (1,0) -- (4,3) -- (4,4) -- (3,4) -- (0,1) -- (1,0);
\draw [fill, opacity=0.1, blue, path fading = north] (1,0) -- (4,3) -- (4,4) -- (3,4) -- (0,1) -- (1,0); 
\draw [fill, opacity=0.1, blue, path fading = west] (-1,0) -- (-4,3) -- (-4,4) -- (-3,4) -- (0,1) -- (-1,0);
\draw [fill, opacity=0.1, blue, path fading = north] (-1,0) -- (-4,3) -- (-4,4) -- (-3,4) -- (0,1) -- (-1,0);
\draw [fill, opacity=0.1, blue, path fading = east] (0,-1) -- (1,-2) -- (2,-2) -- (2,-1) -- (1,0) -- (0,-1);
\draw [fill, opacity=0.1, blue, path fading = south] (0,-1) -- (1,-2) -- (2,-2) -- (2,-1) -- (1,0) -- (0,-1);
\draw [fill, opacity=0.1, blue, path fading = west] (0,-1) -- (-1,-2) -- (-2,-2) -- (-2,-1) -- (-1,0) -- (0,-1);
\draw [fill, opacity=0.1, blue, path fading = south] (0,-1) -- (-1,-2) -- (-2,-2) -- (-2,-1) -- (-1,0) -- (0,-1);
\draw [-latex, thick, black] (3,2) -- (3.7,2); \draw [thick, black] (3.7,2) -- (4,2);
\draw [dashed, black] (3,2) -- (1,2);
\draw [-latex, thick, black] (-1,2) -- (0.2,2); \draw[thick, black] (0.2,2) -- (1,2);
\draw [dashed, black] (-1,2) -- (-3,2);
\draw [-latex, thick, black] (-4,2) -- (-3.4,2); \draw [thick, black] (-3.4,2) --  (-3,2);
\draw [-latex, \PAC, ultra thick] (1,-1) -- (1,0); \draw [\PAC, ultra thick] (1,0) -- (1,1);
\draw [-latex, \PAC, ultra thick] (1,1) -- (0,1); \draw [\PAC, ultra thick] (0,1) -- (-1,1);
\draw [-latex, \PAC, ultra thick] (-1,1) -- (-1,0); \draw [\PAC, ultra thick] (-1,0) -- (-1,-1);
\draw [-latex, \GR, ultra thick] (1,-1) -- (0,-1); \draw [\GR, ultra thick] (0,-1) -- (-1,-1);
\draw [\STE, ultra thick, fill] (1,1) circle [radius=0.09];
\draw [\STE, ultra thick, fill] (1,-1) circle [radius=0.09];
\draw [\STE, ultra thick, fill] (-1,1) circle [radius=0.09];
\draw [\STE, ultra thick, fill] (-1,-1) circle [radius=0.09];
\node [anchor = north east] at (-1,-1) {$b$};
\node [anchor = north west] at (1,-1) {$a$};
\node [anchor = south east] at (-1,1) {$a'$};
\node [anchor = south west] at (1,1) {$b'$};
\node [anchor = north east] at (-2.2,3.2) {$a''$};
\node [anchor = north west] at (2.3,3.2) {$b''$};
\node [anchor = north east] at (-3,2) {$\overline{\Phi}$};
\node [anchor = south west] at (-1,2) {$\Phi$};
\node [anchor = south east] at (1,2) {$\overline{\Phi}$};
\node [anchor = north west] at (3,2) {$\Phi$};
\node [anchor = north east] at (-2,-1) {$y'$};
\node [anchor = north east] at (-1,-2) {$y$};
\node [anchor = north west] at (1,-2) {$x'$};
\node [anchor = north west] at (2,-1) {$x$};
\node [left] at (-4,2) {$z$};
\node [right] at (4.7,1) {$=\quad$};
\end{tikzpicture}
\begin{tikzpicture}[scale=0.8]
\draw [-open triangle 45, thin, black] (1,2) -- (-4,-3);
\draw [-open triangle 45, thin, black] (-3,-4) -- (2,1);
\draw [-open triangle 45, thin, black] (-2,1) -- (3,-4);
\draw [-open triangle 45, thin, black] (4,-3) -- (-1,2);
\draw [fill, opacity=0.1, blue, path fading = east] (1,0) -- (2,1) -- (2,2) -- (1,2) -- (0,1) -- (1,0);
\draw [fill, opacity=0.1, blue, path fading = north] (1,0) -- (2,1) -- (2,2) -- (1,2) -- (0,1) -- (1,0);
\draw [fill, opacity=0.1, blue, path fading = west] (-1,0) -- (-2,1) -- (-2,2) -- (-1,2) -- (0,1) -- (-1,0);
\draw [fill, opacity=0.1, blue, path fading = north] (-1,0) -- (-2,1) -- (-2,2) -- (-1,2) -- (0,1) -- (-1,0);
\draw [fill, opacity=0.1, blue, path fading = east] (0,-1) -- (3,-4) -- (4,-4) -- (4,-3) -- (1,0) -- (0,-1);
\draw [fill, opacity=0.1, blue, path fading = south] (0,-1) -- (3,-4) -- (4,-4) -- (4,-3) -- (1,0) -- (0,-1);
\draw [fill, opacity=0.1, blue, path fading = west] (0,-1) -- (-3,-4) -- (-4,-4) -- (-4,-3) -- (-1,0) -- (0,-1);
\draw [fill, opacity=0.1, blue, path fading = south] (0,-1) -- (-3,-4) -- (-4,-4) -- (-4,-3) -- (-1,0) -- (0,-1);
\draw [-latex, thick, black] (3,-2) -- (3.7,-2); \draw [thick, black] (3.7,-2) -- (4,-2);
\draw [dashed, black] (3,-2) -- (1,-2);
\draw [-latex, thick, black] (-1,-2) -- (0.2,-2); \draw[thick, black] (0.2,-2) -- (1,-2);
\draw [dashed, black] (-1,-2) -- (-3,-2);
\draw [-latex, thick, black] (-4,-2) -- (-3.4,-2); \draw [thick, black] (-3.4,-2) --  (-3,-2);
\draw [-latex, \PAC, ultra thick] (1,-1) -- (1,0); \draw [\PAC, ultra thick] (1,0) -- (1,1);
\draw [-latex, \PAC, ultra thick] (1,1) -- (0,1); \draw [\PAC, ultra thick] (0,1) -- (-1,1);
\draw [-latex, \PAC, ultra thick] (-1,1) -- (-1,0); \draw [\PAC, ultra thick] (-1,0) -- (-1,-1);
\draw [-latex, \GR, ultra thick] (1,-1) -- (0,-1); \draw [\GR, ultra thick] (0,-1) -- (-1,-1);
\draw [\STE, ultra thick, fill] (1,1) circle [radius=0.09];
\draw [\STE, ultra thick, fill] (1,-1) circle [radius=0.09];
\draw [\STE, ultra thick, fill] (-1,1) circle [radius=0.09];
\draw [\STE, ultra thick, fill] (-1,-1) circle [radius=0.09];
\node [anchor = north east] at (-1,-1) {$b'$};
\node [anchor = north west] at (1,-1) {$a'$};
\node [anchor = south east] at (-1,1) {$a''$};
\node [anchor = south west] at (1,1) {$b''$};
\node [anchor = south west] at (-3,-3) {$b$};
\node [anchor = south east] at (3,-3) {$a$};
\node [anchor = south east] at (-3,-2) {$\overline{\Phi}$};
\node [anchor = north west] at (-1,-2) {$\Phi$};
\node [anchor = north east] at (1,-2) {$\overline{\Phi}$};
\node [anchor = south west] at (3,-2) {$\Phi$};
\node [anchor = north east] at (-4,-3) {$y'$};
\node [anchor = north east] at (-3,-4) {$y$};
\node [anchor = north west] at (3,-4) {$x'$};
\node [anchor = north west] at (4,-3) {$x$};
\node [left] at (-4,-2) {$z$};
\end{tikzpicture}
\end{center}
\caption{A graphical representation of the $LLR$-relation \eqref{RLL-irf}.}\label{RLL-new-v2}
\end{figure}

We are now ready to give a simple visual proof of the Yang-Baxter
equation \eqref{RLL-irf}. Substituting the factorized expressions 
\eqref{def3} and \eqref{LL2} and using the above graphical notations 
one can represent this equation as shown in Fig.~\ref{RLL-new-v2}. 
\bigskip\smallskip
Consider the diagram on the left side on this figure. Let us move 
the horizontal $z$-line down through the intersection points of the
other lines and then consecutively use: (i) the relation \eqref{peq1},
(ii) the relations \eqref{peq2} and \eqref{peq3} (in any order) and,
finally, (iii) the relation \eqref{peq1} again. In this way the left diagram
in Fig.~\ref{RLL-new-v2} is transformed to the right one, thus proving
\eqref{RLL-irf}. The above calculation completes our proof of the
linear relation \eqref{fourier} 
between $R$-matrices arising from the ``star'' and
``box'' diagrams. 
Relations of this
type commonly appear as (self-) duality transformations in the Ising-type 
lattice model \cite{Wu:1982,Savit:1980}. 
\section{Functional relations for the transfer matrices\label{FR-section}}
\subsection{Definition of the
 ${\bf T}$- and ${\bf Q}$-matrices.\label{TQ-section}}
To discuss the functional relations for the transfer matrices it is 
convenient
to also use the equivalent vertex model, introduced in
Sect.~\ref{irf-vertex}. 
\begin{figure}[h]
\begin{center}
\begin{tikzpicture}[scale=1.2,baseline={(0,-.05)}]
\draw [thick, \STE] (0.5,2.5) -- (2.5,2.5) -- (4.5,2.5)--
(4.5,0.5) -- (2.5,0.5)--(0.5,0.5)--(0.5,2.5);
\draw [thick,dashed, \STE] (4.5,2.5)--(7.5,2.5);
\draw [thick, \STE] (2.5,2.5)--(2.5,.5);
\draw [thick] (0.,1.5) -- (4.1,1.5);
\draw [thick,dashed] (4.9,1.5)--(7.5,1.5);
\draw [thick] (7.5,1.5)--(10.,1.5);
\draw [thick,dashed, \STE] (4.5,.5)--(7.5,.5);
\draw [thick, \STE] (7.5,2.5)--(9.5,2.5);
\draw [thick, \STE] (7.5,.5)--(9.5,.5);
\draw [thick, \STE] (9.5,2.5)--(9.5,.5);
\draw [thick, \STE] (7.5,2.5)--(7.5,.5);
\draw [thick] (1.5,.0)--(1.5,3.0);
\draw [thick] (3.5,.0)--(3.5,3.0);
\draw [thick] (8.5,.0)--(8.5,3.0);
\node [left] at (0.,1.5) {$k_{N+1}$};
\draw [ultra thick, fill,white] (2.5,1.5) circle [radius=0.2]; 
\node  at (2.5,1.5) {${k_{N}}$};
\draw [ultra thick, fill,white] (4.5,1.5) circle [radius=0.2]; 
\node  at (4.5,1.5) {${k_{N-1}}$};
\draw [ultra thick, fill,white] (7.5,1.5) circle [radius=0.2]; 
\node  at (7.5,1.5) {${k_{2}}$};
\node [above] at (1.5,3.0) {$j_{N}$};
\node [below] at (1.5,.0) {$i_{N}$};
\node [above] at (3.5,3.0) {$j_{N-1}$};
\node [below] at (3.5,.0) {$i_{N-1}$};
\node [above] at (8.5,3.0) {$j_{1}$};
\node [below] at (8.5,.0) {$i_{1}$};
\node [right] at (10.0,1.5) {$k_1$};
\draw [ultra thick, fill,white] (0.5,2.5) circle [radius=0.08]; 
\draw [ultra thick, fill,white] (0.5,0.5) circle [radius=0.08];
\draw [ultra thick, fill,white] (2.5,2.5) circle [radius=0.08];
\draw [ultra thick, fill,white] (2.5,0.5) circle [radius=0.08];
\draw [ultra thick, fill,white] (4.5,2.5) circle [radius=0.08];
\draw [ultra thick, fill,white] (4.5,0.5) circle [radius=0.08];
\draw [ultra thick, fill,white] (7.5,2.5) circle [radius=0.08];
\draw [ultra thick, fill,white] (7.5,0.5) circle [radius=0.08];
\draw [ultra thick, fill,white] (9.5,2.5) circle [radius=0.08];
\draw [ultra thick, fill,white] (9.5,0.5) circle [radius=0.08];

\draw [ultra thick, \STE] (0.5,2.5) circle [radius=0.08]; \node[above,\STE] at (0.5,2.55) {$b_{N+1}$};
\draw [ultra thick, \STE] (0.5,0.5) circle [radius=0.08];\node[below,\STE] 
at (0.5,0.45) {$a_{N+1}$};
\draw [ultra thick, \STE] (2.5,2.5) circle [radius=0.08];\node[above,\STE] at (2.5,2.55) {$b_{N}$};
\draw [ultra thick, \STE] (2.5,0.5) circle [radius=0.08];\node[below,\STE] at (2.5,0.45) {$a_N$};
\draw [ultra thick, \STE] (4.5,2.5) circle [radius=0.08];\node[above,\STE] at (4.5,2.55) {$b_{N-1}$};
\draw [ultra thick, \STE] (4.5,0.5) circle [radius=0.08];\node[below,\STE] at (4.5,0.45) {$a_{N-1}$};
\draw [ultra thick, \STE] (7.5,2.5) circle [radius=0.08];\node[above,\STE] at (7.5,2.55) {$b_{2}$};
\draw [ultra thick, \STE] (7.5,0.5) circle [radius=0.08];\node[below,\STE] at (7.5,0.45) {$a_{2}$};
\draw [ultra thick, \STE] (9.5,2.5) circle [radius=0.08];\node[above,\STE] at (9.5,2.55) {$b_{1}$};
\draw [ultra thick, \STE] (9.5,0.5) circle [radius=0.08];\node[below,\STE] at (9.5,0.45) {$a_{1}$};

\draw [ultra thick, fill, black] (1.5,1.5) circle [radius=0.08];
\draw [ultra thick, fill, black] (3.5,1.5) circle [radius=0.08];
\draw [ultra thick, fill, black] (8.5,1.5) circle [radius=0.08];



\end{tikzpicture}
\end{center}
\caption{The arrangement of spins for the transfer matrices in the IRF
  and vertex formulations of the model, defined by \eqref{tmat} and
  \eqref{t-vert}, respectively. The correspondence between the spin
  variables is given by \eqref{index}. Note also, that the 
boundary conditions \eqref{quasi1} imply 
that $k_{N+1}=b_{N+1}-a_{N+1}=b_{1}-a_{1}=k_1$. \label{t-vertex}}
\end{figure}
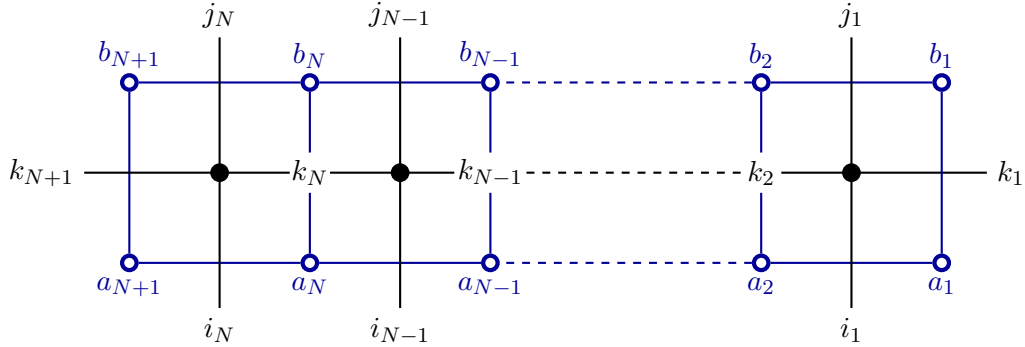
Fig.~\ref{t-vertex} shows the correspondence
between the corner spins of the IRF model and the edge 
indices of the vertex model, 
\be\label{index}
i_n=a_{n+1}-a_n\,,\qquad  j_n=b_{n+1}-b_n\,,\qquad
    k_n=b_n-a_n\,,\qquad n=1,2,\ldots N\,,
\ee
which take arbitrary non-negative integer values.
For the quasi-periodic
boundary condition (involving the horizontal field factor $\omega_h^{k_1}$ at the first
column of the lattice) the elements of the transfer matrix read
\be\label{t-vert}
\begin{array}{l}
\ds
\big({\bf T}^{(+)}(y,y'|x_N,x_N',\ldots,x_1,x'_1)\big)_{i_N,i_{N-1},
\ldots,i_1}^{j_N,j_{N-1},\ldots,j_1}=\rho(y,y')\,
\sum_{k_1,\ldots,k_N=0}^\infty\,
 \omega_h^{k_1} \,\Rc(x_N,x'_N,y,y')_{i_N,k_1}^{j_N,k_N}\,\times\\[.8cm]
\ds\qquad\qquad\qquad\qquad\qquad\qquad\qquad\qquad\times
\,
\Rc(x_{N-1},x'_{N-1},y,y')_{i_{N-1},k_N}^{j_{N-1},k_{N-1}}\ 
\cdots\ 
\Rc(x_1,x'_1,y,y')_{i_1,k_2}^{j_1,k_1}\,,
\end{array}
\ee
where the sum is taken over all nonnegative integer values of
$k_1,\ldots,k_N$. Actually, there is only a one-dimensional summation
with $k_1=0,1,\ldots,\infty$, while the rest of the $k$'s are uniquely
determined by the values of $i$'s, $j$'s and $k_1$ due to 
the conservation laws \eqref{ar-cons} at each vertex.  The superscript
$\scriptstyle{(+)}$ in the notation for the transfer matrix indicates that the sum
over $k_1$ goes over an infinite interval. Using the correspondence 
\eqref{index} and the relation \eqref{R-def} it is easy to see that
the expression inside the sum in \eqref{t-vert} exactly coincides with
the IRF transfer matrix \eqref{tmat} (with $\omega_v=1$).

The transfer matrix \eqref{t-vert} is an operator acting in the
``quantum'' space, formed by the direct
product 
\be\label{qspace}
{\mathcal H}^{\rm (vertex)}
=\Cbb^{(\infty)}_N\otimes\Cbb^{(\infty)}_{N-1}\otimes\cdots
\otimes   \Cbb^{(\infty)}_1\,,
\ee
of $N$ identical infinite-dimensional vector spaces $\Cbb^{(\infty)}$, 
each of which is spanned by the basis vectors $|j\rangle$ with
$j=0,1,2,\ldots,\infty$.  
The lower indices in \eqref{qspace} enumerate the columns of the lattice, 
as in Fig.~\ref{t-vertex}.
Below we will
regard the variables $(x_N,x'_N,\ldots,x_1,x'_1)$ as fixed and 
will write the transfer matrix simply as ${\bf T}^{(+)}(y,y')$ 
assuming the
dependence on the fixed variables implicitly. 
The Yang-Baxter equation \eqref{Rybe} implies that the transfer matrices ${\bf
  T}^{(+)}(y,y')$ with different values of $y,y'$ form a commuting
family,
\be
\big[{\bf T}^{(+)}(y,y'),{\bf T}^{(+)}(z,z')\big]=0\,,\qquad 
\forall \qquad y,y',z,z'\,.\label{tcomm2}
\ee
Moreover, they preserve the number of particles $M$, 
previously introduced in \eqref{down}.
Indeed, the matrix elements \eqref{t-vert} vanish, unless
\be\label{down2}
i_1+i_2+\cdots+i_N=j_1+j_2+\cdots+j_N\equiv M\,,\qquad M=0,1,2\ldots \,.
\ee
So, the transfer matrices act invariantly in each subspace of
\eqref{qspace} with a fixed particle number $M=0,1,2,\ldots,\infty$. 
It is convenient then to define diagonal matrices 
\be
w\;=\; q^{2M}\, \prod_{i=1}^N\, \frac{x_i}{x_i'}\,,\qquad 
Z_0=\sum_{k=0}^\infty\,(w\,\omega_h)^k=\big(1-\omega_h\,w\big)^{-1}
\,,\label{zw-def}
\ee
which also belong to the commuting family \eqref{tcomm2}. The sum in
$Z_0$ converges provided the horizontal field parameter $\omega_h$ 
is such that 
\be \label{ow1}
|\omega_h w|<1\,.
\ee

At this point it is worth noting that the
definition of the $R$-matrix \eqref{R-def2} only involves a finite
summation, so it is well defined for arbitrary values of its parameters, in
particular, for an arbitrary value of the ``anisotropy'' parameter
$q$. However the convergence of the infinite sum in the definition of
the transfer matrix \eqref{t-vert} requires a special investigation. 
It is plausible that for $|q|<1$ or $|q|=1$ the restriction
\eqref{ow1} for the field 
factor is sufficient for convergence of the sum \eqref{t-vert}. On the
other hand, the
considered model is extremely general, so that
a more detailed analysis is, of course, needed for considerations of
its particular cases. For simplicity, in what follows, we assume
$|q|<1$.   

The normalization factor in \eqref{t-vert} is chosen as
\be\label{t-norm}
\rho(y,y')=(y'/y)^M\,\prod_{n=1}^N\frac{\big(\,(y/x'_n)^2\,;q^2\big)_\infty}
{\big(\,(y'/x'_n)^2\,;q^2\big)_\infty}\,,\
\ee
where the $q$-products are defined in \eqref{qprod0}.
Then, taking into account \eqref{R-def2} one concludes\footnote{%
The $R$-matrix \eqref{R-def2} is a Laurent
polynomial in the spectral variable $y'$ and a meromorphic function of
$y$, having poles at $y^2={x'}^2 q^{-2n}$,\  $n=0,1,2,\ldots$\ .} that
the transfer matrix \eqref{t-vert} is a meromorphic function in
the variables $y$ and $y'$. The reason for taking the normalization
\eqref{t-norm} will be clarified below. 
Next, define the factor
\be\label{phi-def}
\varphi(y,y')\,=\,
\prod_{n=1}^N\frac{\big(\,(y'/x'_n)^2\,;q^2\big)_\infty}
{\big(\,(y/x_n)^2\,;q^2\big)_\infty}\;,
\ee
and introduce two new transfer matrices 
\be\label{Q-def}
\Qib(y')=Z^{-1}_0\,\lim_{y\to 0} 
\,\varphi(y,y')\,{\bf T}^{(+)}(y,y')\,,
\qquad
\overline{\Qib}(q^{-1}\,y)=Z^{-1}_0\,
\lim_{y'\to 0} \,\varphi(y,y')\,{\bf T}^{(+)}(y,y')\,,
\ee
which, as we shall see below, play the role of the Baxter ${\bf
  Q}$-matrices \cite{Baxter:1972}. 
Evidently, they belong to the same commuting family \eqref{tcomm2},
\be\label{tq-comm}
[\Tib^{(+)}(y,y'),\Qib(z)]=[\Tib^{(+)}(y,y'),\overline{\Qib}(z)]=
[\Qib(z),\Qib(z')]=[\overline{\Qib}(z),\overline{\Qib}(z')]=
[{\Qib}(z),\overline{\Qib}(z')]=0\,,
\ee
where $y,y',z,z'$ take arbitrary values.

Recall, that $\Tib^{(+)}(y,y')$ as well as $\Qib(y')$ 
and $\overline{\Qib}(y)$ implicitly depend on the set of fixed variables 
$(x_N,x'_N,\ldots,x_1,x'_1)$. Note also, that if a pair of these
variables is constrained as  
\be\label{x-const}
x_n'=x_n\, q^{2s_n}\,,\qquad 2s_n\in {\Zbb}_{\ge0}
\ee
for some particular $n$, then the above transfer matrices admit a
$(2s_n+1)$-dimensional reduction in the $n$-th factor of the 
product \eqref{qspace}.

\subsection{Functional relations}
The elements of the ${\bf Q}$-matrices \eqref{Q-def} 
can be written explicitly by taking required limits of
\eqref{t-vert}. The details are presented in Appendix~\ref{appD}. 
It is worth noting that in the case of $U_q(\widehat{sl}(2))$ 
considered here the algebraic origin of the ${\bf Q}$-matrices 
(or the ${\bf Q}$-operators) is very well understood \cite{Bazhanov:1996dr,
Bazhanov:1998dq}.  
Their main algebraic properties can be concisely expressed by a single
factorization relation, which in our case takes the form
\be\label{factor}
\varphi(y,y')\,\Tib^{(+)}(y,y')=
Z_0\,\Qib(y')\,\overline{\Qib}(q^{-1}\,y)
\,,
\ee
derived in Appendix~\ref{appD}. Let us also mention 
the normalization conditions 
\be \label{Q-norm}
\Qib(0)=\overline{\Qib}(0)=1\,,
\ee
where the ${\bf Q}$-operators reduce to the identity operator. 

Consider now the case when the spectral variables $y$ and $y'$ are
related as
\be\label{findim}
y'=y\, q^{2s}\,,\qquad 2s\in{\mathbb
  Z}_{\ge0}\,. 
\ee
This is precisely the finite-dimensional 
reduction case \eqref{twos2}-\eqref{inf}, considered in 
Sect.~\ref{red} (see also
\eqref{RR-rel}). This reduction implies that the sum in \eqref{t-vert} splits
into two parts: 
\begin{enumerate}[(i)]
\item
the first part, where all the indices
$k_1,\ldots,k_N\in\,{\mathcal I}_s=\{0,1,\ldots,2s\}$
take the finite set of
values. Assuming the relation \eqref{findim} we denote this part as 
${\bf T}_s(y)$ without the superscript
$\scriptstyle(+)$. Explicitly, for $2s\in{\mathbb
  Z}_{\ge0}$ one obtains 
\be\label{ts-def}
\begin{array}{l}
\ds
\big({\bf T}_s(y)\big)_{i_N,i_{N-1},
\ldots,i_1}^{j_N,j_{N-1},\ldots,j_1}=\rho(y,q^{2s}\,y)\,
\sum_{k_1,\ldots,k_N\in \,{\mathcal I}_s}\,
 \omega_h^{k_1}
 \,\Rc(x_N,x'_N,y,q^{2s}\,y)_{i_N,k_1}^{j_N,k_N}\,\times
\\[.8cm]
\ds\ \qquad\qquad\qquad\qquad \times
\,
\Rc(x_{N-1},x'_{N-1},y,q^{2s}\,y)_{i_{N-1},k_N}^{j_{N-1},k_{N-1}}
\cdots
\Rc(x_1,x'_1,y,q^{2s}\,y)_{i_1,k_2}^{j_1,k_1}\,
\end{array}
\ee
\item
the second part where all the indices 
$k_1,\ldots,k_N\in
\overline{\mathcal I}_s=\{2s+1,2s+2,\ldots,\infty\}$
take the infinite set
of values. Using \eqref{twos2inf} one can reduce this part to ${\bf
  T}^{(+)}(y' q,y q^{-1})$. 
\end{enumerate}
In this way one obtains
\be
\varphi(y,y')\,{\bf T}^{(+)}(y,y')=
\varphi(y,y')\,{\bf T}_s(y)+(\omega_h\,w)^{2s+1}\, 
\varphi(y'q, \,y q^{-1})\, {\bf
  T}^{(+)}(y'q,\,yq^{-1})\,,\qquad y'=y q^{2s}\,,
\ee
where the relation \eqref{findim} is assumed.
The above reduction for the
transfer matrices is well known, see \cite{Bazhanov:1998dq}. 
Combining the last
equation with \eqref{factor} one gets
\be\label{T-QQ}
Z_0^{-1}\,\varphi(y,yq^{2s})\, {\bf T}_s(y)={\bf Q}(y')
\overline{{\bf Q}}(y\,q^{-1})
-\big(\omega_h\,w\big)^{2s+1}\,
{\bf Q}(y\,q^{-1})
\overline{{\bf Q}}(y')\,.
\ee

From \eqref{spin0} it follows that ${\bf T}_0(y)\equiv 1$. Thus,
setting $s=0$ in \eqref{T-QQ} one obtains the quantum Wronskian
relation for the ${\bf Q}$-operators 
\be\label{qwrons}
Z_0^{-1}\,\varphi(y,y)={\bf Q}(y)
\overline{{\bf Q}}(y\,q^{-1})
-\big(\omega_h\,w\big)\,
{\bf Q}(y\,q^{-1})
\overline{{\bf Q}}(y)\,.
\ee
Next, setting $s=1/2$ in \eqref{T-QQ} one gets
\be
Z_0^{-1}\,\varphi(y,yq)\, {\bf T}_{\frac{1}{2}}(y)={\bf Q}(yq)
\overline{{\bf Q}}(y\,q^{-1})
-\big(\omega_h\,w\big)^2\,
{\bf Q}(y\,q^{-1})
\overline{{\bf Q}}(y q)\,.
\ee
Combining the last relation with \eqref{qwrons} one obtains the famous
$TQ$-equations \cite{Baxter:1972}
\be\label{tqeq}
\begin{array}{l}
{\bf T}_{\frac{1}{2}}(y){\bf Q}(y)=f(y)\,{\bf Q}(yq)
+(\omega_h\,w)\,g(y)\,{\bf Q}(yq^{-1})\,, 
\\[.4cm]
{\bf T}_{\frac{1}{2}}(y)\overline{{\bf Q}}(y)=(\omega_h\,w)\,
f(y)\,\overline{{\bf Q}}(yq)
+g(y)\,\overline{{\bf Q}}(yq^{-1})\,, 
\end{array}
\ee
where
\be\label{fg-def}
f(y)=\frac{\varphi(y,y)}{\varphi(y,qy)}=\prod_{n=1}^N
\big(1-(y/{x'_n})^{2}
\big)\,,\qquad
g(y)=\frac{\varphi(qy,qy)}{\varphi(y,qy)}=\prod_{n=1}^N
\big(1-(y/{x_n})^{2}
\big)\,.
\ee

To conclude our derivation of the functional relations  
note, that the seemingly artificial normalization factors \eqref{t-norm}
and \eqref{phi-def}, entering the definitions of the transfer matrices
and ${\bf Q}$-operators were specially chosen to keep the functional
relations in the standard form, with a minimal number of scalar
factors. If necessary (for instance, when $|q|=1$) the above  
normalization factors could be removed and the functional
relations then be easily modified. Finally, note that essentially the
same ${\bf Q}$-operators were defined by Mangazeev in
\cite{Mangazeev:2014gwa,Mangazeev:2014bqa} (see Appendix~\ref{appC}
for more details). 

\subsection{Six-vertex model}

By construction the eigenvalues of
$\varphi(y,y')\,\Tib^{(+)}(y,y')$, $\Qib(y')$ and $\overline{\Qib}(y)$
are polynomials in the
variable $(y')^2$, and, in general, meromorphic functions of $y^2$ 
with poles in the finite part of the complex plane.
An important exception is the finite-dimensional reduction case when
the constraint \eqref{x-const} holds for all columns of the lattice 
$m=1,2,\ldots,N$. 
For an illustration consider the case of the six-vertex model with
two-dimensional quantum space representations for all sites of the
chain; that is possible if the spectral parameters obey the relations
\be
x_n'=q x_n\,,\qquad s_n=\textstyle{\frac{1}{2}}\,,\qquad m=1,\ldots,N\,.
\ee
Using the definitions \eqref{t-vert}, \eqref{zw-def}, \eqref{t-norm}
and \eqref{ts-def} one obtains
\be
{\bf T}_{\frac{1}{2}}(y)=z^{-1}\,{\bf
  T}^{(6v)}(q^{-\frac{1}{2}}\,y)\,,\qquad
z=(\omega_h w)^{-\frac{1}{2}}\,,
\ee
where
\bea
\big({\bf T}^{(6v)}(q^{-\frac{1}{2}}\,y)\big)_{i_N,i_{N-1},
\ldots,i_1}^{j_N,j_{N-1},\ldots,j_1}&=&\\[.4cm]
&&\kern-11em=(\omega_h)^{-\frac{1}{2}}
\sum_{k_1,\ldots,k_N=0,1}
 \omega_h^{k_1} \,\Rc(x_N/y)_{i_N,k_1}^{j_N,k_N}
\,
\Rc(x_{N-1}/y)_{i_{N-1},k_N}^{j_{N-1},k_{N-1}}\ 
\cdots\ 
\Rc(x_1/y)_{i_1,k_2}^{j_1,k_1}\,,
\eea
where $\Rc(\lambda)$ is given by \eqref{R6v} with
$\rho(\lambda)=q^{-\frac{1}{2}}\,\lambda^{-1}$. 

To present the expressions for the ${\bf Q}$-operators \eqref{Q-def}
for this case consider the $q$-oscillator algebra 
\beq\label{qosc}
[\Hcal,\Ecal_\pm]=\pm2\,\Ecal_\pm\,,\qquad q\, \Ecal_+\Ecal_--
q^{-1}\,\Ecal_-\Ecal_+=q-q^{-1}
\eeq
and its two representations $\rho_\pm$, defined by the action on the vectors 
$|j\rangle$, $j=0,1,2,\ldots,\infty$,
\beq\label{qreps}
\rho_\pm(\Hcal)|j\rangle=\mp 2j\,|j\rangle\,,\qquad
\rho_\pm(\Ecal_\pm)|j\rangle=q^{\mp\frac{1}{2}}
(q^{-j}-q^{j})\,|j-1\rangle\,,\qquad
\rho_\pm(\Ecal_\mp)|j\rangle=\mp q^{\mp (j+ 1/2)}|j+1\rangle\,.
\eeq
Define operators 
\beq\label{Apm-def}
{\bf A}_\pm(y)=Z^{-1}_0\,{\rm Tr}_{\rho_\pm}\Big\{\omega_h^{\mp\Hcal/2}\,
L^{(N)}_\pm(y/x_N)\, L^{(N-1)}_\pm(y/x_{N-1})\cdots L^{(1)}_\pm(y/x_1)\Big\}\,,\qquad {\bf A}_\pm(y)=1+O(y^2)\,,
\eeq
where 
\beq\label{Lpm-def}
L_+(y)=\begin{pmatrix}
q^{\frac{\Hcal}{2}}&y\,\Ecal_-\\
y\, \Ecal_+& \ \ q^{-\frac{\Hcal}{2}}-q^{-1}\,y^2\,q^{\frac{\Hcal}{2}}
\end{pmatrix}\,,\qquad
L_-(y)=\begin{pmatrix}
q^{-\frac{\Hcal}{2}}-q\,y^2\,q^{\frac{\Hcal}{2}}\ \ &y\,\Ecal_+\\[.3cm]
y\,\Ecal_-& q^{\frac{\Hcal}{2}}
\end{pmatrix}
\eeq
are two by two matrices acting in the quantum space, whose entries are
elements of the $q$-oscillator algebra \eqref{qosc}. 

Using \eqref{Qelem}, \eqref{Qelem4} and \eqref{Qelem5} from
Appendix~\ref{appD} it is not difficult to show that 
\be
{\bf Q}(y)={\bf A}_+(q^{-\frac{1}{2}}\,y)\,,
\qquad
\overline{{\bf Q}}(y)={\bf A}_-(q^{-\frac{1}{2}}\,y)\,.
\ee
The equations \eqref{tqeq} can now be rewriten as
\beq
{\bf T}^{(6v)}(y) {\bf A}_\pm(y)=z^{\pm1} \,g(q^{-\frac{1}{2}}\,y)\, 
{\bf A}_\pm(q\,y)+
z^{\mp1} \,g(q^{+\frac{1}{2}}\,y)\, {\bf A}_\pm(q^{-1}\,y)\,,\qquad
z=(\omega_h w)^{-\frac{1}{2}}\,.
\eeq
Note, that apart from a trivial change $y\to y^2=\zeta$ of the arguments of the
transfer matrices this relation exactly coincides with Eq.(3.3) of 
ref.~\cite{Bazhanov:2020fbp}. 
\section{The homogeneous case}
Consider the column homogeneous case 
\be\label{homo}
x_1=x_2=\ldots=x_N\equiv x\,,\qquad 
x_1'=x_2'=\ldots=x_N'\equiv x'\,,\qquad x'=q^{2s}\,x\,,
\ee
so that we can abbrevate the arguments of the transfer matrix \eqref{t-vert} 
simply as ${\bf T}^{(+)}(y,y'\,|\,x,x')$.
Obviously, the homogeneous model possesses the translational invariance
\be\label{TK-com}
\big[{\bf T}^{(+)}(y,y'\,|\,x,x'),{\bf K}\big]=0\,,\qquad
\ee
where ${\bf K}$ is the one-site  translation operator,
whose matrix elements are given by
\be\label{K-def}
\big({{\bf K}}\big)_{i_{N}\, i_{N-1}\,\ldots
\,\,  i_1}^{j_{N} \,j_{N-1}\,\ldots \,j_1}=(\omega_h)^{i_1}\,
\delta_{i_N}^{j_{N-1}}\,\ldots\,
\delta_{i_2}^{j_1}\,\delta_{i_1}^{j_N} \,.
\ee
In writing this formula we have taken 
into account the quasi-periodic boundary conditions, involving
the horizontal field factor at the first column of the lattice. 

\subsection{Hamiltonians}
Apart from the translation operator the transfer matrix 
commuting family now contains spin chain
Hamiltonians that are given by a sum of terms, each of which acts
non-trivially only in two consecutive factors of the product \eqref{qspace}.
First consider a generic case, when $2s$ is not taking non-negative
integer values. 
Using the definition \eqref{t-vert} and \eqref{R-def2} 
it is not difficult to check that 
\be\label{T-expan}
\Big\{\rho(y,y')^{-1}\,{\bf T}^{(+)}(y,y'\,|\,x,x')\Big\}{\Big|_{y=x e^\varepsilon,\,
y'=x'e^{\varepsilon'}}}
={{\bf K}}\Big(1+\varepsilon\,{\bf H}^{(1)}+
\varepsilon'\,{\bf H}^{(2)}\Big)+O(\varepsilon^2,{\varepsilon'}^2,
\varepsilon \varepsilon')
\ee
where the Hamiltonians ${\bf H}^{(1)}$ and  ${\bf H}^{(2)}$ are
defined as  
\be\label{hamdef1}
\big({\bf H}^{(\ell)})_{i_{N}\, i_{N-1}\,\ldots
\,\,  i_1}^{j_{N} \,j_{N-1}\,\ldots \,j_1}=
\sum_{m=1}^{N} \omega_h^{(j_N-i_N)\delta_{m,N}}\,\,
\delta_{i_N}^{j_N}\cdots\,\delta_{i_{m+2}}^{j_{m+2}}\,\,
\big(H^{(\ell)}\big)_{i_{m+1},i_m}^{j_{m+1},j_m}\,\,
\delta_{i_{m-1}}^{j_{m-1}}\cdots\,\delta_{i_{1}}^{j_{1}}\,,
\ee
with $\ell=1,2$ and 
\begin{equation}\label{hamdef2}
\big(H^{(1)}\big)_{i\,,\,i'}^{j\,,\,j'}\;=\;
\delta_{i}^{j}\,\delta_{i'}^{j'} \,F_{i}(s) \;+\;
2\delta_{i+i',j+j'} \,\frac{\theta(i>j)}{[q^{i-j}]} \,\frac{V_{q^{-2s}}(j)}{V_{q^{-2s}}(i)}\,,\qquad \big(H^{(2)}\big)_{i\,,\,i'}^{j\,,\,j'}\;=\;
\big(H^{(1)}\big)_{i'\,,\,i}^{j'\,,\,j}\,.
\end{equation}
Here we use the notations 
\be
\theta(a>b)=\left\{\begin{array}{ll} 1, 
 &\mbox{if\ } a>b\,,\\[.3cm]
0,& \mbox{otherwise}\,,\end{array}\right.
\ee 
and 
\begin{equation}
F_i(s)\;=\;-\sum_{k=0}^{i-1} \frac{\ds q^{k-2s}+q^{-k+2s}}{\ds q^{k-2s}-q^{-k+2s}}\;,\qquad
[q^n]\;=\;q^n-q^{-n}\;,
\end{equation}
where $i=0,1,2,\ldots,\infty$.
The weight function $V_x(n)$, appearing in \eqref{hamdef2}, is defined in
\eqref{Vdef}. 
Moreover, the boundary conditions imply that $i_{N+1}=i_1$ and  
$j_{N+1}=j_1$. 

It is worth remembering that the transfer matrix ${\bf
  T}^{(+)}(y,y'\,|\,x,x')$ has two spectral parameters $y$ and
$y'$. This is why there are two independent commuting Hamiltonians in
the homogeneous case.

\subsection{{\bf Q}-operators}
For the homogeneous case it is convenient to define 
different ${\bf Q}$-operators
\begin{equation}\label{Qnew}
\Qib^{(h)}(y')\;=\;\rho(y,y')^{-1}\Tib^{(+)}(y,y')|_{y=x}\;,\quad \overline{\Qib}^{(h)}(q^{-1} y)\;=\;\rho(y,y')^{-1}\Tib^{(+)}(y,y')|_{y'=x}\;,
\end{equation}
with matrix elements,
\bea
\Qib^{(h)}(y')_{i_N i_{N-1} \dots i_1}^{j_N j_{N-1} \dots j_1}&=&
\sum_{k_1,\dots, k_N} 
\omega_h^{k_1} \prod_{n=1}^N \delta_{i_n+k_{n+1},j_n+k_n}^{}
\frac{V_{x/y'}(k_n) V_{y'/x'}(i_n-k_n)}{V_{x/x'}(i_n)}\;,\\[.4cm]
\overline{\Qib}^{(h)}(q^{-1}y)_{i_N i_{N-1} 
\dots i_1}^{j_N j_{N-1} \dots j_1}&=&
\sum_{k_1,\dots, k_N} 
\omega_h^{k_1} \prod_{n=1}^N \delta_{i_{n}+k_{n+1},j_n+k_{n}}^{}
\frac{V_{x/x'}(j_n) V_{y/x}(k_n)}{V_{y/x'}(j_n+k_{n})}\;.
\eea
They satisfy the factorization relation (similar to \eqref{factor}) 
\begin{equation}
\rho(y,y') \,\overline{\Qib}^{(h)}(q^{-1} y) \,
\Qib^{(h)}(y')\;=\;\Tib^{(+)}(y,y')\,
\end{equation}
and normalization conditions
\begin{equation}
\Qib^{(h)}(x' e^{\varepsilon'})={\bf K}\big(1+\varepsilon' {\bf H}^{(2)}+
O({\varepsilon'}^2)\big)
\,,\qquad
\Qib^{(h)}(x\,\re^{-\varepsilon}) = 1 + \varepsilon {\bf H}^{(1)}+O(\varepsilon^2)\,,
\ee
\be
\overline{\Qib}^{(h)}(q^{-1} x e^{\varepsilon})=
1+\varepsilon {\bf H}^{(1)}+O(\varepsilon^2)\;,
\end{equation}
where the Hamiltonians are defined in \eqref{hamdef1}, \eqref{hamdef2} and 
``$1$'' stands for the identity operator. We have added the
superscript $(h)$ to the notations for the new ${\bf Q}$-operators
\eqref{Qnew} to distinguish them from the previously defined operators
\eqref{Q-def}, normalized as \eqref{Q-norm}. The latter can, of
course, be specialized to the homogeneous case \eqref{homo} and
connected with \eqref{Qnew},
\begin{equation}\label{Qhnorm1}
\Qib^{(h)}(y')\;=\;\Big(\frac{x}{y'}\Big)^M\,
\Qib(y')\,\Qib(x)^{-1}\;=\;
\Big(\frac{x'}{y'}\Big)^M\, {\bf K} \,\Qib(y')\,\Qib(x')^{-1} \;,
\end{equation}
and
\begin{equation}\label{nQhnorm1}
\overline{\Qib}^{(h)}(q^{-1}y) \;=\; \Big(\frac{y}{x}\Big)^M
\,\varphi(y,y)^{-1}\,\overline{\Qib}(q^{-1}y) \,
\Big(\lim_{y\to x}\,\varphi(y,y)^{-1} \,
\overline{\Qib}(q^{-1}y) \Big)^{-1}\;.
\end{equation}

For completeness let us mention also the IRF version of the new 
operators \eqref{Qnew} 
\begin{equation}\label{2a}
\mathbb{Q}^{(h)}(y')_{a_N,\dots,a_1}^{b_N,\dots,b_1} \;=\; \prod_{i=1}^N
\frac{V_{y'/x'}(a_{i+1}-b_i) V_{x/y'}(b_i-a_i)}{V_{x/x'}(a_{i+1}-a_i)}\,,
\end{equation}
\begin{equation}\label{3}
\overline{\mathbb{Q}}^{(h)}(q^{-1}y)_{a_N,\dots,a_1}^{b_N,\dots,b_1} \;=\; \prod_{i=1}^N
\frac{V_{y/x}(b_{i}-a_i) V_{x/x'}(b_{i+1}-b_i)}{V_{y/x'}(b_{i+1}-b_i)}\,,
\end{equation}
which satisfy the factorization relation
\begin{equation}\label{4}
\mathbb{T}(y,y') \;=\; \mathbb{Q}^{(h)}(y')\, \overline{\mathbb{Q}}^{(h)}(q^{-1} y)\;,
\end{equation}
with the IRF transfer matrix \eqref{tmat} (where the field factors are set
as $\omega_h=\omega_v=1$). Note, that the operator \eqref{2a}
possesses quite remarkable locality properties. It is a rather sparse
matrix (similar to the Hamiltonians \eqref{hamdef2}), which is very 
convenient for numerical calculations. 

\subsection{Finite-dimensional reduction }
Now consider the case when the parameter $2s$ in \eqref{homo} is 
an integer, $2s\in{\mathbb Z}_{\ge0}$. The corresponding Hamiltonians cannot be
obtained from \eqref{hamdef1}, \eqref{hamdef2} just by setting $2s$ to
such a value, because this operation does not commute with taking the
limit \eqref{T-expan}.  Actually, for $2s\in{\mathbb Z}_{\ge0}$ 
the limit \eqref{T-expan} exists only
when $\varepsilon=\varepsilon'$. So, there is only one Hamiltonian in
this case. The quantum space of states \eqref{qspace} becomes
reducible. Each factor of the product \eqref{qspace} splits into two
components
\be 
{\mathbb C}^{(\infty)}={\mathbb C}^{(2s+1)}\oplus{\mathbb
  C}^{(\infty)}
\ee
spanned by the vectors $|j\rangle$, $j\in {\cal
  I}_{s}=\{0,1,2,\ldots,2s\}$ and 
$j\in \overline{\cal I}_{s}=\{2s+1,2s+2,\ldots,\infty\}$, respectively,
which according to Sect.~\ref{red} 
support the $(2s+1)$-dimensional representations $\pi_s$ and
the infinite-finite dimensional representation $\pi^{+}_{-s-1}$ of the
$U_q(sl(2))$ algebra \eqref{uqsl2}. The Hamiltonian takes a block
triangular form, acting invariantly in two blocks where one takes
either the same $(2s+1)$-dimensional or the same infinite-finite dimensional
components in all factors of the product \eqref{qspace}.  

First, consider the $(2s+1)^N$ -dimensional block. The Hamiltonian is
given by the above formula \eqref{hamdef1}, but with the superscipt
$(\ell)$ omitted and 
\begin{equation}\label{H0}
\begin{array}{ll}
\ds H_{i,i'}^{j,j'}\;=\; &
\delta_{i}^{j} \delta_{i'}^{j'}  (F_{i}(s) + F_{i'}(s) ) \\
[5mm]
& \ds
+ 2\delta_{i+i',j+j'} \left(
\frac{\theta(i>j)}{[q^{i-j}]} \frac{V_{q^{-2s}}(j)}{V_{q^{-2s}}(i)}
+
\frac{\theta(i'>j')}{[q^{i'-j'}]} \frac{V_{q^{-2s}}(j')}{V_{q^{-2s}}(i')}
\right)\;.
\end{array}
\end{equation}
for $0\leq i+i'\leq 2s$ and 
\begin{equation}\label{H1}
\begin{array}{ll}
\ds H_{i,i'}^{j,j'}\;=\; &
\delta_{i}^{j} \delta_{i'}^{j'}  (F_{2s-i}(s) + F_{2s-i'}(s) ) \\
[5mm]
& \ds
+ 2\delta_{i+i',j+j'} \left(
\frac{\theta(j>i)}{[q^{j-i}]} \frac{V_{q^{-2s}}(j)}{V_{q^{-2s}}(i)}
+
\frac{\theta(j'>i')}{[q^{j'-i'}]} \frac{V_{q^{-2s}}(j')}{V_{q^{-2s}}(i')}
\right)\;.
\end{array}
\end{equation}
for $2s\leq i,j\leq 4s$. Note, that (\ref{H0}) and (\ref{H1}) display 
the reflection symmetry 
\begin{equation}
H_{i,i'}^{j,j'} \;=\; H_{2s-i,2s-i'}^{2s-j,2s-j'}\;.
\end{equation}

\bigskip

For the infinite dimensional block the local Hamiltonian is given by
the same formula \eqref{H1}, but with $i,i',j,j'> 2s$. 
It is worth noting that for $i>2s$ 
\begin{equation}\label{Fneg}
F_{2s-i}(s)\;=\;-\; 
\sum_{k=2s+1}^{i} \frac{q^k+q^{-k}}{q^k-q^{-k}} \;=\;F_{i-2s-1}(-s-1)\,-\,\frac{q^{2s+1}+q^{-2s-1}}{q^{2s+1}-q^{-2s-1}}\;.
\end{equation}
Moreover, with the l'H\^opitale rule one obtains 
\begin{equation}\label{lH}
\frac{V_{q^{-2s}}(j)}{V_{q^{-2s}}(i)}\;=\;\frac{V_{q^{2s+2}}(i-2s-1)}{V_{q^{2s+2}}(j-2s-1)}\;.
\end{equation}
Taking this into account it is easy to see that upon the shift of
all indices as $i\to i-2s-1$ the resulting Hamiltonian up to a
constant coincides with ${\bf H}^{(1)}+{\bf H}^{(2)}$ defined by
\eqref{hamdef1}, \eqref{hamdef2} with $s$ replaced by $-s-1$.

\section{The Bethe Ansatz}
\subsection{Coordinate Bethe ansatz}
The commuting transfer matrices and ${\bf Q}$-operators 
\eqref{tcomm2}, \eqref{tq-comm} act in the infinite-dimensional quantum space 
${\mathcal H}^{\rm (vertex)}$ defined by \eqref{qspace}. As noted
before (see \eqref{down2}) 
all these matrices act invariantly in all finite dimensional
subspaces of ${\mathcal H}^{\rm (vertex)}$ with a fixed number of
particles $M=0,1,2,\ldots,\infty$. 
Their diagonalization problem can be solved via the coordinate Bethe ansatz 
by generalizing the results of
\cite{Bethe:1931,Lieb:1967zz,Sutherland:1967a,CPYang:1967,Sutherland:1967zz,
Baxter:1971cs}. Let $\widehat\sigma^+$ and $\widehat h$ 
be the ``particle creation'' and ``particle number''  
operators acting on the basis vectors
$|j\rangle \in{\mathbb C}^{\infty}$ as 
\be\label{creation}
\widehat\sigma^+\,|j\rangle=|j+1\rangle\,,\qquad
\widehat h\,|j\rangle=j\,|j\rangle\,,
 \qquad j=0,1,2,\ldots,\infty\,.
\ee
The eigenvectors describing $M$-particle states have the form 
\be\label{cba-state}
{\bf \Psi}=\sum_{1\le r_1\le r_2 \le\cdots \le r_M \le N}
\Psi(r_1,r_2,\ldots ,r_M)\,\, \widehat\sigma_{r_M}^+\,\cdots\,
\,\widehat\sigma_{r_1}^+\,{\bf \Psi}_0\,,\qquad M\ge0\,,
\ee
where the (unnormalized) Bethe ansatz wave function reads as
\bea\label{wave}
\Psi(r_1,\ldots,r_M)=\sum_{\hat P}A_{\hat P}\ \prod_{m=1}^M \phi_{{\hat P}m}(r_m)
\eea
with suitably chosen coefficients $A_{\hat  P}$ and functions $\phi_m(r)$. 
The vector ${\bf \Psi}_0$ is the bare vacuum state 
\be 
{\bf \Psi}_0=
\underbrace{|0\rangle\otimes|0\rangle\otimes\cdots\otimes|0\rangle}_N
\,,
\ee
which
is the $M=0$ eigenstate of all operators in the commuting family, 
and $\widehat\sigma^+_r$ 
denotes the particle creation operator acting 
in the $r$-th factor of the tensor product \eqref{qspace}.
The summation is taken over all $M!$ permutations ${\hat  P}$ of the
integers $(1,2,\ldots,M)$. Recall that the 
number of particles $M$ is a conserved quantity, preserved by all
operators in the commuting family.
\medskip
The coefficients $A_{\hat P}$ read 
\be\label{ap-def}
A_{\hat  P}=
\prod_{1\le j < m\le M}
\frac{q\,y^2_{\hat Pj\vphantom{{|}^{|}}}-q^{-1}\,y^2_{\hat Pm}}
{y^2_{\hat Pj}-y^2_{\hat Pm}}\ \,,
\ee
where the set of complex numbers $\{y_m\}_{m=1}^M$ 
satisfies the system of algebraic equations (called the Bethe ansatz equations)
\be\label{BA-eq}
\prod_{i=1}^N \frac{1-y_m^2/x_i^2{}_{\phantom{|}}\,}
{1-y_m^2/{x'}_i^{2}{}^{\phantom{|}}}
=-(\omega_h w)^{-1}\, \prod_{j=1}^M 
\frac{y_j^2-q^{+2}\,y_m^2}{y_j^2-q^{-2}\,y_m^2}\,,\qquad m=1,2,\ldots,M\,,
\ee
and the functions $\phi_m(r)$ in \eqref{wave} have the form 
\be\label{one-p}
\phi_m(r)=-[q]\,\left(\prod_{i=1}^{r-1} [y_m/x_i]\right)\ 
\left(\prod_{i=r+1}^{N} [y_m/x'_i]\right)\,.
\ee
Note, that for $M=1$ the coefficient $A_{\hat P}\equiv1$.

The corresponding eigenvalues of 
the ${\bf Q}$-operators 
defined by \eqref{Q-def} 
\be 
{\bf Q}(y)\,{\bf \Psi} =Q(y)\, {\bf
  \Psi}\,,\qquad \overline{{\bf Q}}(y)\,{\bf \Psi} =\overline{Q}(y)\,
{\bf \Psi}\,, 
\ee 
(see also \eqref{Qelem},
\eqref{Qelem2} and \eqref{Qelem3} for explicit expressions of their
matrix elements) 
can be expressed via the Bethe roots
$\{y_m\}_{m=1}^M$, satisfying \eqref{BA-eq}. 
The eigenvalues of ${\bf Q}(y)$ are polynomials 
in the variable $y^2$,
\be Q(y)=\prod_{m=1}^M
\Big(1-{y^2}/{y_m^2}\Big)\,,\label{Q-eig} 
\ee 
while the eigenvalues of $\overline{\bf Q}(y)$ are, in general,
meromorphic functions of this variable.  They can be expressed through
$Q(y)$ by solving the quantum Wronskian relation
\eqref{qwrons},
\begin{equation}\label{Qbar-eig}
\overline{Q}(y)\;=\;Z_0^{-1}\,Q(y)\,
 \sum_{j=1}^\infty \big(\omega_h\,w\big)^{j-1} 
\frac{\varphi(q^jy,q^jy)}{{Q}(q^{j-1} y)\, {Q}(q^{j} y)}\,,\qquad |q|<1\,.
\end{equation}
Note, that separate terms in the above sum contain 
poles arising from the zeroes of $Q(q^j y)$,
$j=1,2,\ldots,\infty$. However, it is easy to see that all such 
poles cancel out between 
two successive terms of the sum by virtue of the Bethe ansatz
equations \eqref{BA-eq}. Thus, the poles in \eqref{Qbar-eig} could only  
originate from
the function $\varphi(y,y)$, defined in \eqref{phi-def}.
For arbitrary values of 
$\{s_n\}_{n=1}^N$ in \eqref{x-const} 
is has the poles located
at\footnote{%
Of course, these fixed ``kinematic'' 
poles could easily be removed by multiplying $\overline{Q}(y)$ by a
suitable factor. The functional relations should then
be modified accordingly.}  
\be
y^2=q^{-2j} x_n^2\,,\qquad n=1,\ldots,N\,,\qquad j=1,2,\ldots,\infty\,.   
\ee
However, if $2s_n\in{\mathbb
  Z}_{\ge0}$ for all $n=1,2,\ldots,N$ the function $\varphi(y,y)$
simplifies to 
\be
\varphi(y,y)=\prod_{n=1}^N\, \prod_{j=1}^{2s_n}\,\Big(1-(y/q^{j}x_n)^2\Big)\,.
\ee
and the eigenvalue $\overline{Q}(y)$ also becomes a polynomial in $y^2$.

The above construction for the eigenvectors \eqref{cba-state}-\eqref{one-p} 
is very similar to Baxter's results \cite{Baxter:1971cs} 
for the eigenvectors of the inhomogeneous six-vertex model. The
difference is that the single particle wave function \eqref{one-p} and
the Bethe ansatz equations \eqref{BA-eq} in our case are slightly more
general than those for the 
spin $s=\hf$ six-vertex model where $x_n'=q\,x_n$  ($n=1,2,\ldots,N$).  
Moreover, the sum in \eqref{cba-state} allows any number of the 
particle positions $\{r_1,r_2,\ldots,r_N\}$ to coincide, whereas for
  the six-vertex model the inequalities in \eqref{cba-state} are
  strict, due to the ``exclusion principle'', that no more than one ``down
  arrow'' can be at the same site.   

\medskip\noindent
\subsection{Algebraic Bethe ansatz}
The Bethe state \eqref{cba-state}
may also be constructed within the framework of the QISM
\cite{Faddeev:1979gh,Takhtajan:1979iv}. Introduce the so-called
monodromy matrix 
\be\label{M-in}
\boldsymbol{{\cal M}}(y)\equiv
\boldsymbol{{\cal M}}(y\,|\,x_N,x_N',\ldots,x_1,x'_1)=
\boldsymbol{{\cal L}}_{N}(x_N,x_N',y)\,
\boldsymbol{{\cal L}}_{N-1}(x_{N-1},x_{N-1}',y)
\cdots\boldsymbol{{\cal L}}_{1}(x_{1},x_{1}',y)\,,
\ee
where $\boldsymbol{{\cal L}}_{n}(x_n,x'_n,y)$ 
is the two-by-two matrix 
\eqref{Lop}, whose entries are elements of the $U_q(sl(2))$
algebra \eqref{uqsl2}, acting in the $n$-th
component of the tensor product ${\cal H}^{\rm(vertex)}$,  
defined by \eqref{qspace}.
Each of these components realizes an infinite-dimensional highest weight
representation of this algebra given by \eqref{piplus_s1} 
(where $s_1$ for the $n$-th component is 
replaced by $s_n$, such that $q^{2s_n}=x'_n/x_n$).
It is convenient to denote the entries of the monodromy matrix as
\be\label{Mmat}
\bm{{\cal M}}(y)=
\left(\begin{array}{cc} 
{ \hat{ {\mathsf A}}}(y) & { \hat{ {\mathsf B}}} (y)\\[0.2cm] 
{ \hat{ {\mathsf C}}} (y) & { \hat{ {\mathsf D}}}(y)
\end{array}\right)\,,
\ee
where $ {\hat{ {\mathsf A}}},\, {\hat{ {\mathsf B}}},\, 
{ \hat{ {\mathsf C}}},\,{ \hat{ {\mathsf D}}}$
are operators acting in the quantum space \eqref{qspace}.
With the above notations, the transfer matrix ${\bf T}_{\hf}(y)$,
defined by \eqref{ts-def} with $s=\hf$, 
is given by
\be\label{Tmat1}
{\bf T}_{\hf}(y)=\Big(q^M \prod_{n=1}^N (y/x'_n)\Big)\,
\Big({ \hat{ {\mathsf A}}}(y) +\omega_h\, { \hat{ {\mathsf D}}}(y)\Big)\  ,
\ee
while
the Bethe state \eqref{cba-state}-\eqref{one-p},
with the set $\{y_m\}_{m=1}^M$ solving the Bethe ansatz equations, is expressed as 
\be\label{aba-state}
\boldsymbol{\Psi}={ \hat{ {\mathsf B}}}(y_M)\,\cdots\,
{ \hat{ {\mathsf B}}}(y_2)\, { \hat{ {\mathsf B}}}(y_1)\,{\bf {\Psi}}_0\,.
\ee
The corresponding eigenvalue of the transfer matrix \eqref{Tmat1},
\be
{\bf T}_{\hf}(y)\,{\bf \Psi}={T}_{\hf}(y)\,{\bf \Psi}\,,
\ee
reads \cite{Takhtajan:1979iv,Korepin:1982gg} 
\begin{equation}
T_{\hf}(y)\;=\;f(y) \prod_{j=1}^M \frac{1-q^2 y^2/y_j^2}{1-y^2/y_j^2} \;+\;
(\omega_h w)\, g(y) \prod_{j=1}^M \frac{1-q^{-2} y^2/y_j^2}{1-y^2/y_j^2} \;.
\end{equation}
where $f(y)$ and $g(y)$ are defined in \eqref{fg-def}.
Combining the last formula with
\eqref{Q-norm} and \eqref{tqeq} one immediately deduces
the expression \eqref{Q-eig} for the eigenvalue of the operator ${\bf Q}(y)$. 

Finally, from the definitions \eqref{M-in}, \eqref{Mmat} it follows that 
\be
\hat{{\mathsf B}}(y)=-\sum_{r=1}^N
\left(\prod_{i=1}^{r-1} [q^{-\widehat{h}_i}\,y_m/x_i]\right)
\,[q^{\widehat{h}_r}]\,\widehat
\sigma_r^+\,
\left(\prod_{i=r+1}^{N} [q^{\widehat{h}_i}\,y_m/x'_i]\right)\,,
\ee
where $\widehat h_i$ is the particle number operator \eqref{creation}
acting at the
$i$-th component of the product \eqref{qspace}.
Substituting this into 
\eqref{aba-state} and rearranging similar terms in a special way 
one obtains the coordinate
Bethe ansatz expression for the eigenvectors given by
\eqref{cba-state}-\eqref{ap-def}. 
In the context of the XXX related models 
this statement first appeared in \cite{Izergin:1987} 
(see eq.(4.8) therein). For generalizations to the XXZ case 
and derivations see \cite{Slavnov:2007,Fuksa:2017}.
Note that this equivalence 
between the two forms of the ``ansatz'' for the eigenvectors holds 
for arbitrary values of $\{y_m\}_{m=1}^M$, in particular, it does not   
require them to satisfy the Bethe ansatz equations \eqref{BA-eq}.

\section{Generalizations to other models}

Apparently, all vertex models 
associated with quantized affine Lie algebras \cite{Bazhanov:1984gu,Jimbo:1986}
and superalgebras \cite{Bazhanov:1986av} can
be reformulated as Ising-type models. 
Indeed, the most important algebraic structure required for such a
reformulation --- the 3-dimensional interpretation based on the
tetrahedron equation --- is known for a majority of these models, 
\cite{Bazhanov:2005as,Sergeev:2009,
Bosnjak:2016oze,KunS1,KunS2,KunS3}.  
In support of our statement we present here a new class of  
Ising-type models related to the $U_q(\widehat{sl}(n))$ algebra, 
extending the results of this 
paper which was devoted to the $n=2$ case. 
The formulation of these new
models follows
exactly the same description as in Sect.~\ref{isingtype}, except
that each site of the lattice now carries an $(n-1)$-component integer
spin 
variable $\boldsymbol{a}=(a_1,a_2,\ldots , a_{n-1})\in{\mathbb
  Z}^{(n-1)}$ and the Boltzmann weight function $V$, entering the definition 
of the edge weights \eqref{twotypes}, is now replaced with 
\begin{equation}\label{Veight}
V_x(\boldsymbol{a},\boldsymbol{b}) \;=\; q^{2Q(\boldsymbol{a},\boldsymbol{b})-Q(\boldsymbol{a},\boldsymbol{a})-Q(\boldsymbol{b},\boldsymbol{b})}\;
\left(\frac{q}{x}\right)^{a_0-b_0}\frac{(x^2;q^2)_{a_0-b_0}}{\ds\prod_{i=1}^{n-1}(q^2;q^2)_{a_i-b_i}}\;.
\end{equation}
Note, that this function 
depends on both spin variables $\boldsymbol{a}$ and 
$\boldsymbol{b}$ at the ends of the edge, rather than their
difference, and 
\begin{equation}
a_0=\sum_{i=1}^{n-1}a_i\,,\qquad
\qquad Q(\boldsymbol{a},\boldsymbol{b}) \;=\; \sum_{1\le i<j\le n-1}
a_ib_j\,.
\qquad 
\end{equation}
Similar to \eqref{W-def1}, \eqref{W-def2} one can define 
the corresponding IRF weight 
\begin{subequations}\label{SS2}
\bea
\label{W-def1-2}
\ds \mathcal{W}^{sl(n)}(\boldsymbol{a},\boldsymbol{b},\boldsymbol{c},
\boldsymbol{d}\,|\,x,x',y,x')&=&
\frac{\ds V_{y/y'}(\boldsymbol{b},\boldsymbol{d})}
{\ds V_{y/y'}(\boldsymbol{a},\boldsymbol{c})}\;
\sum_{\boldsymbol{n}=\max(\boldsymbol{b},\boldsymbol{c})}^{\boldsymbol{a}}
\frac{\ds 
V_{x/y'}(\boldsymbol{a},\boldsymbol{n})
V_{y'/x'}(\boldsymbol{n},\boldsymbol{b})
V_{y/x}(\boldsymbol{n},\boldsymbol{c})}
{\ds V_{y/x'}(\boldsymbol{n},\boldsymbol{d})}\,,\\[.3cm]
\label{W-def2-2}
&=&
\frac{\ds V_{x/x'}(\boldsymbol{a},\boldsymbol{b})}{\ds V_{x/x'}(\boldsymbol{c},\boldsymbol{d})}\;
\sum_{\boldsymbol{n}=\boldsymbol{d}}^{\min(\boldsymbol{b},\boldsymbol{c})}
\frac{\ds 
V_{x/y'}(\boldsymbol{n},\boldsymbol{d})
V_{y'/x'}(\boldsymbol{c},\boldsymbol{n})
V_{y/x}(\boldsymbol{b},\boldsymbol{n})}
{\ds V_{y/x'}(\boldsymbol{a},\boldsymbol{n})}\;.
\eea
\end{subequations}
where the summation is taken over $(n-1)$ independent integer variables $n_i$, 
such that $\max(b_i,c_i)\leq n_i \leq a_i$ in (\ref{W-def1-2}) and,
similarly, in \eqref{W-def2-2}. The corner spins 
should obey the relations 
$a_i\geq b_i,c_i\geq d_i$  $(i=1,2,\ldots,n-1)$, otherwise the weights  
$\mathcal{W}^{sl(n)}(\boldsymbol{a},\boldsymbol{b},\boldsymbol{c},
\boldsymbol{d})$ 
are assumed to vanish identically.
The equality of two different
expressions in \eqref{SS2} is the star-star relation, which ensures the
integrability of the model. It has exactly the same graphical
representation as in Fig.~\ref{sspic} and reduces to \eqref{ssrel} for
$n=2$. For $n>3$  
the star-star relation \eqref{SS2} is a new identity which we
currently claim as a conjecture (though we have thoroughly verified 
it for many particular cases).  

It is not difficult to check that the IRF weights
$\mathcal{W}^{sl(n)}(\boldsymbol{a},\boldsymbol{b},
\boldsymbol{c},\boldsymbol{d}\,|\,x,x',y,x')$
only depend on the differences between the corner spins, so one can
trivially convert these weights into a vertex $R$-matrix (similar to
\eqref{R-def}). The resulting solutions of
the Yang-Baxter equation are associated with infinite-dimensional
highest weight evaluations representations of the $U_q(\widehat{sl}(n))$
algebra, namely, the symmetric tensor representations
\cite{Bosnjak:2016oze}.      

The vertex $R$-matrix, corresponding to \eqref{SS2} reads
\be
{\mathcal
  R}^{sl(n)}(x,x',y,y')^{\bm{j}_1,\bm{j}_2}_{\bm{i}_1,\bm{i}_2}=\delta_{\bm{{i}}_1+\bm{i}_2,
\bm{j}_1+\bm{j}_2}\, 
{\Wc}^{sl(n)}(\bm{i}_1+\bm{i}_2,\bm{j}_2,\bm{i}_1,0\,|x,x',y,y')\,,
\label{Rsln-def}
\ee
where $\bm{j}_1,\bm{j}_2,\bm{i}_1,\bm{i}_2\in {\mathbb
  Z}_{\ge0}^{(n-1)}$.
Note, if $x'=q^{\mu_1}x$ and $y'=q^{\mu_2}y$, with $\mu_1,\mu_2\in{\mathbb
  Z}_{\ge0}$, then the above $R$-matrix admits a finite-dimensional
reduction. For instance, 
the Cherednik-Perk-Shultz $R$-matrix \cite{Cherednik:1980ey,Perk:1981}
(vector representations of $U_q(sl(n))$) 
appears in the case $x'=qx$, $y'=qy$.
Let
\begin{equation}
\boldsymbol{e}_0\;=\;(0,\ldots,0)\;,\quad \boldsymbol{e}_i\;=\;(0,\ldots,\underbrace{1}_{i^{th} \;\textrm{place}},\ldots,0)\;,
\end{equation}
and $\varepsilon_{i,j}=-\varepsilon_{j,i}=1$ for $i<j$, except the
case $i=0$ or $j=0$ where $\varepsilon_{i,j}\;=\;0$. 
Then, 
\begin{equation}
\mathcal{R}_{\boldsymbol{e}_i,\boldsymbol{e}_i}^{\boldsymbol{e}_i,\boldsymbol{e}_i}=1\;,\qquad
\mathcal{R}_{\boldsymbol{e}_i,\boldsymbol{e}_j}^{\boldsymbol{e}_i,\boldsymbol{e}_j}\;=\;
q^{\varepsilon_{i,j}} \frac{[x/y]}{[qx/y]}\;,\qquad
\mathcal{R}_{\boldsymbol{e}_i,\boldsymbol{e}_j}^{\boldsymbol{e}_j,\boldsymbol{e}_i}\;=\;
\left(\frac{y}{x}\right)^{\varepsilon_{i,j}}  \frac{[q]}{[qx/y]}\;,\qquad
i\not=j\,,
\end{equation}
where we have omitted the superscript ``$sl(n)$''.
      
\section{Conclusion}
In this paper we have presented a new approach to the six-vertex model and
its higher spin generalizations. We have reformulated these models as
Ising-type models with only a two-spin interaction across edges of the
square lattice. This reformulation leads to a significant simplification 
of the algebraic theory of the higher spin six-vertex model, since 
the edge Boltzmann weights are given by a very simple formula
\eqref{Vdef}. As a result explicit expressions for associated 
$R$-matrices, as well as their properties, become much more transparent.

An Ising-type model can be equivalently converted into a
vertex model in (at least) two different ways. In the first way, 
the $R$-matrix defining the vertex weights is deduced from the
Bolzmann weights of the four-edge star (like the one shown in
Fig.~\ref{fig-star}) by using the star-square transformation followed
by the IRF-vertex correspondence. In our case 
this procedure leads to Eq.\eqref{R-def2}. 
In the second way the vertex-type $R$-matrix is identified with the ``box
diagram'', shown in Fig.~\ref{fig-box}, that leads to Eq.\eqref{def3}.
In Sect.~\ref{descen} we have shown these two $R$-matrices are related
by a linear relation \eqref{fourier}, which is not unnatural to expect,
since they both describe exactly the same model. 

From the quantum group point of view these $R$-matrices 
intertwine equivalent 
representations of the $U_q(\widehat{sl}(2))$ algebra, connected 
by a similarity transformation in the representation space. 
%
For the first of these representation (corresponding to \eqref{Lmat4}) 
the Cartan element is diagonal, while for the other one (corresponding to
\eqref{Lop-new}) it is realised as a shift operator.  
It should be stressed, that 
it is the shift operator realization of the Cartan elements
that leads to the factorized $R$-matrices. This fact has been previously
observed \cite{Bazhanov:1989nc} for the chiral Potts model and its
generalizations \cite{BKMS}, which are related to the
$U_q(\widehat{sl}(n))$ algebra. We expect a similar phenomenon to take place
for all other quantized affine algebras.  
One disadvantage of the factorized $R$-matrices 
is the necessity to work with
infinite-dimensional representations, whereas the models with a finite
number of discrete spin states only appear via a reduction. In
practice, however, this
does not create real problems. Actually, it
would be fair to say that certain infinite-dimensional representations (in
particular, the $q$-oscillator representations) 
cannot be avoided, since they 
are required for the transfer matrix
construction of the Baxter ${\bf Q}$-operators 
\cite{Bazhanov:1996dr,Bazhanov:1998dq}. 
Moreover, they play the important role of ``prefundamental'' 
representations in the theory of quantized affine algebras
\cite{Hernandez:2012,Frenkel:2012}.

As is well known, the six-vertex model has a vast number of important
applications to various problems of physics and mathematics (for reviews
see \cite{Lieb:1980ix,Reshetikhin:2010}). 
We mention, in particular, the enumeration of
the alternating sign matrices \cite{RazStr,BGN}, asymmetric stochastic
exclusion processes 
and stochastic lattice models \cite{Borodin:2014,Garbali:2016}.    
Most recently the (inhomogeneous) six-vertex model was involved in
fascinating connections to black hole sigma models in quantum field
theory %
\cite{Jacobsen:2005xz,Ikhlef:2008zz,Ikhlef:2011ay,Frahm:2012eb,Candu:2013fva,Frahm:2013cma,Bazhanov:2020,Bazhanov:2020uju}.
It would be interesting to revisit these problems and
connections in view of the Ising-type structures studied in our work.  

\section{Acknowledgements}
The authors thank S.L.Lukyanov, V.V.Mangazeev and N.Y.Reshetikhin for 
stimulating discussions. Special thanks to R.J.Baxter and G.A.Kotousov 
for reading the manuscript and important comments. 
One us (VVB) thanks the Sydney Mathematical
Research Institute and especially Prof. A.Molev 
for hospitality at Sydney when some parts of this work were completed.

\medskip
\noindent
SMS acknowledges the support of the Australian Research Council grant 
DP190103144. 

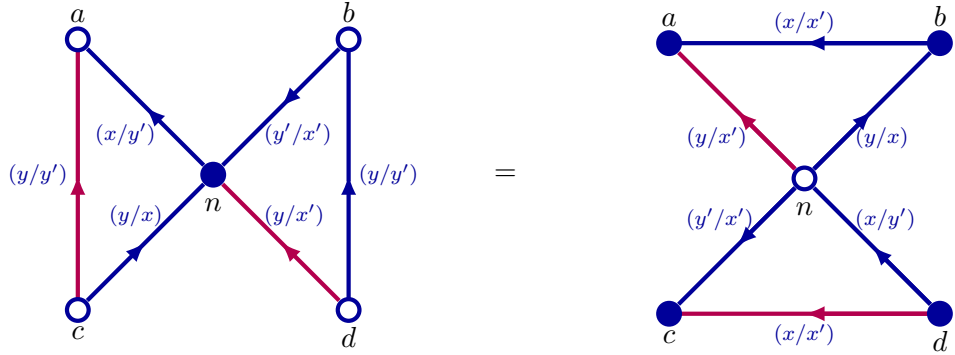
\begin{figure}[ht]
\begin{center}
\begin{tikzpicture}[scale=1.8,baseline={(0,-.05)}]
\draw [-latex, \PAC, ultra thick] (1.43,1.57) -- (1,2); \draw [\PAC, ultra thick] (1.05,1.95) -- (0.57,2.43);
\draw [-latex, \PAC, ultra thick] (2.43,2.43) -- (2,2); \draw [\PAC, ultra thick] (2.05,2.05) -- (1.57,1.57);
\draw [-latex, \PAC, ultra thick] (0.57,0.57) -- (1,1); \draw [\PAC, ultra thick] (.95,.95) -- (1.43,1.43);
\draw [-latex, \GR, ultra thick] (2.43,0.57) -- (2.,1.); \draw [\GR, ultra thick] (2.05,.95) -- (1.57,1.43);
\draw [-latex, \GR, ultra thick] (0.5,0.58) -- (0.5,1.5); \draw [\GR, ultra thick] (0.5,1.45) -- (0.5,2.42);
\draw [-latex, \PAC, ultra thick] (2.5,0.58) -- (2.5,1.5); \draw [\PAC, ultra thick] (2.5,1.45) -- (2.5,2.43);

\node[\PAC,left] at (.5,1.5) {\scriptsize$(y/y')$};
\node[\PAC,right] at (2.5,1.5) {\scriptsize$(y/y')$};
\node[\PAC,left] at (1.15,1.8) {\scriptsize$(x/y')$};
\node[\PAC,left] at (1.2,1.2) {\scriptsize$(y/x)$};
\node[\PAC,right] at (1.8,1.2) {\scriptsize$(y/x')$};
\node[\PAC,right] at (1.8,1.8) {\scriptsize$(y'/x')$};

\draw [ultra thick, \STE] (0.5,2.5) circle [radius=0.08]; \node[above] at (0.5,2.55) {$a$};
\draw [ultra thick, \STE] (0.5,0.5) circle [radius=0.08];\node[below] at (0.5,0.45) {$c$};
\draw [ultra thick, \STE] (2.5,2.5) circle [radius=0.08];\node[above] at (2.5,2.55) {$b$};
\draw [ultra thick, \STE] (2.5,0.5) circle [radius=0.08];\node[below] at (2.5,0.45) {$d$};
\draw [ultra thick, fill, \STE] (1.5,1.5) circle [radius=0.08];\node[below] at (1.5,1.4) {$n$};
\node [right] at (3.5,1.5) {$=\qquad\qquad$};
\end{tikzpicture}
\begin{tikzpicture}[scale=1.8,baseline={(0,0)}]

\node[\PAC,below] at (1.5,.5) {\scriptsize$(x/x')$};
\node[\PAC,above] at (1.5,2.5) {\scriptsize$(x/x')$};
\node[\PAC,left] at (1.15,1.8) {\scriptsize$(y/x')$};
\node[\PAC,left] at (1.2,1.2) {\scriptsize$(y'/x')$};
\node[\PAC,right] at (1.8,1.2) {\scriptsize$(x/y')$};
\node[\PAC,right] at (1.8,1.8) {\scriptsize$(y/x)$};

\draw [-latex, \GR, ultra thick] (1.43,1.57) -- (1,2); \draw [\GR, ultra thick] (1.43,1.57)-- (0.57,2.43);
\draw [-latex, \PAC, ultra thick] (1.57,1.57) -- (2,2); \draw [\PAC, ultra thick] (1.57,1.57) -- (2.43,2.43);
\draw [-latex, \PAC, ultra thick] (1.43,1.43) -- (1,1); \draw [\PAC, ultra thick] (1.43,1.43) -- (0.57,0.57);
\draw [-latex, \PAC, ultra thick] (2.43,0.57) -- (2,1); \draw [\PAC, ultra thick] (2.43,0.57) -- (1.57,1.43);
\draw [-latex, \PAC, ultra thick] (2.42,2.5) -- (1.5,2.5); \draw [\PAC, ultra thick] (2.42,2.5) -- (0.58,2.5);
\draw [-latex, \GR, ultra thick] (2.42,0.5) -- (1.5,0.5); \draw [\GR, ultra thick] (2.42,0.5) -- (0.58,0.5);
\draw [ultra thick, fill,\STE] (0.5,2.5) circle [radius=0.08]; \node[above] at (0.5,2.55) {$a$};
\draw [ultra thick,fill, \STE] (0.5,0.5) circle [radius=0.08];\node[below] at (0.5,0.45) {$c$};
\draw [ultra thick, fill, \STE] (2.5,2.5) circle [radius=0.08];\node[above] at (2.5,2.55) {$b$};
\draw [ultra thick, fill,\STE] (2.5,0.5) circle [radius=0.08];\node[below] at (2.5,0.45) {$d$};
\draw [ultra thick,  \STE] (1.5,1.5) circle [radius=0.08];\node[below] at (1.5,1.4) {$n$};
\end{tikzpicture}
\caption{The star-star relation \eqref{ssrel2} 
with alternative graphical notations
  for the edge weight, as in \eqref{gnote2}.
\label{fig-star2}}
\end{center}
\end{figure}

\app{The star-star and Yang-Baxter relations}\label{appA}
\addcontentsline{toc}{section}{A. The star-star and Yang-Baxter relations}
\sapp{Proof of the star-star relation.}
In this Appendix we will prove the star-star \eqref{ssrel} and 
the Yang-Baxter \eqref{ybe}
relations. 
For the readers' convenience, let us reproduce here the definitions of the
IRF-type weights given by \eqref{W-def} and \eqref{W-def2}, 
\be
\Wc(a,b,c,d\,|\,x,x',y,y')=
\frac{V_{y/y'}(b-d)}{V_{y/y'}(a-c)}\,
\ 
\sum_{n=\max(b,c)}^{a}\frac{V_{x/y'}(a-n)V_{y'/x'}(n-b)
V_{y/x}(n-c)}{V_{y/x'}(n-d)}\,,\label{star1}
\ee
\be\label{star2}
{\overline \Wc}(a,b,c,d\,|\,x,x',y,y')
=\frac{V_{x/x'}(a-b)}{V_{x/x'}(c-d)}
\,\ \sum_{{n}=a}^{\min(b,c)} \ \ \frac{V_{x/y'}({n}-d)
V_{y'/x'}(c-{n}) 
V_{y/x}(b-{n})}
{V_{y/x'}(a-{n})}\,.
\ee
We need to prove that the two definitions lead to the same result
\be\label{ssrel2}
\Wc(a,b,c,d\,|\,x,x',y,y')=\overline{\Wc}(a,b,c,d\,|\,x,x',y,y')\,.
\ee
This is the star-star relation, which states the equality of the Boltzmann
weights of the two ``stars'' shown in Fig.~\ref{sspic}. 
Below it will be
more convenient to modify the graphical representations
\eqref{twotypes} for the edge weights. We will abolish the
rapidity lines, but instead indicate the rapidity ratio arguments of the
weights for all edges of the lattice. These arguments are shown
near the edges, as in the pictures below,
\be\label{gnote2}
\begin{tikzpicture}
\node [left] at (-0.,1) {$({i}):$};
\draw [\STE, ultra thick] (1,0) circle [radius=0.09];
\draw [\STE, ultra thick] (1,2) circle [radius=0.09];
\node [\PAC,right] at (1,1) {
$(x)$};
\draw [-latex, \PAC, ultra thick] (1,0.12) -- (1,1);
\draw [\PAC, ultra thick] (1,.12) -- (1,1.88);
\node [above] at (1,2.1) {$a$};
\node [below] at (1,-0.1) {$b$};
\node [right] at (2.1,1) {$
\ds =\;V_{x}(a-b)\;;\qquad\qquad
$};
\end{tikzpicture}
\qquad
\begin{tikzpicture}
\node [left] at (-0.5,1) {$({ii}):$};
\draw [\STE, ultra thick] (1,0) circle [radius=0.09];
\draw [\STE, ultra thick] (1,2) circle [radius=0.09];
\node [\PAC,right] at (1,1) {
$(x)$};
\draw [-latex, \GR, ultra thick] (1,0.12) -- (1,1); 
\draw [\GR, ultra thick] (1,.12) -- (1,1.88);
\node [above] at (1,2.1) {$a$};
\node [below] at (1,-0.1) {$b$};
\node [right] at (2.1,1) {$
\ds =\;\frac{1}{V_{x}(a-b)}\;.
$};
\end{tikzpicture}
\ee
Otherwise, the rules remain unchanged: the edge arrow pointing from
$b$ to $a$ means that the spin difference argument of the weight
function is 
$(a-b)$ rather than $(b-a)$, 
moreover, the colour distiguishes the type $(i)$
(blue) and the type $(ii)$ (red) edges with the Boltzmann weights
$V_x(a-b)$ and $(V_x(a-b))^{-1}$, respectively. 
Recall, that the
function $V_x(n)$ is defined in \eqref{Vdef}. 
With the new 
convensions the star-star relation \eqref{ssrel2} is presented in
Fig.~\ref{fig-star2}. 

Below we will use the second Sears's transformation formula for the basic
hypergeometric series (see \cite{Sears:1951}), 
\begin{equation}\label{S2}
\phantom{|}_4\phi_3\left(\begin{array}{cccc|}
q^{-2n} & \as & \bs & \cs\\ & \dds & \es & \fs\end{array}\,\,q^2,q^2\right) \;=\;
\frac{\ds \left(\as,\,\frac{\es\, \fs}{\as\, \bs},\,\frac{\es\, \fs}{\as\,
    \cs};\,q^2\right)_n}{\ds \left(\es,\,\fs,\,\frac{\es\, \fs}{\as\, \bs\, \cs};\,q^2\right)_n}\ \phantom{|}_4\phi_3\left(\begin{array}{cccc|}
\ds q^{-2n} & \ds \frac{\es}{\as} & \ds\frac{\fs}{\as} & \ds \frac{\es\,
  \fs}{\as\, \bs\, \cs} 
\\&&&  \\
\ds & \ds \frac{\es\, \fs}{\as\, \bs} & \ds \frac{\es\, \fs}{\as\, \cs}  & \ds
\frac{q^{2-2n}}{\as}\end{array} \,\,q^2,q^2\right)\;,
\end{equation}
where the integer $n\ge0$ and the other
parameters satisfiy the constraint
\begin{equation}
\as\, \bs \,\cs \;=\;q^{2n-2}\, \dds\, \es\, \fs\;.
\end{equation}
Here we use the standard notations 
\begin{equation}\label{phi43}
\phantom{|}_{r+1}\phi_r\left(\begin{array}{ccccc|}
q^{-2n}, & \as_1, &\as_2, & \ldots\ , & \as_r\\ & \bs_1, &\bs_2 & \ldots\ , &
\bs_r\end{array}\,\,q^2,q^2\right) 
\;=\;
\sum_{k=0}^n \frac{(q^{-2n},\as_1, \as_2, \ldots,
  \as_r;q^2)_k}{(q^2,\bs_1,\bs_2, 
  \ldots, \bs_r;q^2)_k}\, q^{2k}\;,
\end{equation}
and 
\be\label{qprod}
(\as;q^2)_k=\prod_{\ell=0}^{k-1}(1-\as\, q^{2\ell})\,;\qquad 
(\as_1,\as_2,\ldots,\as_m;q^2)_k=(\as_1;q^2)_k\,(\as_2;q^2)_k\,\cdots\,
(\as_m;q^2)_k\,.
\ee

We are now ready to prove the star-star relation
\eqref{ssrel2}. First, consider the case 
$b\ge c$.
From the definition \eqref{Vdef} it is easy to verify, that
\begin{equation}\label{cor}
V_x(a-n) \;=\; V_x(a) \left(\frac{q}{x}\right)^n
\frac{(q^{-2a};q^2)_n}
{(q^{2-2a}\,x^{-2};q^2)_n}\;.
\end{equation}
Using this formula and the definition \eqref{phi43} 
one can rewrite \eqref{star1} and \eqref{star2} in the form
\begin{equation}\label{SL}
\begin{array}{l}
\ds \Wc\;=\;\frac{V_{y/y'}(b-d)}{V_{y/y'}(a-c)} 
\frac{V_{y'/x'}(a-b) V_{y/x}(a-c)}{V_{y/x'}(a-d)}\\
[7mm]
\ds \times
\phantom{|}_4\phi_3\left(\begin{array}{rrrr|}
\ds q^{-2(a-b)}, & \ds q^{-2(a-c)}, & \ds (x/y')^2, & \ds q^{2(1-a+d)} 
(x'/y)^{2}\\
[5mm]
\ds  & \ds q^{2(1-a+b)} (x'/y')^{2}, & \ds q^{-2(a-d)}, & 
\ds q^{2(1-a+c)} (x/y)^{2} \end{array}\,\,q^2,q^2\right) \;,
\end{array}
\end{equation}
and
\begin{equation}\label{SR}
\begin{array}{l}
\ds \overline{\Wc}\;=\;\frac{V_{x/x'}(a-b)}{V_{x/x'}(c-d)} \frac{V_{y'/x'}(c-d) 
V_{y/x}(b-d)}{V_{y/x'}(a-d)}\\
[7mm]
\ds \times
\phantom{|}_4\phi_3\left(\begin{array}{rrrr|}
\ds q^{-2(c-d)}, & \ds q^{-2(b-d)}, & \ds (x/y')^2, & \ds q^{2(1-a+d)} 
(x'/y)^{2}\\
[5mm]
\ds  & \ds q^{2(1-c+d)} (x'/y')^{2}, & \ds q^{-2(a-d)}, & \ds
q^{2(1-b+d)} 
(x/y)^{2} \end{array}\,\,q^2,q^2\right) \;.
\end{array}
\end{equation}
where we have omitted the arguments of $\Wc$ and $\overline{\Wc}$ since
they do not change during these calculations.
Applying the second Sears transformation (\ref{S2}) to both (\ref{SL}) and (\ref{SR}), one obtains 
\begin{equation}
\begin{array}{l}
\ds \Wc\;=
\;\frac{V_{x/x'}(a-b)\, V_{y/y'}(a-d)\, V_{y/x}(b-c) }
{ V_{y/y'}(a-c)\, V_{y/x'}(a-d)} \\
[7mm]
\ds \times
\phantom{|}_4\phi_3\left(\begin{array}{rrrr|}
\ds q^{-2(a-b)}, & \ds q^{-2(c-d)}, & \ds q^2\,(x/y)^{2}, & \ds (y'/x')^2\\
[5mm]
\ds  & \ds q^{2(1-a+d)}\, (y'/y)^{2}, & \ds (x/x')^2, & \ds
q^{2(1-c+b)} 
\end{array}\,\,q^2,q^2\right) \;,
\end{array}\label{SL2}
\end{equation}
and
\begin{equation}
\begin{array}{l}
\ds \overline{\Wc}\;=\;\frac{V_{x/x'}(a-b)\, V_{y/y'}(a-d) \,
V_{y/x}(b-c) }{ V_{y/y'}(a-c)\, V_{y/x'}(a-d)} \\
[7mm]
\ds \times
\phantom{|}_4\phi_3\left(\begin{array}{rrrr|}
\ds q^{-2(c-d)}, & \ds q^{-2(a-b)}, & \ds q^2\,(x/y)^{2}, & \ds (y'/x')^2\\
[5mm]
\ds  & \ds q^{2(1-a+d)}\, (y'/y)^{2}, & \ds (x/x')^2, & \ds q^{2(1-c+b)} \end{array}\,\,q^2,q^2\right) \;,
\end{array}\label{SR2}
\end{equation}

The resulting expressions coincide due to the obvious symmetry of
${}_4\varphi_3$ upon permutations of its upper line
arguments. Evidently, this proves the star-star relation \eqref{ssrel2}.
The above computation is valid for $b\geq c$. 
The case $b<c$ can be obtained by the transformation  $b\leftrightarrow c$,
$x'\leftrightarrow y^{-1}$ and $y'\leftrightarrow x^{-1}$, which maps
the star-star relation into itself.

\sapp{Proof of the IRF-type Yang-Baxter equation\label{app-irf-ybe}}
Let us now prove the Yang-Baxter relation \eqref{ybe}. 
The star-star
relation \eqref{ssrel2} could be used in two ways. Firstly, one could
express the sum in \eqref{star1} via the sum in \eqref{star2} (of
course, taking into account the pre-factors in front of the sums). 
We will call this
operation as the ``{\bf direct star-star map centered at $n$}'', where $n$
denotes the summation index in the first sum \eqref{star1}, which is
the central spin in the left star depicted in Fig.~\ref{fig-star2}. 
Similarly, one could
express the second sum \eqref{star2} via the sum in \eqref{star1}.  
We will call this
operation as the ``{\bf reverse star-star map centered at ${n}$}'',
where ${n}$ now refers to the central spin in the right star 
in Fig.~\ref{fig-star2}.

Return now to the Yang-Baxter equation \eqref{ybe}. For convenience we
reproduce it here
\begin{equation}\label{ybe2}
\begin{array}{l}
\ds \sum_{a\in{\mathbb Z}} 
\Wc(g,a,b,f\,|\,{\boldsymbol x},{\boldsymbol y})\ 
\Wc(c,e,g,a\,|\,\xb,\zb)\ 
\Wc(e,d,a,f\,|\,\yb,\zb)\;=\;\\
[3mm]
\ds \qquad\qquad\qquad\qquad\qquad=\;
\sum_{h\in{\mathbb Z}} 
\Wc(c,h,g,b\,|\,\yb,\zb)\ 
\Wc(h , d ,b , f\,|\,\xb,\zb)\ 
\Wc(c , e ,h , d\,|\,\xb,\yb)
\end{array}
\end{equation}
where the bold symbols denote the pairs of spectral variables 
$\xb=(x,x')$, $\yb=(y,y')$ and $\zb=(z,z')$. Consider the LHS of this
equation and use the expression \eqref{star1} for each of the weights 
$\Wc(g,a,b,f\,|\,{\boldsymbol x},{\boldsymbol y})$,
$\Wc(c,e,g,a\,|\,\xb,\zb)$ and 
$\Wc(e,d,a,f\,|\,\yb,\zb)$
denoting the central spins by $n_1$, $n_2$ and $n_3$,
respectively. Then, with the graphical representation of the weight
\eqref{star1} 
shown on the left side of Fig.~\ref{fig-star2}, the complete LHS of
\eqref{ybe2} will be represented as shown below
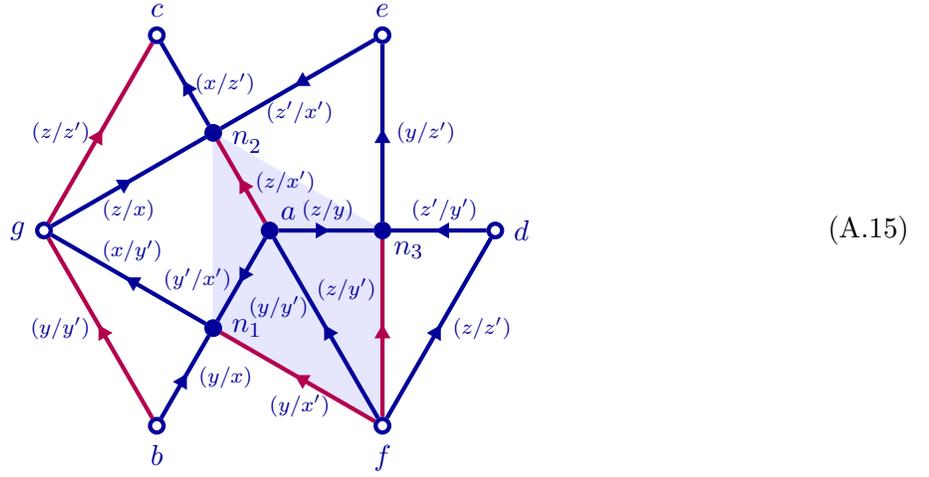
\newcommand{\midarrow}{\tikz \draw[-triangle 60] (0,0) -- +(.1,0);}

\def\pa{(0.,0.)}
\def\pb{(-5,-8.66025)}
\def\pc{(-5,+8.66025)}
\def\pd{(10.,0.)}
\def\pe{(+5,+8.66025)}
\def\pf{(+5,-8.66025)}
\def\pg{(-10.,0.)}

\def\pbb{(-2.5,-4.33013)}
\def\pcc{(-2.5,+4.33013)}
\def\pdd{(5.,0.)}
\def\pee{(+2.5,+4.33013)}
\def\pff{(+2.5,-4.33013)}
\def\pgg{(-5.,0.)}
\def\rad{0.3}
\def\scal{0.3}
\be\label{ybepic1}
\begin{tikzpicture}[scale=\scal,baseline=(current  bounding  box.center)]

\draw [fill, opacity=.1, blue]\pcc -- \pdd --\pf --\pbb--\pcc;

\draw \pg -- \pcc node [\PAC, midway, below=1.mm] {\scriptsize$(z/x)$};
\draw \pg -- \pc node [\PAC, \PAC, midway, left] {\scriptsize$(z/z')$};
\draw \pc -- \pcc node [\PAC, midway, right] {\scriptsize$(x/z')$};
\draw \pg -- \pbb node [\PAC, midway, above=1.mm] {\scriptsize$\,\,(x/y')$};
\draw \pg -- \pb node [\PAC, midway, left] {\scriptsize$(y/y')$};
\draw \pb -- \pbb node [\PAC, midway, right] {\scriptsize$\,(y/x)$};
\draw \pf -- \pbb node [\PAC, midway, below=1mm] {\scriptsize$\,(y/x')$};
\draw \pa -- \pbb node [\PAC, midway, left] {\scriptsize$\,(y'/x')$};
\draw \pa -- \pcc node [\PAC, midway, right] {\scriptsize$\,(z/x')$};
\draw \pa -- \pdd node [\PAC, midway, above] {\scriptsize$\,(z/y)$};
\draw \pdd -- \pd node [\PAC, midway, above] {\scriptsize$\,\,\,(z'/y')$};
\draw \pe -- \pdd node [\PAC, midway, right] {\scriptsize$\,(y/z')$};
\draw \pf -- \pa node [\PAC, midway, above left] {\scriptsize$(y/y')\,\,$};
\draw \pf -- \pdd node [\PAC, pos=.7, left=-.5mm] 
{\scriptsize$(z/y')$};
\draw \pf -- \pd node [\PAC, midway, right] {\scriptsize$\,(z/z')$};
\draw \pe -- \pcc node [\PAC, midway, below=1mm] {\scriptsize$\,(z'/x')$};

\begin{scope}[very thick, every node/.style={sloped,allow upside down}]
\draw [ \PAC, ultra thick] \pg -- node {\midarrow} \pcc; 
\draw [ \PAC, ultra thick] \pe -- node {\midarrow} \pcc; 
\draw [ \PAC, ultra thick] \pcc -- node {\midarrow} \pc; 
\draw [ \GR, ultra thick] \pa -- node {\midarrow} \pcc;

\draw [ \GR, ultra thick] \pg -- node {\midarrow} \pc;
\draw [ \PAC, ultra thick] \pb -- node {\midarrow} \pbb; 
\draw [ \PAC, ultra thick] \pa -- node {\midarrow} \pbb;
\draw [ \PAC, ultra thick] \pbb -- node {\midarrow} \pg;
\draw [ \GR, ultra thick] \pb -- node {\midarrow} \pg;
\draw [ \GR, ultra thick] \pf -- node {\midarrow} \pbb; 
\draw [ \PAC, ultra thick] \pf -- node {\midarrow} \pa;
\draw [ \PAC, ultra thick] \pa -- node {\midarrow} \pdd; 
\draw [ \PAC, ultra thick] \pd -- node {\midarrow} \pdd; 
\draw [ \PAC, ultra thick] \pdd -- node {\midarrow} \pe; 
\draw [ \GR, ultra thick] \pf -- node {\midarrow} \pdd;
\draw [ \PAC, ultra thick] \pf -- node {\midarrow} \pd; 

\draw [white, fill, ultra thick] \pa circle [radius=\rad];
\draw [white, fill, ultra thick] \pb circle [radius=\rad];
\draw [white, fill, ultra thick] \pc circle [radius=\rad];
\draw [white, fill, ultra thick] \pd circle [radius=\rad];
\draw [white, fill, ultra thick] \pe circle [radius=\rad];
\draw [white, fill, ultra thick] \pf circle [radius=\rad];
\draw [white, fill, ultra thick] \pg circle [radius=\rad];
\draw [white, fill, ultra thick] \pdd circle [radius=\rad];
\draw [white, fill, ultra thick] \pbb circle [radius=\rad];
\draw [white, fill, ultra thick] \pcc circle [radius=\rad];

\draw [\PAC, fill, ultra thick] \pa circle [radius=\rad];
\draw [\PAC, ultra thick] \pb circle [radius=\rad];
\draw [\PAC, ultra thick] \pc circle [radius=\rad];
\draw [\PAC, ultra thick] \pd circle [radius=\rad];
\draw [\PAC, ultra thick] \pe circle [radius=\rad];
\draw [\PAC, ultra thick] \pf circle [radius=\rad];
\draw [\PAC, ultra thick] \pg circle [radius=\rad];

\draw [\PAC,  fill, ultra thick] \pdd circle [radius=\rad];
\draw [\PAC,  fill, ultra thick] \pbb circle [radius=\rad];
\draw [\PAC,  fill, ultra thick] \pcc circle [radius=\rad];
\node [\PAC,below right] at \pdd {$n_3$};
\node [\PAC,right=1mm] at \pcc {$n_2^{\phantom{|}}$};
\node [\PAC,right=1mm] at \pbb {$n_1$};
\node [\PAC,above right] at \pa {$a$};
\node [\PAC,below=1mm] at \pb {$b$};
\node [\PAC,above=1mm] at \pc {$c$};
\node [\PAC,right=1mm] at \pd {$d$};
\node [\PAC,above=1mm] at \pe {$e$};
\node [\PAC,below=1mm] at \pf {$f$};
\node [\PAC,left=1mm] at \pg {$g$};

\end{scope}
\end{tikzpicture}
\ee
Note, that the external sites, corresponding to fixed spins in
\eqref{ybe2}, are represented by open circles, while the internal
sites, corresponding to the summation spin indices, are shown by
filled circles. 

Now, make the reverse star map centered at $a$. This transformation
involves six edges, the four edges surrounding the site $a$ and two boundary
edges $(f, n_1)$ and $(f, n_3)$ (see the shaded
area on the figure \eqref{ybepic1} above). The result is shown in the
left side of \eqref{ybepic2}. 

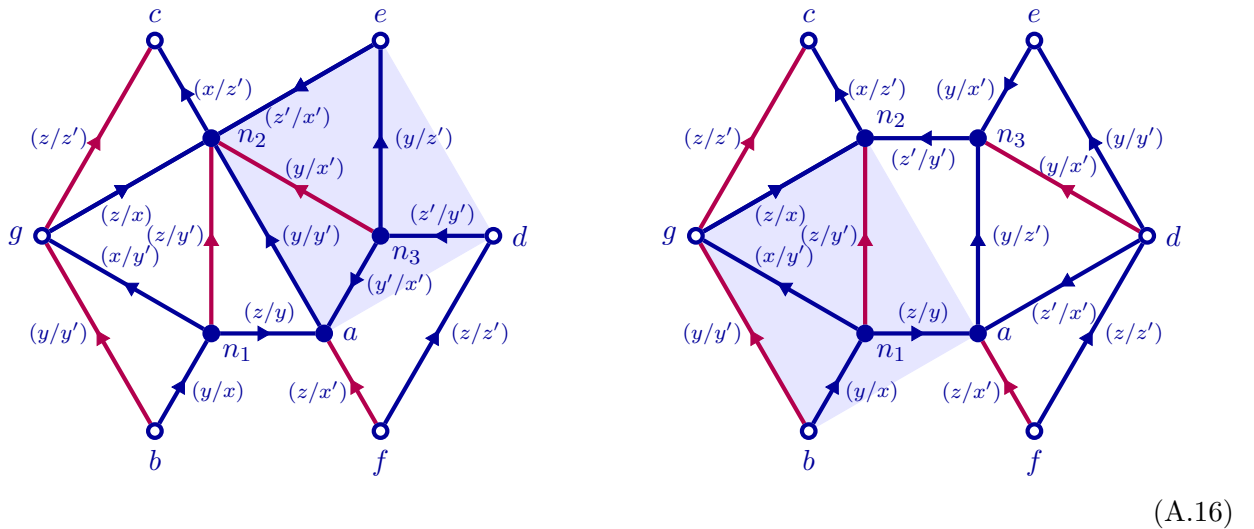
\def\pdd{(5.,0.)}

\def\pa{(+2.5,-4.33013)}
\def\pddd{(+2.5,+4.33013)}

\be\label{ybepic2}
\begin{tikzpicture}[scale=\scal]
\draw [fill, opacity=.1, blue]\pcc -- \pe --\pd --\pa--\pcc;
\draw \pg -- \pcc node [\PAC, midway, below=1.mm] {\scriptsize$(z/x)$};
\draw \pg -- \pc node [\PAC, \PAC, midway, left] {\scriptsize$(z/z')$};
\draw \pc -- \pcc node [\PAC, midway, right] {\scriptsize$(x/z')$};
\draw \pg -- \pbb node [\PAC, midway, above=1.mm] {\scriptsize$\,\,(x/y')$};
\draw \pg -- \pb node [\PAC, midway, left] {\scriptsize$(y/y')$};

\draw \pb -- \pbb node [\PAC, pos=.4, right] {\scriptsize$\,(y/x)$};
\draw \pdd -- \pcc node [\PAC, pos=0.4, above=1mm] {\scriptsize$\,(y/x')$};
\draw \pa -- \pbb node [\PAC, midway, above] {\scriptsize$\,(z/y)$};
\draw \pa -- \pcc node [\PAC, midway, right] {\scriptsize$\,(y/y')$};
\draw \pa -- \pdd node [\PAC, midway, right] {\scriptsize$\,(y'/x')$};
\draw \pdd -- \pd node [\PAC, midway, above] {\scriptsize$\,\,\,(z'/y')$};
\draw \pe -- \pdd node [\PAC, midway, right] {\scriptsize$\,(y/z')$};
\draw \pf -- \pa node [\PAC, pos=.4, left] {\scriptsize$(z/x')$};
\draw \pbb -- \pcc node [\PAC, pos=.5, left=-.5mm] 
{\scriptsize$(z/y')$};
\draw \pf -- \pd node [\PAC, midway, right] {\scriptsize$\,(z/z')$};
\draw \pe -- \pcc node [\PAC, midway, below=1mm] {\scriptsize$\,(z'/x')$};

\begin{scope}[very thick, every node/.style={sloped,allow upside down}]
\draw [ \PAC, ultra thick] \pg --  node (xx) {\midarrow} \pcc;

\draw [ \PAC, ultra thick] \pe -- node {\midarrow} \pcc; 
\draw [ \PAC, ultra thick] \pcc -- node {\midarrow} \pc; 
\draw [ \PAC, ultra thick] \pa -- node {\midarrow} \pcc;
\draw [ \GR, ultra thick] \pg -- node {\midarrow} \pc;
\draw [ \PAC, ultra thick] \pg -- node {\midarrow} \pcc;
\draw [ \PAC, ultra thick] \pe -- node {\midarrow} \pcc;
\draw [ \PAC, ultra thick] \pb -- node {\midarrow} \pbb; 
\draw [ \PAC, ultra thick] \pbb -- node {\midarrow} \pa;
\draw [ \PAC, ultra thick] \pbb -- node {\midarrow} \pg;
\draw [ \GR, ultra thick] \pb -- node {\midarrow} \pg;
\draw [ \GR , ultra thick] \pf -- node {\midarrow} \pa;
\draw [ \GR, ultra thick] \pbb -- node {\midarrow} \pcc; 
\draw [ \GR, ultra thick] \pdd -- node {\midarrow} \pcc; 
\draw [ \PAC, ultra thick] \pdd -- node {\midarrow} \pa; 
\draw [ \PAC, ultra thick] \pd -- node {\midarrow} \pdd; 
\draw [ \PAC, ultra thick] \pdd -- node {\midarrow} \pe; 
\draw [ \PAC, ultra thick] \pf -- node {\midarrow} \pd; 

\draw [white, fill, ultra thick] \pa circle [radius=\rad];
\draw [white, fill, ultra thick] \pb circle [radius=\rad];
\draw [white, fill, ultra thick] \pc circle [radius=\rad];
\draw [white, fill, ultra thick] \pd circle [radius=\rad];
\draw [white, fill, ultra thick] \pe circle [radius=\rad];
\draw [white, fill, ultra thick] \pf circle [radius=\rad];
\draw [white, fill, ultra thick] \pg circle [radius=\rad];
\draw [white, fill, ultra thick] \pdd circle [radius=\rad];
\draw [white, fill, ultra thick] \pbb circle [radius=\rad];
\draw [white, fill, ultra thick] \pcc circle [radius=\rad];

\draw [\PAC, fill, ultra thick] \pa circle [radius=\rad];
\draw [\PAC, ultra thick] \pb circle [radius=\rad];
\draw [\PAC, ultra thick] \pc circle [radius=\rad];
\draw [\PAC, ultra thick] \pd circle [radius=\rad];
\draw [\PAC, ultra thick] \pe circle [radius=\rad];
\draw [\PAC, ultra thick] \pf circle [radius=\rad];
\draw [\PAC, ultra thick] \pg circle [radius=\rad];

\draw [\PAC,  fill, ultra thick] \pdd circle [radius=\rad];
\draw [\PAC,  fill, ultra thick] \pbb circle [radius=\rad];
\draw [\PAC,  fill, ultra thick] \pcc circle [radius=\rad];

\node [\PAC,below right] at \pdd {$n_3$};
\node [\PAC,right=2mm] at \pcc {$n_2$};
\node [\PAC,below right] at \pbb {$n_1$};
\node [\PAC,right=1mm] at \pa {$a$};
\node [\PAC,below=1mm] at \pb {$b$};
\node [\PAC,above=1mm] at \pc {$c$};
\node [\PAC,right=1mm] at \pd {$d$};
\node [\PAC,above=1mm] at \pe {$e$};
\node [\PAC,below=1mm] at \pf {$f$};
\node [\PAC,left=1mm] at \pg {$g$};

\end{scope}
\end{tikzpicture}
%
%
\qquad\qquad
\begin{tikzpicture}[scale=\scal]
\draw [fill, opacity=.1, blue]\pg -- \pcc --\pa --\pb--\pg;
\draw \pg -- \pcc node [\PAC, midway, below=1.mm] {\scriptsize$(z/x)$};
\draw \pg -- \pc node [\PAC, \PAC, midway, left] {\scriptsize$(z/z')$};
\draw \pc -- \pcc node [\PAC, midway, right] {\scriptsize$(x/z')$};
\draw \pg -- \pbb node [\PAC, midway, above=1.mm] {\scriptsize$\,\,(x/y')$};
\draw \pg -- \pb node [\PAC, midway, left] {\scriptsize$(y/y')$};

\draw \pb -- \pbb node [\PAC, pos=.4, right] {\scriptsize$\,(y/x)$};
\draw \pddd -- \pcc node [\PAC, midway, below] {\scriptsize$\,(z'/y')$};
\draw \pa -- \pbb node [\PAC, midway, above] {\scriptsize$\,(z/y)$};
\draw \pd -- \pe node [\PAC, midway, right] {\scriptsize$\,(y/y')$};
\draw \pa -- \pddd node [\PAC, midway, right] {\scriptsize$\,(y/z')$};
\draw \pddd -- \pd node [\PAC, midway, above] {\scriptsize$\,\,\,(y/x')$};
\draw \pe -- \pddd node [\PAC, midway, left] {\scriptsize$\,(y/x')$};
\draw \pf -- \pa node [\PAC, pos=.4, left] {\scriptsize$(z/x')$};
\draw \pbb -- \pcc node [\PAC, pos=.5, left=-.5mm] 
{\scriptsize$(z/y')$};
\draw \pf -- \pd node [\PAC, midway, right] {\scriptsize$\,(z/z')$};
\draw \pa -- \pd node [\PAC, midway, below=1mm] {\scriptsize$\,(z'/x')$};

\begin{scope}[very thick, every node/.style={sloped,allow upside down}]
\draw [ \PAC, ultra thick] \pg -- node {\midarrow} \pcc; 
\draw [ \PAC, ultra thick] \pcc -- node {\midarrow} \pc; 
%
\draw [ \GR, ultra thick] \pg -- node {\midarrow} \pc;
\draw [ \PAC, ultra thick] \pg -- node {\midarrow} \pcc;
\draw [ \PAC, ultra thick] \pb -- node {\midarrow} \pbb; 
\draw [ \PAC, ultra thick] \pbb -- node {\midarrow} \pa;
\draw [ \PAC, ultra thick] \pbb -- node {\midarrow} \pg;
\draw [ \GR, ultra thick] \pb -- node {\midarrow} \pg;
\draw [ \GR , ultra thick] \pf -- node {\midarrow} \pa;
\draw [ \GR, ultra thick] \pbb -- node {\midarrow} \pcc; 
\draw [\PAC, ultra thick] \pddd -- node {\midarrow} \pcc; 
\draw [ \PAC, ultra thick] \pa -- node {\midarrow} \pddd; 
\draw [ \GR, ultra thick] \pd -- node {\midarrow} \pddd; 
\draw [ \PAC, ultra thick] \pe -- node {\midarrow} \pddd; 
\draw [ \PAC, ultra thick] \pf -- node {\midarrow} \pd; 
\draw [ \PAC, ultra thick] \pd -- node {\midarrow} \pe;
\draw [ \PAC, ultra thick] \pd -- node {\midarrow} \pa;

\draw [white, fill, ultra thick] \pa circle [radius=\rad];
\draw [white, fill, ultra thick] \pb circle [radius=\rad];
\draw [white, fill, ultra thick] \pc circle [radius=\rad];
\draw [white, fill, ultra thick] \pd circle [radius=\rad];
\draw [white, fill, ultra thick] \pe circle [radius=\rad];
\draw [white, fill, ultra thick] \pf circle [radius=\rad];
\draw [white, fill, ultra thick] \pg circle [radius=\rad];
\draw [white, fill, ultra thick] \pddd circle [radius=\rad];
\draw [white, fill, ultra thick] \pbb circle [radius=\rad];
\draw [white, fill, ultra thick] \pcc circle [radius=\rad];

\draw [\PAC, fill, ultra thick] \pa circle [radius=\rad];
\draw [\PAC, ultra thick] \pb circle [radius=\rad];
\draw [\PAC, ultra thick] \pc circle [radius=\rad];
\draw [\PAC, ultra thick] \pd circle [radius=\rad];
\draw [\PAC, ultra thick] \pe circle [radius=\rad];
\draw [\PAC, ultra thick] \pf circle [radius=\rad];
\draw [\PAC, ultra thick] \pg circle [radius=\rad];

\draw [\PAC,  fill, ultra thick] \pddd circle [radius=\rad];
\draw [\PAC,  fill, ultra thick] \pbb circle [radius=\rad];
\draw [\PAC,  fill, ultra thick] \pcc circle [radius=\rad];

\node [\PAC, right=1.mm] at \pddd {$n_3$};
\node [\PAC,above right] at \pcc {$n_2$};
\node [\PAC,below right] at \pbb {$n_1$};
\node [\PAC,right=1mm] at \pa {$a$};
\node [\PAC,below=1mm] at \pb {$b$};
\node [\PAC,above=1mm] at \pc {$c$};
\node [\PAC,right=1mm] at \pd {$d$};
\node [\PAC,above=1mm] at \pe {$e$};
\node [\PAC,below=1mm] at \pf {$f$};
\node [\PAC,left=1mm] at \pg {$g$};

\end{scope}
\end{tikzpicture}
\ee
\noindent
The shading on this figure now 
indicates the area of the next transformation, which
is the reverse star map centered at $n_3$. The result is 
shown on the right side of \eqref{ybepic2}. Next, make the 
reverse star map centered at $n_1$ followed by the direct star map 
centred at $n_2$. The results are shown on the left and right sides of 
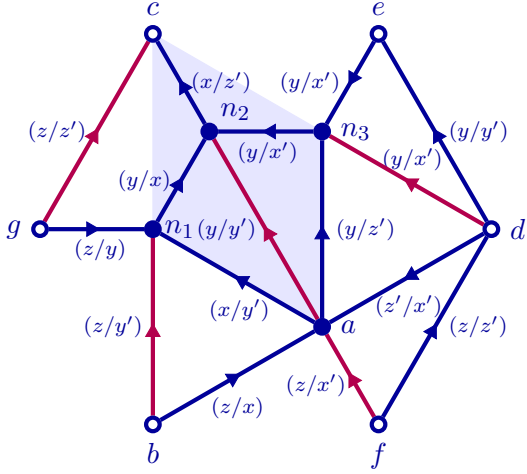
\eqref{ybepic3}, respectively. 

\def\pa{(+2.5,-4.33013)}
\def\pbb{(-2.5,-4.33013)}
\def\pbb{(-5.,0.)}
\def\pccc{(0,0)}

\be\label{ybepic3}
\begin{tikzpicture}[scale=\scal]
\draw [fill, opacity=.1, blue]\pc -- \pddd --\pa --\pbb--\pc;
\draw \pb -- \pa node [\PAC, midway, below=1.5mm] {\scriptsize$(z/x)$};
\draw \pg -- \pc node [\PAC, \PAC, midway, left] {\scriptsize$(z/z')$};
\draw \pc -- \pcc node [\PAC, midway, right] {\scriptsize$(x/z')$};
\draw \pg -- \pbb node [\PAC, midway, below] {\scriptsize$\,\,(z/y)$};
\draw \pa -- \pcc node [\PAC, midway, left=-.1mm] {\scriptsize$(y/y')$};

\draw \pb -- \pbb node [\PAC, midway, left] {\scriptsize$\,(z/y')$};
\draw \pddd -- \pcc node [\PAC, midway, below] {\scriptsize$\,(y/x')$};
\draw \pa -- \pbb node [\PAC, midway, below=1.5mm] {\scriptsize$\,(x/y')$};
\draw \pd -- \pe node [\PAC, midway, right] {\scriptsize$\,(y/y')$};
\draw \pa -- \pddd node [\PAC, midway, right] {\scriptsize$\,(y/z')$};
\draw \pddd -- \pd node [\PAC, midway, above] {\scriptsize$\,\,\,(y/x')$};
\draw \pe -- \pddd node [\PAC, midway, left] {\scriptsize$\,(y/x')$};
\draw \pf -- \pa node [\PAC, pos=.4, left] {\scriptsize$(z/x')$};
\draw \pbb -- \pcc node [\PAC, pos=.5, left] 
{\scriptsize$(y/x)$};
\draw \pf -- \pd node [\PAC, midway, right] {\scriptsize$\,(z/z')$};
\draw \pa -- \pd node [\PAC, midway, below=1mm] {\scriptsize$\,(z'/x')$};

\begin{scope}[very thick, every node/.style={sloped,allow upside down}]
\draw [ \PAC, ultra thick] \pcc -- node {\midarrow} \pc; 
%
\draw [ \GR, ultra thick] \pg -- node {\midarrow} \pc;
\draw [\GR, ultra thick] \pb -- node {\midarrow} \pbb; 
\draw [ \PAC, ultra thick] \pa -- node {\midarrow} \pbb;
\draw [ \PAC, ultra thick] \pg -- node {\midarrow} \pbb;
\draw [ \GR , ultra thick] \pf -- node {\midarrow} \pa;
\draw [\PAC, ultra thick] \pbb -- node {\midarrow} \pcc; 
\draw [\PAC, ultra thick] \pddd -- node {\midarrow} \pcc; 
\draw [\GR, ultra thick] \pa -- node {\midarrow} \pcc; 
\draw [\PAC, ultra thick] \pb -- node {\midarrow} \pa; 
\draw [ \PAC, ultra thick] \pa -- node {\midarrow} \pddd; 
\draw [ \GR, ultra thick] \pd -- node {\midarrow} \pddd; 
\draw [ \PAC, ultra thick] \pe -- node {\midarrow} \pddd; 
\draw [ \PAC, ultra thick] \pf -- node {\midarrow} \pd; 
\draw [ \PAC, ultra thick] \pd -- node {\midarrow} \pe;
\draw [ \PAC, ultra thick] \pd -- node {\midarrow} \pa;

\draw [white, fill, ultra thick] \pa circle [radius=\rad];
\draw [white, fill, ultra thick] \pb circle [radius=\rad];
\draw [white, fill, ultra thick] \pc circle [radius=\rad];
\draw [white, fill, ultra thick] \pd circle [radius=\rad];
\draw [white, fill, ultra thick] \pe circle [radius=\rad];
\draw [white, fill, ultra thick] \pf circle [radius=\rad];
\draw [white, fill, ultra thick] \pg circle [radius=\rad];
\draw [white, fill, ultra thick] \pddd circle [radius=\rad];
\draw [white, fill, ultra thick] \pbb circle [radius=\rad];
\draw [white, fill, ultra thick] \pcc circle [radius=\rad];

\draw [\PAC, fill, ultra thick] \pa circle [radius=\rad];
\draw [\PAC, ultra thick] \pb circle [radius=\rad];
\draw [\PAC, ultra thick] \pc circle [radius=\rad];
\draw [\PAC, ultra thick] \pd circle [radius=\rad];
\draw [\PAC, ultra thick] \pe circle [radius=\rad];
\draw [\PAC, ultra thick] \pf circle [radius=\rad];
\draw [\PAC, ultra thick] \pg circle [radius=\rad];

\draw [\PAC,  fill, ultra thick] \pddd circle [radius=\rad];
\draw [\PAC,  fill, ultra thick] \pbb circle [radius=\rad];
\draw [\PAC,  fill, ultra thick] \pcc circle [radius=\rad];

\node [\PAC, right=1.mm] at \pddd {$n_3$};
\node [\PAC,above right] at \pcc {$n_2$};
\node [\PAC,right=.1mm] at \pbb {$n_1$};
\node [\PAC,right=1mm] at \pa {$a$};
\node [\PAC,below=1mm] at \pb {$b$};
\node [\PAC,above=1mm] at \pc {$c$};
\node [\PAC,right=1mm] at \pd {$d$};
\node [\PAC,above=1mm] at \pe {$e$};
\node [\PAC,below=1mm] at \pf {$f$};
\node [\PAC,left=1mm] at \pg {$g$};
\end{scope}
\end{tikzpicture}
%
\qquad\qquad
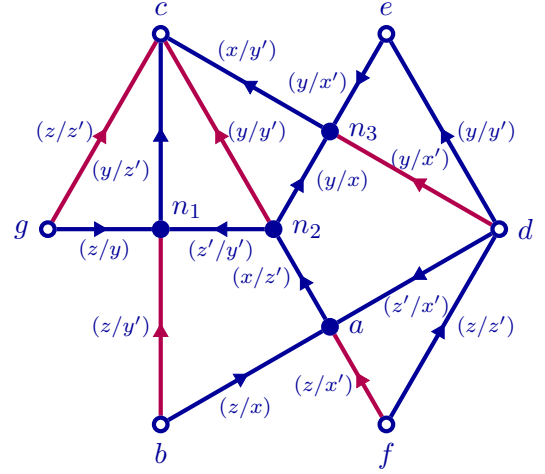
\begin{tikzpicture}[scale=\scal]
\draw \pb -- \pa node [\PAC, midway, below=1.5mm] {\scriptsize$(z/x)$};
\draw \pg -- \pc node [\PAC, \PAC, midway, left] {\scriptsize$(z/z')$};
\draw \pc -- \pccc node [\PAC, midway, right] {\scriptsize$(y/y')$};
\draw \pccc -- \pbb node [\PAC, midway, below] {\scriptsize$\,\,(z'/y')$};
\draw \pa -- \pccc node [\PAC, midway, left] {\scriptsize$(x/z')$};

\draw \pb -- \pbb node [\PAC, midway, left] {\scriptsize$\,(z/y')$};
\draw \pddd -- \pccc node [\PAC, midway, right] {\scriptsize$\,(y/x)$};
\draw \pc -- \pddd node [\PAC, midway, above=1.5mm] {\scriptsize$\,(x/y')$};
\draw \pd -- \pe node [\PAC, midway, right] {\scriptsize$\,(y/y')$};
\draw \pc -- \pbb node [\PAC, pos=.7, left] {\scriptsize$\,(y/z')$};
\draw \pddd -- \pd node [\PAC, midway, above] {\scriptsize$\,\,\,(y/x')$};
\draw \pe -- \pddd node [\PAC, midway, left] {\scriptsize$\,(y/x')$};
\draw \pf -- \pa node [\PAC, pos=.4, left] {\scriptsize$(z/x')$};
\draw \pbb -- \pg node [\PAC, pos=.5, below] 
{\scriptsize$(z/y)$};
\draw \pf -- \pd node [\PAC, midway, right] {\scriptsize$\,(z/z')$};
\draw \pa -- \pd node [\PAC, midway, below=1mm] {\scriptsize$\,(z'/x')$};

\begin{scope}[very thick, every node/.style={sloped,allow upside down}]
\draw [ \GR, ultra thick] \pccc -- node {\midarrow} \pc; 
\draw [ \PAC, ultra thick] \pa -- node {\midarrow} \pccc;
\draw [ \GR, ultra thick] \pg -- node {\midarrow} \pc;
\draw [\GR, ultra thick] \pb -- node {\midarrow} \pbb; 
\draw [ \PAC, ultra thick] \pbb -- node {\midarrow} \pc;

\draw [ \PAC, ultra thick] \pg -- node {\midarrow} \pbb;
\draw [ \GR , ultra thick] \pf -- node {\midarrow} \pa;
\draw [\PAC, ultra thick] \pccc -- node {\midarrow} \pbb; 
\draw [\PAC, ultra thick] \pccc -- node {\midarrow} \pddd; 
\draw [\PAC, ultra thick] \pb -- node {\midarrow} \pa; 
\draw [ \PAC, ultra thick] \pddd -- node {\midarrow} \pc; 
\draw [ \GR, ultra thick] \pd -- node {\midarrow} \pddd; 
\draw [ \PAC, ultra thick] \pe -- node {\midarrow} \pddd; 
\draw [ \PAC, ultra thick] \pf -- node {\midarrow} \pd; 
\draw [ \PAC, ultra thick] \pd -- node {\midarrow} \pe;
\draw [ \PAC, ultra thick] \pd -- node {\midarrow} \pa;

\draw [white, fill, ultra thick] \pa circle [radius=\rad];
\draw [white, fill, ultra thick] \pb circle [radius=\rad];
\draw [white, fill, ultra thick] \pc circle [radius=\rad];
\draw [white, fill, ultra thick] \pd circle [radius=\rad];
\draw [white, fill, ultra thick] \pe circle [radius=\rad];
\draw [white, fill, ultra thick] \pf circle [radius=\rad];
\draw [white, fill, ultra thick] \pg circle [radius=\rad];
\draw [white, fill, ultra thick] \pddd circle [radius=\rad];
\draw [white, fill, ultra thick] \pbb circle [radius=\rad];
\draw [white, fill, ultra thick] \pccc circle [radius=\rad];

\draw [\PAC, fill,ultra thick] \pa circle [radius=\rad];
\draw [\PAC, ultra thick] \pb circle [radius=\rad];
\draw [\PAC, ultra thick] \pc circle [radius=\rad];
\draw [\PAC, ultra thick] \pd circle [radius=\rad];
\draw [\PAC, ultra thick] \pe circle [radius=\rad];
\draw [\PAC, ultra thick] \pf circle [radius=\rad];
\draw [\PAC, ultra thick] \pg circle [radius=\rad];

\draw [\PAC,  fill, ultra thick] \pddd circle [radius=\rad];
\draw [\PAC,  fill, ultra thick] \pbb circle [radius=\rad];
\draw [\PAC,  fill, ultra thick] \pccc circle [radius=\rad];

\node [\PAC, right=1.mm] at \pddd {$n_3$};
\node [\PAC,right=1mm] at \pccc {$n_2$};
\node [\PAC,above right] at \pbb {$n_1$};
\node [\PAC,right=1mm] at \pa {$a$};
\node [\PAC,below=1mm] at \pb {$b$};
\node [\PAC,above=1mm] at \pc {$c$};
\node [\PAC,right=1mm] at \pd {$d$};
\node [\PAC,above=1mm] at \pe {$e$};
\node [\PAC,below=1mm] at \pf {$f$};
\node [\PAC,left=1mm] at \pg {$g$};

\end{scope}
\end{tikzpicture}
\ee
Finally, redenoting the summation indices
\be
n_3\to n_1,\qquad a\to n_2,\qquad n_1\to n_3,\qquad  n_2\to h\,,
\ee
one converts the LHS of \eqref{ybe2} to its RHS. This proves the
Yang-Baxter relation \eqref{ybe} and commutativity of transfer
matrices \eqref{tcomm} in the main text.
\sapp{Proof of the vertex-type Yang-Baxter equation\label{box-YBE}}
Now consider the Yang-Baxter equation \eqref{ybe4} 
for the $R$-matrix \eqref{def3}. For convenience let us reproduce it here,
\begin{equation}\label{ybe4a}
\sum_{a',\,b',\,c'}\Rbb_{a\,,\,b}^{a',\,b'}(\xb,\yb) \,
\Rbb_{a'\;,\,c}^{a'',\,c'}(\xb,\zb)\,
\Rbb_{b'\;,\,c'}^{b'',\,c''}(\yb,\zb)
\;=\;
\sum_{a',\,b',\,c'}
\Rbb_{b\,,\,c}^{b',\,c'}(\yb,\zb)\,
\Rbb_{a\,,\,c'}^{a',\,c''}(\xb,\zb)\,
\Rbb_{a'\;,\,b'}^{a'',\,b''}(\xb,\yb)\,.
\end{equation}
With the graphical notations \eqref{gnote2} the LHS of this equation can be represented as
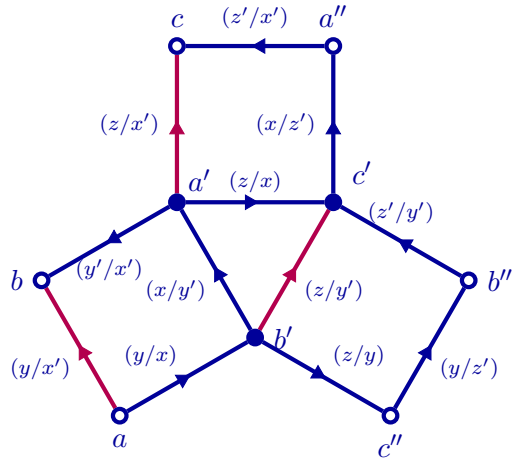
\def\pa{(0.,0.)}
\def\pb{(-6,-7.4641)}
\def\pc{(-3.4641,+8.9282)}
\def\pd{(+9.4641,-1.4641)}
\def\pe{(+3.4641,+8.9282)}
\def\pf{(+6,-7.4641)}
\def\pg{(-9.4641,-1.4641)}

\def\px{(-3.4641,2)}
\def\py{(0,-4)}
\def\pz{(+3.4641,2)}
\def\rad{0.3}
\def\scal{0.3}
\be\label{ybepic4}
\begin{tikzpicture}[scale=\scal,baseline=(current  bounding  box.center)]

\draw \pb -- \pg node [\PAC, midway, below left] {\scriptsize $(y/x')$};
\draw \pb -- \py node [\PAC, midway, above left] {\scriptsize $(y/x)$};
\draw \py -- \px node [\PAC, midway, below left] {\scriptsize $(x/y')$};
\draw \px -- \pg node [\PAC, midway, below=1.mm] {\scriptsize $(y'/x')$};

\draw \px -- \pc node [\PAC, midway, left=1.mm] {\scriptsize $(z/x')$};
\draw \px -- \pz node [\PAC, midway, above] {\scriptsize $(z/x)$};
\draw \pz -- \pe node [\PAC, midway, left=1.mm] {\scriptsize $(x/z')$};
\draw \pe -- \pc node [\PAC, midway, above=1.mm] {\scriptsize $(z'/x')$};

\draw \py -- \pz node [\PAC, midway, below right] {\scriptsize $(z/y')$};
\draw \py -- \pf node [\PAC, midway, above right] {\scriptsize $(z/y)$};
\draw \pf -- \pd node [\PAC, midway, below right] {\scriptsize $(y/z')$};
\draw \pd -- \pz node [\PAC, midway, above=1.mm] {\scriptsize $(z'/y')$};

\begin{scope}[very thick, every node/.style={sloped,allow upside down}]
\draw [ \PAC, ultra thick] \px -- node {\midarrow} \pg; 
\draw [ \PAC, ultra thick] \py -- node {\midarrow} \px; 
\draw [ \PAC, ultra thick] \pb -- node {\midarrow} \py; 
\draw [ \GR, ultra thick] \pb -- node {\midarrow} \pg;
\draw [ \PAC, ultra thick] \px -- node {\midarrow} \pz; 
\draw [ \PAC, ultra thick] \pz -- node {\midarrow} \pe; 
\draw [ \PAC, ultra thick] \pe -- node {\midarrow} \pc; 
\draw [ \GR, ultra thick] \px -- node {\midarrow} \pc;
\draw [ \PAC, ultra thick] \py -- node {\midarrow} \pf; 
\draw [ \PAC, ultra thick] \pf -- node {\midarrow} \pd; 
\draw [ \PAC, ultra thick] \pd -- node {\midarrow} \pz; 
\draw [ \GR, ultra thick] \py -- node {\midarrow} \pz;

\draw [white, fill, ultra thick] \pb circle [radius=\rad];
\draw [white, fill, ultra thick] \pc circle [radius=\rad];
\draw [white, fill, ultra thick] \pd circle [radius=\rad];
\draw [white, fill, ultra thick] \pe circle [radius=\rad];
\draw [white, fill, ultra thick] \pf circle [radius=\rad];
\draw [white, fill, ultra thick] \pg circle [radius=\rad];
\draw [white, fill, ultra thick] \px circle [radius=\rad];
\draw [white, fill, ultra thick] \py circle [radius=\rad];
\draw [white, fill, ultra thick] \pz circle [radius=\rad];

\draw [\PAC, ultra thick] \pb circle [radius=\rad];
\draw [\PAC, ultra thick] \pc circle [radius=\rad];
\draw [\PAC, ultra thick] \pd circle [radius=\rad];
\draw [\PAC, ultra thick] \pe circle [radius=\rad];
\draw [\PAC, ultra thick] \pf circle [radius=\rad];
\draw [\PAC, ultra thick] \pg circle [radius=\rad];

\draw [\PAC,  fill, ultra thick] \px circle [radius=\rad];
\draw [\PAC,  fill, ultra thick] \py circle [radius=\rad];
\draw [\PAC,  fill, ultra thick] \pz circle [radius=\rad];
\node [\PAC,above right] at \px {$a'$};
\node [\PAC,right=1mm] at \py {$b'$};
\node [\PAC,above right=1mm] at \pz {$c'$};

\node [\PAC,below=1mm] at \pb {$a$};
\node [\PAC,above=1mm] at \pc {$c$};
\node [\PAC,right=1mm] at \pd {$b''$};
\node [\PAC,above=1mm] at \pe {$a''$};
\node [\PAC,below=1mm] at \pf {$c''$};
\node [\PAC,left=1mm] at \pg {$b$};

\end{scope}
\end{tikzpicture}
\ee  
\noindent
The proof of this equation consists of the repeated application of
the star-star relation \eqref{ssrel2}, represented graphically in
Fig.~\ref{fig-star2}. 
The working is very similar to the proof of
\eqref{ybe2} in the previous subsection. In particular we will use the
notions of the ``direct'' and ``reverse'' star-star maps introduced in
the first paragraph of Sect.~\ref{app-irf-ybe}.

Start from the left hand side of (\ref{ybe4a}) and then 
\begin{enumerate}
\item make the reverse star map centred at $a'$,
\item make the reverse star map centred at $b'$,
\item make the reverse star map centred at $c'$,
\item make the direct star map centred at $a'$ again,
\end{enumerate}
and then making the change
\begin{equation}
a'\to a'\;,\quad b'\to c'\;,\quad c'\to b'\;,
\end{equation}
one converts the left hand side of (\ref{ybe4a}) to its right hand side.

\app{Star-triangle relation as a primary integrability condition\label{appB}}
\addcontentsline{toc}{section}{B. Star-triangle relation as a primary
  integrability condition} 
\sapp{Star-triangle relation.}
Here we prove the star-triangle relation \eqref{str} by reducing it  
to the Pfaff-Saalshch\"utz-Jackson summation formula
(see eq.(3.5.1) in \cite{Gasper:1995}),
\be\label{pfaff}
\phantom{|}_3\phi_2\left(\begin{array}{ccc|} \ds q^{-2n}, & \ds \as, &
  \ds \bs\\ \ds & \ds \cs, & \ds \ q^{2-2n}\,\as\,\bs /{\cs}\end{array} \,\,
q^2, q^2\right) 
\;=\; 
\frac{\ds \left({\cs}/{\as},\,{\cs}/{\bs}\,;\,q^2\right)_n}
{\ds \left(\,\cs,\ {\cs}/{\as\,\bs}\,;\,q^2\right)_n}
\ee 
where the notations are defined in \eqref{phi43} and \eqref{qprod}.
Here $n$ is a non-negative integer $n\ge 0$, 
and $q$, $\as$, $\bs$ and $\cs$ stand for
arbitrary complex parameters. 

For the readers' convenience we reproduce the relations \eqref{str} here
\begin{subequations}
\label{stra}
\begin{equation}
\sum_d \,V_{y/z}(d-a)\, W_{z,x}(b-d)\, V_{x/y}(c-d)  \;=\;
W_{z,y}(b-c) \,V_{x/z}(c-a)  \,W_{y,x}(b-a)\,,\label{str1a}
\end{equation}
and
\begin{equation}
\sum_d V_{y/z}(a-d)\,W_{z,x}(d-b)\,V_{x/y}(d-c)   \;=\;
W_{z,y}(c-b)\, V_{x/z}(a-c)\, W_{y,x}(a-b)\,.\label{str2a}
\end{equation}
\end{subequations}
The above two relations are equivalent (see 
\eqref{str-tr}), so it is enough to
consider one of them. For definiteness 
we choose the second one \eqref{str2a}. Due to \eqref{vanish} both
sides of this relation vanish identically unless $a\ge c$. So in the
following we assume $a\ge c$. Next, we change the summation
variable there from $d$ to $n=a-d$, which takes values
in the interval $0\le n\le a-c$ (since otherwise the summand
in \eqref{str2a} vanishes). Using the above definitions one can 
transform the LHS of \eqref{str2a} as follows
\begin{subequations}
\bea
\mbox{LHS\ of\ \eqref{str2a}}=&&\nonumber\\[.4cm]
&&\hskip -3cm
=V_{x/y}(a-c)\,W_{z,x}(a-b)\phantom{|}_3\varphi_2
\left(\begin{array}{ccc|}q^{-2(a-c)},&(y/z)^2,&q^{2(1-a+b)} x^{-2}\\[.2cm]
&q^{2(1-a+c)}(y/x)^2,&q^{2(1-a+b)} z^{-2}
\end{array}\,q^2,q^2\right)\label{pf1}\\[.3cm]
&&\hskip -3cm
=\ds V_{x/y}(a-c)\,W_{z,x}(a-b)\,
 \frac{\left(q^{2(1-a+c)}\,(z/x)^2,\,q^{2(c-b)}\,y^2\,;q^2\right)_{a-c}}
{\left(q^{2(1-a+c)}\,(y/x)^2,\,q^{2(c-b)}\,x^2\,;q^2\right)_{a-c}}\label{pf2}
\\[.4cm]
&&\hskip -3cm
=W_{z,y}(c-b)\, V_{x/z}(a-c)\, W_{y,x}(a-b)\,,\label{pf3}
\eea
\end{subequations}
where in the first line we used \eqref{Vdef}, \eqref{Wxy} and
\eqref{cor} as well as the formula 
\begin{equation}\label{Wcor}
W_{x,y}(a-n) \;=\; W_{x,y}(a) \left(\frac{y}{x}\right)^n
\frac{(q^{2-2a}\,y^{-2};q^2)_n}
{(q^{2-2a}\,x^{-2};q^2)_n}\;.
\end{equation}
In the next line \eqref{pf2} we simply 
substituted the summation formula \eqref{pfaff}. Finally 
Eqs.\eqref{cor} and \eqref{Wcor} were  used again to rewrite 
the result in the form 
\eqref{pf3}, which is identical to the RHS of \eqref{str2a}, thus
completing the proof of this relation.  

\sapp{Star-star relation.} 
Next, we will show how to derive the star-star relation \eqref{ssrel} from
the star-triangle relation. Setting $y=q$ in  \eqref{stra}
one obtains
\begin{subequations}
\label{tr}
\bea
\frac{V_{z}(b-c)\, V_{x/z}(c-a)}{V_x(b-a)}&=&
\sum_k V_{q/z}(k-a)\, W_{z,x}(b-k)\, V_{x/q} (c-k)\,,\label{tr1}\\[.3cm]
&=& 
\sum_k W_{x/z,x}(k-a)\, V_{qz/x}(b-k)\,  V_{x/q} (k-c)\,,\label{tr2}
\eea 
\end{subequations}
where the first equality follows from \eqref{str1a} while the second
one follows from \eqref{str2a} after an additional substitution
$a\leftrightarrow b$ and $z\to x/z$.

The LHS of \eqref{ssrel} can be transformed as 
\begin{equation}
\begin{array}{l}
\ds \mbox{LHS\ of\ \eqref{ssrel}}=\ds \sum_n 
\left(\frac{\ds V_{y'/x'}({n-b})\, V_{y/y'}(b-d)}
{\ds V_{y/x'}(n-d)}\right)\;
\frac{\ds V_{x/y'}(a-n)\, V_{y/x}(n-c)}{\ds V_{y/y'}(a-c)}\\
[6mm]
\ds =\;\sum_{k,n} V_{qx'/y'}(k-d)\,V_{y/qx'}(b-k)\,
W_{y'/x',\,y/x'}(n-k)\,\,
\frac{\ds V_{x/y'}(a-n)\, V_{y/x}(n-c)}{\ds V_{y/y'}(a-c)}\\
[8mm]
\ds =\;\sum_k V_{qx'/y'}(k-d)\,V_{y/qx'}(b-k)\, 
W_{x/x',y/x'}(a-k) \, W_{y'/x',x/x'}(c-k)\;.
\end{array}\label{SS1}
\end{equation}
where at first \eqref{tr1} is applied to the product of three $V$'s
enclosed by the parenthesis and
then the summation over $n$ is performed with help of \eqref{str2a}. 
Consider now the RHS of \eqref{ssrel}, given by \eqref{W-def2}, 
\begin{equation}
\begin{array}{l}
\ds \mbox{RHS\ of\ \eqref{ssrel}}
=\ds \sum_n \left(\frac{\ds V_{x/x'}(a-b) \,V_{y/x}(b-n)}{\ds
  V_{y/x'}(a-n)}\right)\;
\left(\frac{\ds V_{y'/x'}(c-n)\, V_{x/y'}(n-d)}{\ds V_{x/x'}(c-d)}\right)\\
[6mm]
\ds =\;\sum_{n,k,k'} V_{q x'/x}(k-n)\, V_{y/q x'}(b-k)\,
W_{x/x',y/x'}(a-k) 
\,V_{qx'/y'}(k'-d) \, V_{x/q x'}(n-k') \, W_{y'/x',x/x'}(c-k')\\
[6mm]
\ds =\; \sum_k \,
V_{y/q x'}(b-k)\,
W_{x/x',y/x'}(a-k) 
\,V_{qx'/y'}(k-d) \,  W_{y'/x',x/x'}(c-k)\,,
\end{array}\label{SS4}
\end{equation}
where at first \eqref{tr1} is applied to both products of three $V$'s
grouped in the parentheses and then the summation over $n$ is
removed by using the inversion relation \eqref{inv}. The resulting
expression coincides with \eqref{SS1}, thereby proving \eqref{ssrel}.
Thus the star-triangle relation \eqref{str} 
implies the star-star relation \eqref{ssrel} and, consequently, the
Yang-Baxter equation \eqref{ybe} and the commutativity of the transfer
matrices \eqref{tcomm}.
Therefore, the star-triangle relation
\eqref{str} does, indeed, plays the role of a primary integrability condition
in the model.

\sapp{Alternative graphical notations.}
To present the star-triangle relation graphically we need to introduce
a new variant of graphical notations for the edge Bolzmann
weights. They involve the edges of the type $(i)$ and type $(iii)$
\be\label{type3}
\begin{tikzpicture}[scale=1.,baseline]
\node [left] at (-0.5,1) {$({i}):$};
\draw [-open triangle 45, black, thin, dashed] (0,0) -- (2,2);
\draw [-open triangle 45, black, thin, dashed] (2,0) -- (0,2);
\draw [\STE, ultra thick] (1,0) circle [radius=0.09];
\draw [\STE, ultra thick] (1,2) circle [radius=0.09];
\draw [-latex, \PAC, ultra thick] (1,0.12) -- (1,1);
\draw [\PAC, ultra thick] (1,1) -- (1,1.88);
\node [above] at (2,2) {$x$};
\node [above] at (0,2) {$y$};
\node [above] at (1,2.1) {$a$};
\node [below] at (1,-0.1) {$b$};
\node [right] at (2.1,1) {$
\ds =\;V_{x/y}(a-b)\;,\qquad
$};
\end{tikzpicture}
\newcommand{\midarr}{\tikz \draw[-{Latex[length=3.5mm]}](0,0) -- +(.1,0);}
\begin{tikzpicture}[scale=1.,baseline]
\node [left] at (-1,1) {$({iii}):$};
\draw [-open triangle 45, black, thin, dashed] (0,0) -- (2,2);
\draw [-open triangle 45, black, thin, dashed] (2,0) -- (0,2);
\begin{scope}[very thick, every node/.style={sloped,allow upside down}]
\draw[double=none, double distance=0.5pt, line join=round, very thick, \STE] (2.,1)--node [pos=.4] {\midarr} (-0,1);
\end{scope}
\draw [white, fill, ultra thick] (-.0,1.) circle [radius=0.09];
\draw [white, fill, ultra thick] (2.0,1.) circle [radius=0.09];
\draw [\STE, ultra thick] (-.,1.) circle [radius=0.09];
\draw [\STE, ultra thick] (2.,1.) circle [radius=0.09];
\node [above] at (2.,2) {$y$};
\node [above] at (-0.,2) {$x$};
\node [left=1mm] at (-0.,1) {$a$};
\node [right=1mm] at (2.,1) {$b$};
\node [right] at (2.1,1) {$
\ds \qquad=\;W_{x,y}(a-b)\;.
$};
\end{tikzpicture}
\ee
which are distiguished by relative orientation of the edge and associated
spectral parameter (dashed) lines. 
The corresponding Boltzmann weights $V_{x/y}(a-b)$ 
and $W_{x,y}(a-b)$ are defined by \eqref{Vdef} and \eqref{Wxy}, respectively. 
Note that for the type $(iii)$ edges the directed rapidity lines
cross the edge from the same side, unlike the type $(i)$ edges where they
come from different sides. For better visibility the type $(iii)$
edges will be shown by double lines. 

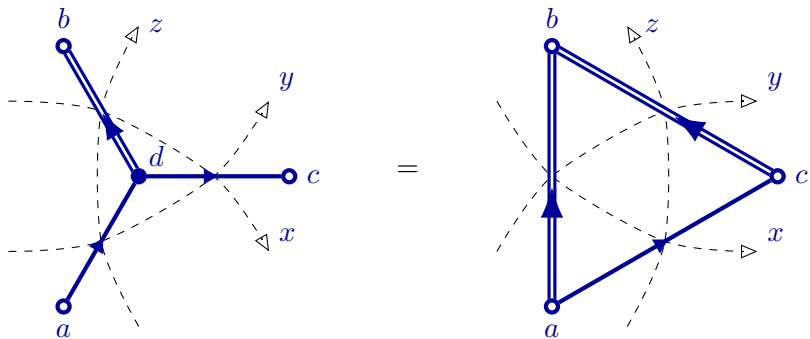
\begin{figure}[ht]
\begin{center}
\newcommand{\midarr}{\tikz \draw[-{Latex[length=3.5mm]}](0,0) -- +(.1,0);}
\newcommand{\midar}{\tikz \draw[-{open triangle 45}];}
\renewcommand{\midarrow}{\tikz \draw[-triangle 60] (0,0) -- +(.1,0);}
\def\pa{(0.,0.)}
\def\pb{(-5,-8.66025)}
\def\pc{(-5,+8.66025)}
\def\pd{(10.,0.)}
\def\pe{(+5,+8.66025)}
\def\pf{(+5,-8.66025)}
\def\pg{(-10.,0.)}
\def\pbb{(-2.5,-4.33013)}
\def\pcc{(-2.5,+4.33013)}
\def\pdd{(5.,0.)}
\def\pee{(+2.5,+4.33013)}
\def\pff{(+2.5,-4.33013)}
\def\pgg{(-5.,0.)}
\def\rad{0.4}
\def\pxf{(8.66025, -5.)}
\def\pxi{(-8.66025, 5.)}
\def\pxxf{(8.66025, 5.)}
\def\pxxi{(-8.66025,-5.)}
\def\pxxxf{(0.,10.)}
\def\pxxxi{(0.,-10.)}
\def\scal{0.2}
\begin{tikzpicture}[scale=\scal,baseline]
\begin{scope}[very thick, every node/.style={sloped,allow upside down}]
\draw [-open triangle 45,black,thin,dashed] 
plot [smooth,tension=.5] coordinates {\pxi \pcc \pdd \pxf};
\draw [-open triangle 45,black,thin,dashed] 
plot [smooth,tension=.5] coordinates {\pxxi \pbb \pdd \pxxf};
\draw [-open triangle 45,black,thin,dashed] 
plot [smooth,tension=.5] coordinates {\pxxxi \pbb \pcc \pxxxf};
\draw[double=none, double distance=1.pt, line join=round, very thick, \STE]
\pa --node [pos=.4] {\midarr} \pc;
\draw [ \PAC, ultra thick] \pa -- node {\midarrow} \pd; 
\draw [ \PAC, ultra thick] \pb -- node {\midarrow} \pa; 
\end{scope}
\draw [white, fill, ultra thick] \pa circle [radius=\rad];
\draw [white, fill, ultra thick] \pb circle [radius=\rad];
\draw [white, fill, ultra thick] \pc circle [radius=\rad];
\draw [white, fill, ultra thick] \pd circle [radius=\rad];
\draw [\PAC, fill, ultra thick] \pa circle [radius=\rad];
\draw [\PAC, ultra thick] \pb circle [radius=\rad];
\draw [\PAC, ultra thick] \pc circle [radius=\rad];
\draw [\PAC, ultra thick] \pd circle [radius=\rad];
\node [\PAC,above right] at \pa {$d$};
\node [\PAC,below=1mm] at \pb {$a$};
\node [\PAC,above=1mm] at \pc {$b$};
\node [\PAC,right=1mm] at \pd {$c$};
\node [\PAC,above right] at \pxf {$x$};
\node [\PAC,above right] at \pxxf {$y$};
\node [\PAC,right] at \pxxxf {$z$};
\end{tikzpicture}
\qquad = \qquad 
\renewcommand{\midarr}{\tikz \draw[-{Latex[length=4mm]}](0,0) -- +(.1,0);}
\begin{tikzpicture}[scale=\scal,baseline]
\begin{scope}[very thick, every node/.style={sloped,allow upside down}]
\draw [-open triangle 45,black,thin,dashed] 
plot [smooth,tension=.5] coordinates {\pxi \pgg \pff \pxf};
\draw [-open triangle 45,black,thin,dashed] 
plot [smooth,tension=.5] coordinates {\pxxi \pgg \pee \pxxf};
\draw [-open triangle 45,black,thin,dashed] 
plot [smooth,tension=.5] coordinates {\pxxxi \pff \pee \pxxxf};
\draw[double=none, double distance=1.pt, line join=round, very thick, \STE]
\pd --node [pos=.4] {\midarr} \pc;
\draw[double=none, double distance=1.pt, line join=round, very thick, \STE]
\pb --node [pos=.4] {\midarr} \pc;
\draw [ \PAC, ultra thick] \pb -- node {\midarrow} \pd; 
\end{scope}
\draw [white, fill, ultra thick] \pb circle [radius=\rad];
\draw [white, fill, ultra thick] \pc circle [radius=\rad];
\draw [white, fill, ultra thick] \pd circle [radius=\rad];
\draw [\PAC, ultra thick] \pb circle [radius=\rad];
\draw [\PAC, ultra thick] \pc circle [radius=\rad];
\draw [\PAC, ultra thick] \pd circle [radius=\rad];
\node [\PAC,below=1mm] at \pb {$a$};
\node [\PAC,above=1mm] at \pc {$b$};
\node [\PAC,right=1mm] at \pd {$c$};
\node [\PAC,above right] at \pxf {$x$};
\node [\PAC,above right] at \pxxf {$y$};
\node [\PAC,right=1mm] at \pxxxf {$z$};

\end{tikzpicture}
\end{center}
\caption{Graphical representation of the star-triangle relation
  \eqref{str1}.
\label{figstr}}
\end{figure}

As in \eqref{twotypes} the weights in \eqref{type3} 
depend on the difference of spins at the ends of the edge, but, in
general, they do not have the spectral parameter ``difference
property''. The function $W$ 
depends non-trivially on both rapidities $x$ and $y$,
not just on their ratio,
With the graphical rules \eqref{type3} the first
relation in \eqref{str} is represented graphically as in Fig.~\ref{figstr}. 
For the 
second relation one needs to reverse all arrows (both edge arrows and
the dashed line arrows).

\sapp{New solutions of the Yang-Baxter equations\label{sappb}}
The existence of the star-triangle relation \eqref{str} 
allows one to define new integrable Ising-type models in the standard
way. First, let us present them in the IRF form. To make the notations
uniform it is convenient to redenote weights \eqref{W-def1} as
\be
\mathcal{W}^{(1,1)}(a,b,c,d\,|\,\boldsymbol{x},\boldsymbol{y})\;=\;
\mathcal{W}(a,b,c,d\,|\,\boldsymbol{x},\boldsymbol{y})
\end{equation}
where $\mathcal{W}$ is given by \eqref{W-def1} and 
$\xb=(x,x')$, $\yb=(y,y')$ denote pairs of spectral
variables. 

Define new weight functions,
\begin{equation}
\mathcal{W}^{(2,1)}(a,b,c,d\,|\,\boldsymbol{x},\boldsymbol{y})\;=\;
\theta(a\ge c)\,\sum_{n}
W_{x,y}(a-n) V_{y/x'}(b-n) W_{y',x}(c-n)  V_{x'/y'}(n-d)\,,
\end{equation}
\begin{equation}
\mathcal{W}^{(1,2)}(a,b,c,d\,|\,\boldsymbol{x},\boldsymbol{y})\;=\;
\theta(a\ge b)\,\sum_{n}
W_{x,y}(a-n) W_{y,x'}(b-n) V_{y'/x}(n-c)  V_{x'/y'}(d-n)\,,
\end{equation}
and
\begin{equation}
\mathcal{W}^{(2,2)}(a,b,c,d\,|\,\boldsymbol{x},\boldsymbol{y})\;=\;\sum_{n}
W_{x,y}(a-n) V_{y/x'}(b-n) V_{y'/x}(n-c) W_{x',y'}(n-d)\;.
\end{equation}
Here $\theta(a\ge b)=1$ if $a\ge b$ 
and vanishes otherwise. 
Using the notations \eqref{type3} we can represent these weights graphically
\newcommand{\midarr}{\tikz \draw[-{Latex[length=3.5mm]}](0,0) -- +(.1,0);}
\begin{equation}\label{W1}
\begin{tikzpicture}[scale=0.8,baseline=(current  bounding  box.center)]
\draw [-open triangle 45, black, thin, dashed] (1,-0.5) -- (1,4.5);
\draw [-open triangle 45, black, thin, dashed] (3,-0.5) -- (3,4.5);
\draw [-open triangle 45, black, thin, dashed] (-0.5,1) -- (4.5,1);
\draw [-open triangle 45, black, thin, dashed] (-0.5,3) -- (4.5,3);
\begin{scope}[very thick, every node/.style={sloped,allow upside down}]
\draw[double=none, double distance=0.5pt, line join=round, very thick, \STE] (4,0)--node [pos=.4] {\midarr} (2,2);
\draw[double=none, double distance=0.5pt, line join=round, very thick, \STE] (2,2)--node [pos=.4] {\midarr} (0,4);
\end{scope}
\draw [-latex, \STE, ultra thick] (0,0) -- (1,1); 
\draw [-latex, \STE, ultra thick] (1,1) -- (3,3);
\draw [\STE, ultra thick] (3,3) -- (4,4);  
\draw [white, fill, ultra thick] (0,0) circle [radius=0.09];
\draw [\STE, ultra thick] (0,0) circle [radius=0.09];
\draw [white, fill, ultra thick] (2,2) circle [radius=0.09];
\draw [\STE, ultra thick] (2,2) circle [radius=0.09];
\draw [white, fill, ultra thick] (0,4) circle [radius=0.09];
\draw [\STE, ultra thick] (0,4) circle [radius=0.09];
\draw [white, fill, ultra thick] (4,0) circle [radius=0.09];
\draw [\STE, ultra thick] (4,0) circle [radius=0.09];
\draw [white, fill, ultra thick] (4,4) circle [radius=0.09];
\draw [\STE, ultra thick] (4,4) circle [radius=0.09];
\node [above] at (1,4.5) {$x$};
\node [above] at (3,4.5) {$x'$};
\node [right] at (4.5,3) {$y$};
\node [right] at (4.5,1) {$y'$};
\node [above] at (0,4.1) {$a$};
\node [above] at (4,4.1) {$b$};
\node [below] at (0,-0.1) {$c$};
\node [below] at (4,-0.1) {$d$};
\node [left] at (-1,2) {$\ds \mathcal{W}^{(2,2)}(a,b,c,d| x,x';y,y')\;=$};
\end{tikzpicture}
\end{equation}

\begin{equation}\label{W2}
\begin{tikzpicture}[scale=0.8,baseline=(current  bounding  box.center)]
\draw [-open triangle 45, black, thin, dashed] (1,-0.5) -- (1,4.5);
\draw [-open triangle 45, black, thin, dashed] (3,-0.5) -- (3,4.5);
\draw [-open triangle 45, black, thin, dashed] (-0.5,3) -- (4.5,3);
\draw [-open triangle 45, black, thin, dashed] (4.5,1) -- (-0.5,1);
\begin{scope}[very thick, every node/.style={sloped,allow upside down}]
\draw[double=none, double distance=0.5pt, line join=round, very thick, \STE] (2,2)--node [pos=.4] {\midarr} (0,0);
\draw[double=none, double distance=0.5pt, line join=round, very thick, \STE] (2,2)--node [pos=.4] {\midarr} (0,4);
\end{scope}
\draw [-latex, \STE, ultra thick] (4,0) -- (3,1);
\draw [\STE, ultra thick] (3,1) -- (2,2); 
\draw [-latex, \STE, ultra thick] (2,2) -- (3,3);
\draw [\STE, ultra thick] (3,3) -- (4,4);  
\draw [white, fill, ultra thick] (0,0) circle [radius=0.09];
\draw [\STE, ultra thick] (0,0) circle [radius=0.09];
\draw [white, fill, ultra thick] (2,2) circle [radius=0.09];
\draw [\STE, ultra thick] (2,2) circle [radius=0.09];
\draw [white, fill, ultra thick] (0,4) circle [radius=0.09];
\draw [\STE, ultra thick] (0,4) circle [radius=0.09];
\draw [white, fill, ultra thick] (4,0) circle [radius=0.09];
\draw [\STE, ultra thick] (4,0) circle [radius=0.09];
\draw [white, fill, ultra thick] (4,4) circle [radius=0.09];
\draw [\STE, ultra thick] (4,4) circle [radius=0.09];
\node [above] at (1,4.5) {$x$};
\node [above] at (3,4.5) {$x'$};
\node [right] at (4.5,3) {$y$};
\node [right] at (4.5,1) {$y'$};
\node [above] at (0,4.1) {$a$};
\node [above] at (4,4.1) {$b$};
\node [below] at (0,-0.1) {$c$};
\node [below] at (4,-0.1) {$d$};
\node [left] at (-1,2) {$\ds \mathcal{W}^{(2,1)}(a,b,c,d| x,x';y,y')\;=$};
\end{tikzpicture}
\end{equation}

\begin{equation}\label{W3}
\begin{tikzpicture}[scale=0.8,baseline=(current  bounding  box.center)]
\draw [-open triangle 45, black, thin, dashed] (1,-0.5) -- (1,4.5);
\draw [-open triangle 45, black, thin, dashed] (3,4.5) -- (3,-0.5);
\draw [-open triangle 45, black, thin, dashed] (-0.5,3) -- (4.5,3);
\draw [-open triangle 45, black, thin, dashed] (-0.5,1) -- (4.5,1);
\begin{scope}[very thick, every node/.style={sloped,allow upside down}]
\draw[double=none, double distance=0.5pt, line join=round, very thick, \STE] (2,2)--node [pos=.4] {\midarr} (4,4);
\draw[double=none, double distance=0.5pt, line join=round, very thick, \STE] (2,2)--node [pos=.4] {\midarr} (0,4);
\end{scope}
\draw [-latex, \STE, ultra thick] (0,0) -- (1,1);
\draw [\STE, ultra thick] (1,1) -- (2,2); 
\draw [-latex, \STE, ultra thick] (2,2) -- (3,1);
\draw [\STE, ultra thick] (3,1) -- (4,0);  
\draw [white, fill, ultra thick] (0,0) circle [radius=0.09];
\draw [\STE, ultra thick] (0,0) circle [radius=0.09];
\draw [white, fill, ultra thick] (2,2) circle [radius=0.09];
\draw [\STE, ultra thick] (2,2) circle [radius=0.09];
\draw [white, fill, ultra thick] (0,4) circle [radius=0.09];
\draw [\STE, ultra thick] (0,4) circle [radius=0.09];
\draw [white, fill, ultra thick] (4,0) circle [radius=0.09];
\draw [\STE, ultra thick] (4,0) circle [radius=0.09];
\draw [white, fill, ultra thick] (4,4) circle [radius=0.09];
\draw [\STE, ultra thick] (4,4) circle [radius=0.09];
\node [above] at (1,4.5) {$x$};
\node [above] at (3,4.5) {$x'$};
\node [right] at (4.5,3) {$y$};
\node [right] at (4.5,1) {$y'$};
\node [above] at (0,4.1) {$a$};
\node [above] at (4,4.1) {$b$};
\node [below] at (0,-0.1) {$c$};
\node [below] at (4,-0.1) {$d$};
\node [left] at (-1,2) {$\ds \mathcal{W}^{(1,2)}(a,b,c,d| x,x';y,y')\;=$};
\end{tikzpicture}
\end{equation}
There are eight different Yang-Baxter equations for $\mathcal{W}^{(i,j)}$:
\begin{equation}
\begin{array}{l}
\ds \sum_{a} 
\mathcal{W}^{(i,j)}(g,a,b,f \,|\, \boldsymbol{x},\boldsymbol{y}) \,
\mathcal{W}^{(i,k)}(c,e,g,a \,|\, \boldsymbol{x},\boldsymbol{z}) \,
\mathcal{W}^{(j,k)}(e,d,a,f \,|\, \boldsymbol{y},\boldsymbol{z}) \;=\\
[5mm]
\ds \qquad \qquad
=\; \sum_{h} 
\mathcal{W}^{(j,k)}(c,h,g,b \,|\, \boldsymbol{y},\boldsymbol{z})\,
\mathcal{W}^{(i,k)}(h,d,b,f \,|\, \boldsymbol{x},\boldsymbol{z})\,
\mathcal{W}^{(i,j)}(c,e,h,d \,|\, \boldsymbol{x},\boldsymbol{y})
\end{array}
\end{equation}
for all possible choices of the indices $i,j,k=1,2$. 

All four IRF weights $\mathcal{W}^{(i,j)}$ 
allow a corresponding vertex form $\mathcal{R}^{(i,j)}$ 
by means of \eqref{R-def} \eqref{ij-def}. From the quantum group point
of view these $R$-matrices are related to infinite-dimensional
evaluation representations of the $U_q(\widehat{sl}(2))$ algebra. The
indices $i,j=1$ in the notation $\mathcal{R}^{(i,j)}$
indicate representations with highest weights, while 
$i,j=2$ indicate representations without highest or lowest weights.
As in Sect.~\ref{red} the $R$-matrices, associated with the 
highest weight representations admit
finite-dimensional reductions. For instance, 
when $y'=q\,y$, the $R$-matrix 
$\mathcal{L}\;=\;\mathcal{R}^{(2,1)}(x,x';y,qy)$ has a
block-triangular structure in the second space with a $2$-dimensional
block given by 
\begin{equation}\label{Lop-inf}
\begin{array}{ll}
\ds \mathcal{L}_{i_1,0}^{j_1,0}\;=\;\delta_{i_1,j_2} \frac{[q^{i_1}y]}{[y]}\;,
& \ds \mathcal{L}_{i_1,0}^{j_1,1}\;=\;-\delta_{i_1,j_1+1} \frac{[q^{j_1} x']}{[y]}\;,\\
[7mm]
\ds \mathcal{L}_{i_1,1}^{j_1,0}\;=\;\delta_{i_1+1,j_1}\frac{[q^{i_1}x]}{[y]}\;, &
\ds \mathcal{L}_{i_1,1}^{j_1,1}\;=\;-\delta_{i_1,j_1} \frac{[q^{i_1-1}xx'/y]}{[y]}\;.
\end{array}
\end{equation}
It can be written in the canonical form (the same as \eqref{Lop}) 
\be
\boldsymbol{{\cal L}}=\frac{1}{[\mu\,q^{s+s'}]}\,
\begin{pmatrix}
\mu\, q^{\frac{H}{2}}-\mu^{-1} \,q^{-\frac{H}{2}}&
F\\[3mm]
E& \mu \,q^{-\frac{H}{2}}-\mu^{-1} \,q^{\frac{H}{2}}
\end{pmatrix}\,,\label{Lop1}
\ee
where $E,F,H$ stands for the generators of 
the quantum universal enveloping algebra $U_q(sl_2)$, defined in
\eqref{uqsl2},
and the matrix elements \eqref{Lop-inf} correspond to the
infinite-dimensional representation $\pi^{+}_{s,s'}$ of this algebra
\be
\pi^+_{s,s'}[H]\,|j\rangle=(2s+2s'-2j)\,|j\rangle\,,
\quad \pi^+_{s,s'}[E]\,|j\rangle=[q^{2s+1-j}]\,|j-1\rangle\,,\quad 
\pi^+_{s,s'}[F]\,|j\rangle=[q^{j+1-2s'}]\,|j+1\rangle\,,
\ee
where $j\in {\mathbb Z}$ and the parameters are defined as 
\be
y=\big(\mu q^{s+s'}\big)^{-1}\,,\qquad x=q^{-2s}\,,\qquad x'=q^{1-2s'}\,.
\ee
Note, that for the generic case this representation is irreducible and 
does not have highest or lowest weights.

\app{Mangazeev's $R$-matrix for higher spin 6-vertex model\label{appC}}
\addcontentsline{toc}{section}{C. Mangazeev's $R$-matrix for higher
  spin 6-vertex model} 
In ref.\cite{Mangazeev:2014gwa} Mangazeev obtained a remarkably simple
formula for the most general higher spin $R$-matrix related to
the 6-vertex model, 
expressing it via a ${}_4\varphi_3$ $q$-hypergeometric series. 
The result of \cite{Mangazeev:2014gwa} 
can be represented in various equivalent forms. Here we
will use its variant from \cite{Bosnjak:2016oze}, eq.(3.22) therein,
which we quote below preserving their notations 
\begin{equation}\label{R-Man}
\begin{array}{lll}
\ds \bigl[R_{I,J}(\lambda)\bigr]_{i,j}^{i',j'} &=& \ds \delta_{i+j,i'+j'} q^{i'j'-ij-iJ-Ij'}\, \frac{(q^2,q^2)_{i+j}}{(q^2;q^2)_i\,(q^2,q^2)_j}\\
[6mm]
&\ds \times &\ds  \frac{(\lambda^{-2}q^{I-J};q^2)_{j'}\, 
(\lambda^{-2}q^{J-I};q^2)_i\, (q^{-2J};q^2)_j}
{(\lambda^{-2}q^{-I-J};q^2)_{i+j}\,(q^{-2J};\,q^2)_{j'}}\\
[6mm]
&\times & \ds \sum_{n\geq 0} \frac{\ds (q^{-2i},\, q^{-2j'},\, 
\lambda^2q^{-I-J},\, \lambda^2q^{2+I+J-2i-2j};q^2)_n}{\ds (q^2,\, 
q^{-2i-2j},\, \lambda^2q^{2+I-J-2i},\, \lambda^2q^{2+J-I-2j'};q^2)_n}\, q^{2n}\,.
\end{array}
\end{equation}
Apart from the parameter $q$, this expressions contains three
independent variables $q^I$, $q^J$ and $\lambda^2$.
They only appear in certain combinations, so it is convenient 
to introduce the variables $x,x',y,y'$, such that 
\begin{equation}
\left(\frac{x}{y'}\right)^2\;=\;\lambda^2 q^{-I-J}\;,\quad 
\left(\frac{x'}{y}\right)^2\;=\;\lambda^2 q^{I+J}\;,\quad 
\left(\frac{x}{y}\right)^2\;=\;\lambda^2 q^{J-I}\;,\quad 
\left(\frac{x'}{y'}\right)^2\;=\;\lambda^2 q^{I-J}\;.
\end{equation}
Clearly, there are  
only three independent ratios of these new variables,
but the fourth equation above is a corollary of the other three.
Next, let us make a substitution of the integer indices in \eqref{R-Man},
\begin{equation}
i=a-b\;,\quad j= b-d\;,\quad i'=c-d\;,\quad j'= a-c\,,\qquad
a,b,c,d\in {\mathbb Z}\,,
\end{equation}
which automatically solves the $\delta$-function constraint. 
Further, let us combine all $q$-product factors in front of the sum in
\eqref{R-Man} into a product of the weight functions $V_x(n)$, using the
definition \eqref{Vdef}, and then write the sum in
\eqref{R-Man} as a ${}_4\varphi_3$ series using \eqref{phi43}. Then,
after all these substitutions, the resulting expression for \eqref{R-Man}
almost literally coincides with the RHS of \eqref{SL}. More precisely,
one obtains,
\be
\mbox{\Big[RHS of \eqref{R-Man}\Big]}=q^{i'j'-ij}\, 
\big(\lambda q^{(I-J)/2}\big)^{i'-i}\,\times\,
\mbox{\Big[RHS of \eqref{SL}\Big]}\,.
\ee
Taking now into account the definitions \eqref{R-def}, \eqref{pardef},
\eqref{RR-rel} and the fact that \eqref{SL} coincides with \eqref{star1} 
one concludes that the $R$-matrix \eqref{R-Man} is simply related to that
defined by \eqref{RR-rel}, \eqref{R-def2},  
\be
\bigl[R_{I,J}(\lambda)\bigr]^{i_1,i_2}_{j_1,j_2}=
q^{(j_1j_2-i_1i_2)}\,\big(q^{s_1-s_2}\,\lambda\big)^{i_1-j_1}\,
{\cal R}(\lambda\,\,|\,\,s_1,s_2)_{i_1,i_2}^{j_1,j_2}
\ee
where
\be
I=2s_1\,,\qquad J=2s_2\,,\qquad
\ee
and $\lambda$ is the same in both sides. 
So the two $R$-matrices are related by the matrix
transpostion in both spaces, followed by a multiplication by 
some gauge transformation factors. Note, that the latter 
cancel out in the Yang-Baxter equation. The reason we have dropped these
factors is that it is impossible to 
consistently attribute them to the edge Boltzmann weights
in the Ising type formulation of the model.

Let us now present the connection with Mangazeev's definitions of the
${\bf Q}$-operators. First compare two definitions, our
Eq.\eqref{Qelem3}, defining
$\overline{\mathfrak{R}}_{i_1,i_2}^{j_1,j_2}$, and Eq.(6.21) of
\cite{Mangazeev:2014gwa}, which defines 
$[A_{-}^{(I)}(\lambda)]_{\,n,\,i}^{n',i'}$. 
These quantities are related as follows,
\begin{equation}\label{1}
[A_{-}^{(I)}(\lambda)]_{\,n,\,i}^{n',i'}\;=\;(-\lambda)^{I-i'} \; ((y/x')^2;q^2)_{I} \; \overline{\mathfrak{R}}(y|x,x')_{i',n'}^{\,i,\,n} \; q^{n'i'-ni}\;\frac{c_{n'}}{c_n}\;,
\end{equation}
where 
\begin{equation}\label{22}
\left(\frac{y}{x'}\right)^2\;=\;\lambda^{-2}q^{1-I}\;,\quad
\left(\frac{y}{x}\right)^2\;=\;\lambda^{-2}q^{1+I}\;,\quad
c_n\;=\;q^{-n(1+I)/2} (q^2;q^2)_n\;.
\end{equation}
The extra power $(-\lambda)^{I-i'}$ in (\ref{1}) appears 
because in Mangazeev' formulas $Q_{-}$ is the Laurent polynomial. 

Thus, Mangazeev's $Q_{-}(\lambda)$ is related to $\overline{\mathbf{Q}}(q^{-1}y)$ by
\begin{equation}
Q_{-}(\lambda) \;=\; \lambda^{NI-M} \overline{\mathbf{Q}}(q^{-1}y)\;.
\end{equation}
for integer $I=2s_1$.

\app{{\bf Q}-matrices and functional relations\label{appD}}
\addcontentsline{toc}{section}{D. {\bf Q}-matrices and functional relations}
The elements of the ${\bf Q}$-matrices \eqref{Q-def} 
\begin{equation}
\label{Qelem}
\begin{array}{rcl}
\ds \big(\Qib(y')\big)_{i_N,\ldots,i_1}^{j_N,\ldots,j_1}&=&
\ds Z_0^{-1}\,\sum_{k_1=0}^\infty \omega_h^{k_1} \prod_{m=1}^N
{\mathfrak R}(y'|x_m^{},x'_m)_{i_m,k_{m+1}}^{j_m,k_m}\;,\\[.5cm]
\ds
\varphi(y,y)^{-1}\big(\overline{\Qib}(q^{-1}\,y)\big)_{i_N,\ldots,i_1}^{j_N,\ldots,j_1}&=&
\ds Z_0^{-1}\,\sum_{k_1=0}^\infty \omega_h^{k_1} \prod_{m=1}^N 
 \overline{{\mathfrak R}}(y|x_m^{},x'_m)_{i_m,k_{m+1}}^{j_m,k_m}\;,
\end{array}
\end{equation}
where $k_{N+1}=k_1$, are expressed through the 
$R$-matrices, 
\be\label{Qelem2}
\begin{array}{l}
\ds{\mathfrak R}(y'\,|\,x,x')_{i_1,i_2}^{j_1,j_2}=
\ds q^{i_2-j_2}\,\frac{(q^2;q^2)_{j_2}}{(q^2;q^2)_{i_2}}\,
\lim_{y\to0}\Big(\frac{y'}{y}\Big)^{j_1}\,{\cal R}(x,x',y,y')_{i_1,i_2}^{j_1,j_2}
\\[7mm]
\qquad=\ds\delta_{i_1+i_2,j_1+j_2} \left(q\frac{x'}{y'}\right)^{-i_1}
\sum_{n=0}^{\min(i_2,j_1)}
\left(\frac{x'}{x}\right)^{n-i_2} \frac{(q^2;q^2)_{i_1+i_2-n}}{(q^2;q^2)_{i_2-n}}
V_{x/y'}(n) V_{y'/x'}(j_1-n)
\end{array}
\ee
and 
\be\label{Qelem3}
\begin{array}{l}
\ds\overline{{\mathfrak R}}(y\,|\,x,x')_{i_1,i_2}^{j_1,j_2}=
\ds (-)^{i_2-j_2}\,q^{i_2^2-j_2^2}\,\frac{(q^2;q^2)_{j_2}}{(q^2;q^2)_{i_2}}\,
\lim_{y'\to0}\Big(\frac{y'}{y}\Big)^{i_1}\,{\cal R}(x,x',y,y')_{i_1,i_2}^{j_1,j_2}
\\[7mm]
\qquad=\ds\delta_{i_1+i_2,j_1+j_2}  \left(q\frac{x'}{y}\right)^{j_1}\sum_{n=0}^{\min(i_2,j_1)} 
\left(-q\frac{x'}{x}\right)^{-n} \frac{q^{n^2}}{(q^2;q^2)_{n} (q^2;q^2)_{j_1-n}} \frac{V_{y/x}(i_2-n)}{V_{y/x'}(i_1+i_2-n)}\;.
\end{array}
\ee
obtained by suitable limits of \eqref{R-def2}. 

At $x=q^{-1}x'$, which is $s=1/2$, one has
\be\label{Qelem4}
\begin{array}{ll}
\ds \mathfrak{R}_{0,i_2}^{0,j_2}\;=\;\delta_{i_2,j_2} \, q^{-j_2}\;,& 
\ds \mathfrak{R}_{0,i_2}^{1,j_2}\;=\; - \,\delta_{i_2,j_2+1}\, \frac{y'}{x'} \,q^{-j_2}\;,\\
[7mm]
\ds \mathfrak{R}_{1,i_2}^{0,j_2}\;=\; -\, \delta_{i_2+1,j_2} \, \frac{y'}{x'}\, [q^{j_2}]\;,& 
\ds \mathfrak{R}_{1,i_2}^{1,j_2}\;=\;\delta_{i_2,j_2} \, \frac{y'}{x'} \, [ \frac{x'}{y'} q^{j_2} ] \;,
\end{array}
\ee 
and
\be\label{Qelem5}
\begin{array}{ll}
\ds \overline{\mathfrak{R}}_{0,i_2}^{0,j_2}\;=\;\delta_{i_2,j_2} \, \frac{\ds  [ \frac{y}{x'} q^{j_2} ]}{\ds [\frac{y}{x'}]}\;,& 
\ds \overline{\mathfrak{R}}_{0,i_2}^{1,j_2}\;=\; - \,\delta_{i_2,j_2+1}\, \frac{\ds q^{j_2}}{\ds  [\frac{y}{x'}]}\;,\\
[7mm]
\ds \overline{\mathfrak{R}}_{1,i_2}^{0,j_2}\;=\; \delta_{i_2+1,j_2} \, \frac{\ds [q^{j_2}]}{\ds [\frac{y}{x'}]} \;,& 
\ds \overline{\mathfrak{R}}_{1,i_2}^{1,j_2}\;=\;-\, \delta_{i_2,j_2} \, \frac{\ds \frac{x'}{y} q^{j_2}}{\ds [\frac{y}{x'} ]} \;.
\end{array}
\ee 
In the short notations
\begin{equation}
\mathfrak{R}\;=\;\left(\begin{array}{cc}
\ds q^{H/2} & \ds \mu \mathcal{E}_{-} \\
[5mm]
\ds \mu \mathcal{E}_{+} & \ds q^{-H/2} - q^{-1}\mu^2 q^{H/2}
\end{array}\right)
\end{equation}
where
\begin{equation}
\mu\;=\;q^{1/2}\frac{y'}{x'}\;,\quad \langle i |q^{H/2}| j\rangle \;=\; \delta_{i,j} q^{-j}\;,
\end{equation}
and
\begin{equation}
\langle i | \mathcal{E}_{-} | j \rangle \;=\; -\delta_{i,j+1} q^{-j-1/2}\;,\quad 
\langle i | \mathcal{E}_{+} | j \rangle \;=\; \delta_{i+1,j} (q^{-j} - q^{j}) q^{-1/2}\;.
\end{equation}
Similarly,
\begin{equation}
\overline{\mathfrak{R}}\;=\;\frac{1}{1-q\mu^2} \,
\left(\begin{array}{cc}
\ds q^{-H/2} - q\mu^2 q^{H/2} & \ds \mu\mathcal{E}_{+}\\
[5mm]
\ds \mu\mathcal{E}_{-} & \ds q^{H/2}
\end{array}
\right)\;,
\end{equation}
where
\begin{equation}
\mu\;=\;q^{-1/2}\frac{y}{x'}\;,\quad \langle i | q^{H/2} | j \rangle \;=\; \delta_{i,j} q^j\;,
\end{equation}
and
\begin{equation}
\langle i | \mathcal{E}_{-} | j \rangle \;=\; \delta_{i+1,j} (q^{-j}-q^j) q^{1/2}\;,\quad 
\langle i | \mathcal{E}_{+} | j \rangle \;=\; -\delta_{i,j+1} q^{j+1/2}\;.
\end{equation}
In both cases
\begin{equation}
[H,\mathcal{E}_{\pm}]\;=\;\pm 2 \mathcal{E}_{\pm}\;,\quad
q\mathcal{E}_{+}\mathcal{E}_{-} - q^{-1} \mathcal{E}_{-}\mathcal{E}_{+}\;=\;q-q^{-1}\;.
\end{equation}

Using the correspondence \eqref{index} one can write the IRF version
of the above operators \eqref{Qelem} as
\be\label{Qcal-def}
\begin{array}{l}
\ds
\big({\mathbb
  Q}(
y')\big)_{a_{N+1},a_{N},\ldots,a_{1}}^{b_{N+1},\,b_{N},\ldots,b_{1}}=\\[.5cm]
\qquad=
\ds\prod_{i=1}^N \left(\frac{q\,x'_i}{y'}\right)^{a_i-a_{i+1}}\,
\left(\sum_{m_i} \left(\frac{x'_i}{x_i}\right)^{a_{i+1}-m_i}\,
\frac{(q^2;q^2)_{m_i-a_i}\,
V_{x_i/y'}(b_{i+1}-m_i)\, V_{y'/x'_i}(m_i-b_i)}
{(q^2;q^2)_{m_i-a_{i+1}}
}
\right)\,,\\[.7cm]
\ds\big(\overline{{\mathbb
  Q}}(q^{-1}\,
y)\big)_{a_{N+1},a_{N},\ldots,a_{1}}^{b_{N+1},\,b_{N},\ldots,b_{1}}=\\[.5cm]
\qquad=\ds
\prod_{i=1}^N \left(\frac{q\,x'_i}{y}\right)^{b_{i+1}-b_i}
\left(
 \sum_{n_i}
\left(-q\frac{x'_i}{x_i}\right)^{n_i-b_{i+1}}\,
\frac{q^{(b_{i+1}-n_i)^2}\,V_{y/x_i}(n_i-a_{i+1})}
{(q^2;q^2)_{b_{i+1}-n_i}\, (q^2;q^2)_{n_i-b_i}\, V_{y/x'_i}(n_i-a_i)}
\right)\,,
\end{array}
\end{equation}
where the summations run over the intervals 
\be\label{intervals}
\max(a_{i+1},b_i)\le
n_i\le b_{i+1}\,,\qquad \max(a_{i+1},b_i)\le
m_i\le b_{i+1}\,,
\ee
and 
\be
a_{N+1}=M+a_1\,,\qquad b_{N+1}=M+b_1\,,
\ee
with $M$ being
the conserved number of up spins \eqref{down2}. 
With the same correspondence between indices \eqref{index} the matrix
elements \eqref{Qelem} can now be written as
\be
\begin{array}{rcl}
\ds \big(\Qib( y')\big)_{i_N,\ldots,i_1}^{j_N,\ldots,j_1}&=&
\ds Z_0^{-1}\,\sum_{a_1\le b_1\le \infty}
\omega_h^{b_1-a_1} \,
\big({\mathbb
  Q}(
y')\big)_{a_{N+1},a_{N},\ldots,a_{1}}^{b_{N+1},\,b_{N},\ldots,b_{1}}\\[.7cm]
\ds \varphi(y,y)^{-1}\big(\overline{\Qib}(q^{-1}\, y)\big)_{i_N,\ldots,i_1}^{j_N,\ldots,j_1}&=&
\ds Z_0^{-1}\,\sum_{a_1\le b_1\le \infty}
\omega_h^{b_1-a_1} \,
\big(\overline{{\mathbb
  Q}}(q^{-1}\,
y )\big)_{a_{N+1},a_{N},\ldots,a_{1}}^{b_{N+1},\,b_{N},\ldots,b_{1}}
\end{array}
\ee
where the sum is taken over the spin $b_1$. Consider now the product
\be\label{QQ-prod}
\begin{array}{l}
\ds \varphi(y,y)^{-1}
\big(\overline{{\bf
  Q}}(q^{-1}\,
y)\,{{\bf
  Q}}(
y')\big)_{i_N,\ldots,i_1}^{j_N,\ldots,j_1}=\\[.5cm]
\qquad=\ds
Z_0^{-2}\sum_{a_1\le b_1\le \infty}\,
\omega_h^{b_1-a_1}\sum_{c_1,c_2,\ldots,c_N}
\overline{{\mathbb
  Q}}(q^{-1}\,
y)_{a_{N+1},\ldots,a_{1}}^{c_{N+1},\ldots,c_{1}}
\,
{\mathbb
  Q}(
y')_{c_{N+1},\ldots,c_{1}}^{b_{N+1},\ldots,b_{1}}\,.
\end{array}
\ee
Note that the elements \eqref{Qcal-def} have the same vanishing
conditions as those of the IRF transfer matrix
\eqref{nonzero}. Therefore, the summation over $c_1,c_2,\ldots,c_N$ is
restricted to a finite domain 
\be
c_{i+1}\ge c_i\,,\qquad a_i\le c_i\le b_i\,,\qquad i=1,2,\ldots,N\,, 
\ee
where $c_{N+1}=M+c_1$. Using \eqref{Qcal-def} one obtains 
\be
\mbox{\eqref{QQ-prod}}=
\ds
Z_0^{-2}\left(\frac{y'}{y}\right)^M\sum_{b_1=a_1}^\infty\,
\omega_h^{b_1-a_1}
\sum_{\{n_i,m_i\}}\left(\prod_{i=1}^N
\,\frac{V_{x_i/y'}(b_{i+1}-m_i)\, V_{y'/x_i'}(m_i-b_i)\,
  V_{y/x_i}(n_i-a_{i+1})}{V_{y/x'_i}(n_i-a_i)}G_i \right),
\label{QQ-prod2}
\ee
where 
\begin{equation}\label{G-def}
{G}_i\;=\;
\left(\frac{x'_{i-1}}{x_{i-1}}\right)^{(n_{i-1}-m_{i-1})} 
\ \sum_{c_i=n_{i-1}}^{\min(n_i,m_{i-1})}
\frac{q^{(n_{i-1}-c_i)^2}\,(-q)^{n_{i-1}-c_i}\, 
(q^2;q^2)_{m_i-c_i}}{(q^2;q^2)_{c_i-n_{i-1}} (q^2;q^2)_{n_i-c_i}
  (q^2;q^2)_{m_{i-1}-c_i}}\,,
\end{equation}
with the boundary conditions 
\be\label{n-bound}
n_0=n_N-M\,,\qquad m_0=m_N-M\,.
\ee
One can show that for all the non-vanishing contributions
to \eqref{QQ-prod2} the spins variables $\{n_i\}$ and $\{m_i\}$ are
ordered as
\be\label{m-order}
m_{i+1}\ge n_i\,,\qquad n_{i+1}\ge n_i\,,\qquad m_i\ge n_i\,.
\ee
The considerations are similar to those presented 
in Sect.~\ref{irf-form}, where the ordering of white spins in
Fig.~\ref{fig-lattice} was discussed. Note that spins $\{n_i\}$ and
$\{m_i\}$ correspond to black sites  in the same figure. With 
the substitution 
\be \label{abc}
\as=n_i-n_{i-1}\,,\qquad \bs=m_{i-1}-n_{i-1}\,,\qquad \cs=m_i-n_{i-1}\,,
\ee
the sum in \eqref{G-def} can be written as 
\begin{equation}\label{tozh1}
\Fs(\as,\bs,\cs)=\sum_{n=0}^{\min(\as,\bs)}  \frac{(-)^n
  q^{n(n-1)}\,
(q^2;q^2)_{\cs-n}}{(q^2;q^2)_n(q^2;q^2)_{\as-n}(q^2;q^2)_{\bs-n}}\;=\;  
q^{2\as\bs} 
\frac{(q^2;q^2)_{\cs-\as} (q^2;q^2)_{\cs-\bs}}{(q^2;q^2)_\as (q^2;q^2)_\bs
  (q^2;q^2)_{\cs-\as-\bs}}\;. 
\end{equation}
It is evaluated exactly by using the Heine's $q$-analog
of the Gauss summation formula for the hypergeometric function
\cite{Gasper:2004}.  
It is not difficult to see that
\begin{equation}\label{tozh}
\Fs(\as,\bs,\cs)\;=\;
\left\{\begin{array}{ll}
0\;,\qquad\qquad\quad &\textrm{if}\quad \max(\as,\bs)\leq \cs <
\as+\bs\,,\\[.1cm] 
\ds q^{2\as \bs}\;,\quad &\textrm{if}\quad \cs=\as+\bs\,,\\[.1cm] 
\textrm{non-zero}\;,\quad &\textrm{otherwise}\,.
\end{array}\right.
\end{equation}
Note that with the substitution \eqref{abc} 
the first inequality $\max(\as,\bs)\leq \cs$ in \eqref{tozh}
is satisfied for all sums in \eqref{G-def} by virtue of \eqref{m-order}.
Let us now examine the difference $\Delta=\cs-\as-\bs$. In terms of the
variables \eqref{abc} it reads 
\be
\Delta_i=(m_i-n_i)-(m_{i-1}-n_{i-1})\,.
\ee
Evidently, with the boundary condition \eqref{n-bound} one gets 
\be
\sum_{i=1}^N \, \Delta_i=0\,.
\ee
Therefore, either $\Delta_i=0$ for all $i=1,2,\ldots,N$ or at least
one $\Delta_i<0$ (for some particular value of $i$). In the latter
case the corresponding $G_i$ vanishes and so does the whole product in
\eqref{QQ-prod2}. Thus the non-vanishing contributions to
\eqref{QQ-prod2} could only arise when $\Delta_i=0$ for all
$i=1,2,\ldots,N$. This gives $N-1$ independent constraints between the
spins $\{n_i\}$ and $\{m_i\}$ which allow one to easily perform the
summation over all but one of the spins $\{m_i\}$. Then, instead of $m_1$ we choose the integer 
variable 
\be
\delta=m_1-n_1\,, \qquad 0\le \delta \le b_1-a_1,
\ee
whose allowed values are deternimed by the inequalities
\eqref{intervals} and \eqref{m-order}. Eq.\eqref{QQ-prod2} then
takes the form
\begin{equation}
\mbox{\eqref{QQ-prod}}=\ds
Z_0^{-2}\left(\frac{y'}{y}\right)^M\sum_{b_1=a_1}^\infty
\omega_h^{b_1-a_1} 
\ds\sum_{\delta=0}^{b_1-a_1} w^\delta \prod_{i=1}^N 
{\cal W}(b_{i+1},b_i,a_{i+1}+\delta,a_i+\delta\,|\,x_i,x'_i\,|\,y,y')
\label{QQ-prod3}
\end{equation}
where we used \eqref{W-def1} and
\be
Z_0=\sum_{k=0}^\infty\,(w\,\omega_h)^k\,,\qquad 
w\;=\; q^{2M}\, \prod_{i=1}^N\, \frac{x_i}{x_i'}\,.
\ee
Remembering the definitions \eqref{R-def}, \eqref{t-vert}, \eqref{t-norm}, 
using a simple identity 
\be
(y'/y)^M \varphi(y,y)=\varphi(y,y')\,\rho(y,y')\,,
\ee  
and rearanging the summations
one gets from \eqref{QQ-prod3},
\be\label{QQ-prod4}
\ds
\overline{{\bf
  Q}}(q^{-1}\,
y)\,{{\bf
  Q}}(
y')
=\ds
Z_0^{-1}\,\varphi(y,y')\,
\Tib^{(+)}(y,y')
\ee
as stated in \eqref{factor} in the main text.




\end{document}